\documentclass[reprint,aps,prf,superscriptaddress,onecolumn]{revtex4-2}

\usepackage{natbib}
\usepackage{amsmath}
\usepackage{graphicx}
\usepackage{amssymb}
\usepackage{bm}
\usepackage{color}
\usepackage{hyperref}
\usepackage[capitalize,nameinlink]{cleveref}

\usepackage{cool}% physics macros
\usepackage[caption=false]{subfig}
\usepackage{float}
\usepackage[]{booktabs}
\usepackage{array}

\newcommand{\Ca}{\mathrm{Ca}}
\newcommand{\De}{\mathrm{De}}
\newcommand{\gTrim}[1]{\includegraphics[width=0.30\linewidth,trim={0.5cm 0.5cm 0.8cm 0.2cm},clip]{#1}}
\newcommand{\fTrim}[1]{\includegraphics[width=0.40\linewidth,trim={0.5cm 0.5cm 0.8cm 0.2cm},clip]{#1}}

\usepackage{dcolumn}

\begin{document}

\title{\LARGE Dynamics of vesicles in large amplitude oscillatory extensional 
flow}

\author{Charlie Lin}
\thanks{These authors contributed equally}
\affiliation{
	Davidson School of Chemical Engineering, Purdue University, West Lafayette, 
	IN 47907
}

\author{Dinesh Kumar}
\thanks{These authors contributed equally}
\affiliation
{
	Department of Chemical and Biomolecular Engineering \\ University of 
	Illinois at Urbana-Champaign, Urbana, IL, 61801
}
\affiliation
{
	Beckman Institute for Advanced Science and Technology \\ University of 
	Illinois at Urbana-Champaign, Urbana, IL, 61801
}

\author{Channing M. Richter}
\affiliation{
	Department of Chemical and Biomedical Engineering, University of Illinois 
	at Urbana-Champaign, Urbana, IL 61801
}

\author{Shiyan Wang}
\affiliation{
	Davidson School of Chemical Engineering, Purdue University, West Lafayette, 
	IN 47907
}

\author{Charles M. Schroeder}
\email[To whom correspondence must be addressed: ]{cms@illinois.edu}
\affiliation
{
	Department of Chemical and Biomolecular Engineering \\ University of 
	Illinois at Urbana-Champaign, Urbana, IL, 61801
}
\affiliation
{
	Beckman Institute for Advanced Science and Technology \\ University of 
	Illinois at Urbana-Champaign, Urbana, IL, 61801
}
\affiliation
{
	Department of Materials Science and Engineering \\ University of Illinois 
	at Urbana-Champaign, Urbana, IL, 61801
}

\author{Vivek Narsimhan}
\email{vnarsim@purdue.edu}
\affiliation{
	Davidson School of Chemical Engineering, Purdue University, West Lafayette, 
	IN 47907
}

\date{\today}

\begin{abstract}

	Although the behavior of fluid-filled vesicles in steady flows has been extensively studied, far less is understood regarding the shape dynamics of vesicles in time-dependent oscillatory flows.
	Here, we investigate the nonlinear dynamics of vesicles in large amplitude oscillatory extensional (LAOE) flows using both experiments and boundary integral (BI) simulations.
	Our results characterize the transient membrane deformations, dynamical regimes, and stress response of vesicles in LAOE in terms of reduced volume (vesicle asphericity), capillary number ($\Ca$, dimensionless flow strength), and Deborah number ($\De$, dimensionless flow frequency).
	Results from single vesicle experiments are found to be in good agreement with BI simulations across a wide range of parameters.
	Our results reveal three distinct dynamical regimes based on vesicle deformation: pulsating, reorienting, and symmetrical regimes.
	We construct phase diagrams characterizing the transition of vesicle shapes between pulsating, reorienting, and symmetrical regimes within the two- dimensional Pipkin space defined by $\De$ and $\Ca$.
	Contrary to observations on clean Newtonian droplets, vesicles do not reach a maximum length twice per strain rate cycle in the reorienting and pulsating regimes.
	The distinct dynamics observed in each regime result from a competition between the flow frequency, flow time scale, and membrane deformation timescale.
	By calculating the particle stresslet, we quantify the nonlinear relationship between average vesicle stress and strain rate.
	Additionally, we present results on tubular vesicles that undergo shape transformation over several strain cycles.
	Broadly, our work provides new information regarding the transient dynamics of vesicles in time-dependent flows that directly informs bulk suspension rheology.

\end{abstract}

\maketitle

\section{Introduction}

In recent years, fluid-filled vesicles have been used in a wide array of 
technological applications ranging from food products to bioinspired 
microreactors, and reagent delivery applications in functional 
materials~\cite{huang2004acoustically}.
Moreover, giant vesicles are widely regarded as a model membrane system in 
various biophysical and biochemical 
processes~\cite{dimova2019giant,boal2002mechanics}.
In these applications, precise characterization of the membrane shape dynamics in response to a fluid flow is of fundamental importance.
Despite the increasing prevalence of vesicles in biophysics and materials 
science, we lack a complete understanding of how time-dependent flows influence 
the membrane shape dynamics and overall rheological response of vesicle 
suspensions~\cite{vlahovska2009vesicles,abreu2014fluid}.
Lipid vesicles consist of a small amount of fluid enclosed by a bilayer membrane of thickness $\approx$ 5 nm.
This molecularly thin membrane enables intriguing morphological dynamics for 
vesicles, including complex conformations in linear 
flows~\cite{Deschamps2009a,kumar2020conformational,Dahl2016,lin2019shape}, 
non-linear stretching behavior, and heterogeneous relaxation following 
deformation~\cite{zhou2011stretching,yu2015ellipsoidal,kumar2020double}.

Recent advances in experiments, computations, and theory have largely focused 
on vesicle dynamics in steady shear 
flows~\cite{Deschamps2009a,Kantsler2005,Kantsler2006,Kantsler2008,Zhao2011}.
These studies have revealed three different dynamical regimes: tumbling, trembling, and tank-treading.
Relevant research in shear flow includes investigation of the hydrodynamic lift 
of a single vesicle near a wall~\cite{Callens2008,Podgorski2011,Zhao2011}, pair 
interactions between two vesicles~\cite{Kantsler2008,Vitkova2008}, the 
amplification of thermal fluctuations in the transition regime between tumbling 
and tank treading~\cite{Zabusky2011,Levant2012,Abreu2013}, and characterization 
of tank-treading, vacillating-breathing (trembling), and tumbling motion with 
increasing viscosity ratio between the interior and the exterior of the 
vesicle~\cite{Mader2006,Vlahovska2006, Deschamps2009a}.
Knowledge of single vesicle dynamics has been essential for interpreting the bulk rheological response for dilute vesicle suspensions.
For instance, it is now known that the tank-treading to tumbling behavior of 
vesicles directly affects the bulk viscosity of the suspension, where tumbling 
results in a higher bulk viscosity with the minimum bulk viscosity occurring at 
the tank-treading to tumbling transition~\cite{vlahovska2009vesicles}.

Compared to the vast body of experiments in shear flows, vesicle dynamics in hyperbolic flows even for the canonical case of steady elongational flow are more challenging to understand.
In extensional flow, fluid elements separate exponentially in 
time~\cite{leal1992laminar}, and it is generally not possible to observe a 
single vesicle in flow for long periods of time in the absence of feedback 
controllers.
Automation in flow control techniques using sophisticated feedback algorithms 
has recently enabled the precise characterization of vesicle dynamics in 
elongational 
flows~\cite{Shenoy2016,shenoy2019flow,kumar2020automation,Kumar2019,shenoy2015characterizing}.
In a steady extensional flow, it is known that highly deflated tubular vesicles 
undergo a conformation change to a symmetric dumbbell 
shape~\cite{Kantsler2008a,Narsimhan2014,Narsimhan2015,kumar2020conformational} 
while moderately deflated vesicles transition to an asymmetric dumbbell 
shape~\cite{Dahl2016,kumar2020conformational}.
Precise control over the center-of-mass position of single vesicles led to 
detailed studies of the transient and steady-state stretching dynamics of 
membranes~\cite{kumar2020conformational}, and direct observation of the 
double-mode relaxation following high deformation~\cite{kumar2020double}.
Prior work in unsteady flows has been limited to a one-time reversal of 
elongational flow and reported membrane wrinkling shapes for quasi-spherical 
vesicles~\cite{Kantsler2007}.

Extensional flows are commonly encountered in microfluidic devices that utilise contractions or expansions, porous media, and other complex channel geometries.
Moreover, \textit{in vivo} capillaries and complex microfluidic devices that have many bifurcations and sharp directional changes routinely encounter time-dependent pulsatile flows.
The biomedical community has created several biomimetic capillary designs that 
contain several rows of bifurcations and contractions with small angle zigzags 
in between, resulting in improved flow control and lower fluid flow 
resistance~\cite{Domachuk2010, Lim2003}.
In general, elastic particles traversing through these fluidic systems experience spatially dependent external flows and will not reach a steady-state conformation.
From this view, there is a need for comprehensive studies on how microscopic stretching and compression of vesicles in complex, time-dependent oscillatory flows will affect their shape and bulk rheology.

Recently, the shape dynamics of elastic capsules were studied numerically in 
large amplitude oscillatory extensional (LAOE) flow~\cite{bryngelson2019non}.
However, the non-equilibrium stretching and compression dynamics of lipid vesicles in LAOE flows is largely unexplored.
Vesicle dynamics are strongly governed by membrane bending elasticity; therefore, we anticipate that vesicles will exhibit qualitatively different behavior than capsules in time-dependent extensional flow.
In this paper, we study the dynamics of single vesicles in LAOE using a combination of microfluidic experiments and boundary integral (BI) simulations.
LAOE experiments are performed using the Stokes 
trap~\cite{Shenoy2016,shenoy2019flow,kumar2020automation,Kumar2019}, which is a 
new method for controlling the center-of-mass position, orientation and 
trajectories of freely suspended single and multiple vesicles using only fluid 
flow.
We find that single vesicles experience periodic cycles of compression and extension in LAOE with membrane dynamics governed by the dimensionless flow strength Capillary number (Ca), reduced volume (measure of vesicle asphericity, $\nu$) and flow frequency Deborah number (De).
Experimental results are compared to BI simulations without thermal fluctuations, and our results show that BI simulations accurately capture the dynamics of single quasi-spherical vesicles over a wide range of parameters.
In addition, we identify three distinct dynamical regimes for vesicle dynamics, including the pulsating, reorienting, and symmetrical regimes, based on the amount of deformation occurring in each half cycle of the LAOE flow.
The qualitatively different dynamics observed in each regime results due to a competition between the flow frequency, flow time scale, and membrane deformation timescale.
We further construct precise phase diagrams characterizing the transition of vesicle shapes between pulsating, reorienting, symmetrical regimes.
We find that the relationship between average vesicle stress and strain rate is nonlinear, which is discussed in the context of bulk suspension rheology.
Finally, we present results on the shape dynamics of long tubular vesicles in LAOE which exhibit markedly different behavior in flow compared to their quasi-spherical analogues.
Taken together, our results provide new insights into the direct observation of membrane dynamics during time-dependent oscillatory flows, which opens new avenues for understanding bulk suspension rheology in unsteady flows.

\section{Methods}

\subsection{Vesicle preparation}

\begin{figure}
	\centering
	\includegraphics[width=0.6\linewidth]{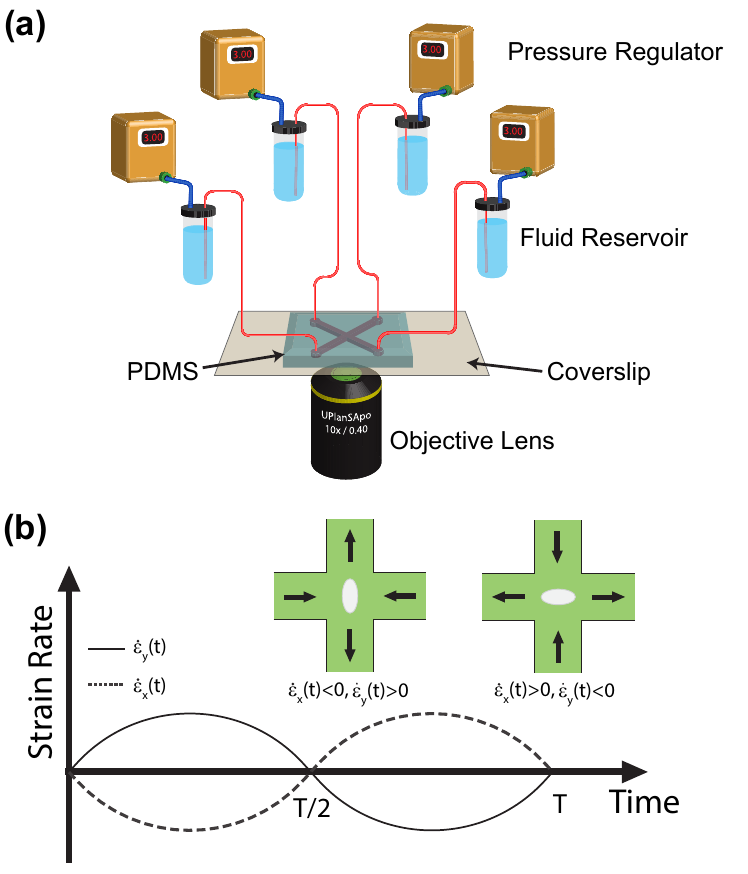}
	\caption{%
		Stokes trap for studying vesicle dynamics in large amplitude oscillatory extensional (LAOE) flow.
		(a) Schematic of the experimental setup used to generate planar extensional flow.
		Inlet/outlet channels in the microfluidic device are connected to fluidic reservoirs containing the vesicle suspension and pressurized by regulators controlled by a custom LabVIEW program, thereby generating pressure-driven flow in the cross-slot.
		(b) Schematic of the sinusoidal strain rate input function for one full cycle.
		Inset: schematics showing the oscillatory extensional flow profile in the microfluidic cross-slot device during the first half ($0<t<T/2$), and second half period ($T/2<t<T$) of the cycle.
	}%
	\label{fig:stokes_trap}
\end{figure}

A mixture of 1,2-dioleoyl-sn-glycero-3-phosphocholine (DOPC, Avanti Polar Lipids) and 0.12 mol\% of 1,2-dioleoyl-sn-glycero-3-phosphoethanolamine-N-(lissamine rhodamine B sulfonyl) (DOPE-Rh, Avanti Polar Lipids) is used to generate giant unilamellar vesicles (GUVs) with the electroformation process described by~\citet{Angelova1992}.
For electroformation of GUVs, a stock lipid solution in chloroform is prepared with 25 mg/mL DOPC and 0.04 mg/mL DOPE-Rh for fluorescent imaging.
Next, 10 $\mu$L of the lipid solution in chloroform is spread on a conductive indium tin oxide (ITO) coated glass slide (resistance 5 $\Omega$, 25$\times 50 \times$1.1 mm, Delta Technologies) and dried under vacuum overnight.
The pair of ITO slides are sandwiched together using a 1.5 mm Teflon spacer, forming a chamber with a volume of $\approx$ 2.4 mL and coupled to a function generator (Agilent 33220 A).
The electroformation chamber is filled with a mixture of 100 mM sucrose solution (Sigma-Aldrich), and glycerol-water is added to achieve a total viscosity of 0.030 Pa-s measured using a benchtop viscometer (Brookfield) at 22$^\circ$C.
An alternating current (AC) electric field of 2 V/mm is then applied at 10 Hz for 120 min at room temperature (22$^\circ$C).
Under these conditions, DOPC lipid remains in the fluid 
phase~\cite{Kantsler2005}.
Most of the vesicles prepared by this method are quasi-spherical and unilamellar with few defects in the size range of 5--25 $\mu m$ in radius.

\subsection{Stokes trap for large amplitude oscillatory extension}\label{sec:Stokes_trap}

It is challenging to observe vesicle dynamics in time-dependent extensional flow for long periods of time while simultaneously imposing precisely controlled flow rates.
To achieve this, we used the Stokes trap~\cite{Shenoy2016,kumar2020automation} 
to precisely position the center-of-mass of single vesicles near the center of 
a cross-slot microfluidic device for long times using model predictive control 
(\cref{fig:stokes_trap}a).
Briefly, the centroid of a single vesicle is determined in real-time using image processing and fluorescence microscopy and is communicated to the controller.
The controller determines the optimal flow rates through four-channels of the device to maintain a fixed vesicle position with desired strain rate.
The flow rates are then applied through four independent pressure regulators (Elveflow). During this process, the device operates at a net positive pressure so that each of the four ports can act as inlet or outlet.
This whole procedure requires $\approx$ 30 ms in a single cycle, as previously 
described~\cite{Shenoy2016,zhou2016transient,zhou2016single}.
In this work, a sinusoidal strain rate input is imposed (\cref{fig:stokes_trap}b) while simultaneously trapping a single vesicle such that:
\begin{equation*}
	\begin{aligned}
		\dot{\epsilon}_x(t)=-\dot{\epsilon}_0 \sin\left(\frac{2\pi}{T}t\right) \\
		\dot{\epsilon}_y(t)=\dot{\epsilon}_0 \sin\left(\frac{2\pi}{T}t\right)
		\label{eq:strain_rate}
	\end{aligned}
\end{equation*}
where $T$ is the period of the sinusoidal cycle and $\dot{\epsilon}_0$ is the maximum strain rate in one cycle.
During the first half-cycle for $0<t<T/2$, the x-axis is the compressional axis and y-axis is the elongational axis ($\dot{\epsilon}_x(t)<0,\dot{\epsilon}_y(t)>0$), and the fluid is delivered from the two horizontal inlets of the microfluidic device by the pressure regulators (\cref{fig:stokes_trap}b).
During the second half-cycle for $T/2<t<T$, the direction of flow reverses, and fluid is delivered by the two vertical ports in the cross-slot device as shown in \cref{fig:stokes_trap}b.
We note that during vesicle trapping, the correctional pressure required for 
controlling the vesicle's position is small compared to the magnitude of the 
base pressure used to generate the oscillatory extensional flow 
~\cite{Shenoy2016}.
Thus, the strain rate is well defined during the LAOE cycle, which is determined
as a function of the input pressure using particle tracking velocimetry (PTV) 
as previously described ~\cite{kumar2020conformational}.
We also determined the characteristic response time for actuating fluid flow in the microfluidic device in response to a step change in pressure.
For an extreme change in pressure from 0 to 4 psi (strain rate jump from 0 to $\sim$ 30 $s^{-1}$), the rise time and settling time are $\sim$20 ms and $\sim$300 ms respectively(Fig. S1).
However, the maximum value of pressure used in our experiments is 0.4 psi, which is continuously varied with small incremental changes during the LAOE cycle, for which we generally expect much smaller characteristic response times.
Nevertheless, the lowest cycle time $T$ in our experiments is 2 seconds which is much larger than the maximum characteristic response time for actuating flow in the device corresponding to a step input pressure.

For all experiments, single vesicles are first trapped and imaged for 10--30 s under zero flow conditions to allow for equilibration, followed by LAOE flow for at least 2 strain rate cycles.
During the equilibration step, the vesicle reduced volume $\nu$ and equivalent 
radius $a$ are determined, as previously 
described~\cite{Dahl2016,kumar2020conformational}.
Reduced volume $\nu$ is a dimensionless quantity that measures the amount of osmotic deflation, and is described as:

	\begin{equation}
		\nu=\frac{3V\sqrt{4\pi}}{A^{3/2}}
	\end{equation}

where $V$ and $A$ are the vesicle volume and surface area, respectively.
The equivalent radius $a$ of the vesicle is obtained as $a=\sqrt{A/4\pi}$.
Specifically, $\nu$ is a measure of vesicle asphericity such that $\nu=1$ represents a perfectly spherical shape.
For the experiments in this paper, the typical range of reduced volume is 
$0.75<\nu<1$.

The maximum strain rate $\dot{\epsilon}_0$ experienced by a vesicle in a half-cycle is non-dimensionalized to define a capillary number $\Ca = \mu_{out}\dot{\epsilon}_0 a^3 / \kappa$ where $\mu_{out}$ is the suspending medium viscosity, $a$ is the equivalent vesicle radius, and $\kappa$ is the membrane bending modulus.
Prior to vesicle experiments in LAOE flow, we determined the average bending 
modulus of nearly spherical vesicles to be $\kappa=(22.3 \pm 0.5) k_B T$ using 
contour fluctuation spectroscopy \cite{kumar2020conformational}.
Similarly, the cycle period is rendered dimensionless by the bending time scale to define the Deborah number $\De = \mu_{out} a^3 / \kappa T$.
Single vesicle experiments are generally performed in the range $10 < \Ca < 1000$ and $0.1 < \De <100$ by adjusting the input pressures and strain rate cycle periods.
Only vesicles near the center plane of the microchannel (with respect to the z-direction) are considered during experiments.
Single vesicle trajectories are analyzed using a custom MATLAB program that allows for determination of the vesicle deformation parameter in flow.

\subsection{Numerical methods}

\subsubsection{Governing equations and non-dimensionalization}
The system is modeled as a droplet surrounded by a two-dimensional incompressible membrane with a bending resistance.
At the length scale of a GUV ($a \approx$ 10 $\mu$m) with a strain rate at $\dot{\epsilon} \approx$ 1 $s^{-1}$ the Reynolds number is $\mathrm{Re} = \dot{\epsilon} \rho a^2 / \mu \approx 10^{-4}$, allowing us to model the inner and outer velocity fields using the Stokes equations.
Due to the nature of the time-dependent flow, it is also important to check the Womersley number to assess whether the time-dependent Stokes equations are required.
At a flow frequency of $\omega$ = 10 $s^{-1}$, the Womersley number is $\alpha$ = $\sqrt{\omega \rho a^2 / \mu} \approx 0.03$.
In this work, the flow frequencies are $\omega <$ 10 $s^{-1}$, therefore the time-dependent Stokes equations are not necessary.
The Stokes equations are:
\begin{equation}
	\nabla \cdot \bm{u} = 0, \quad	\nabla p = \mu \nabla^2 \bm{u}.
\end{equation}
where $\bm{u}$ is fluid velocity, $p$ is the pressure, and $\mu$ is the fluid viscosity ($\mu_{in}$ for the inner fluid and $\mu_{out}$ for the outer fluid).
The system is subject to continuity of velocity across the interface and a traction balance across the phospholipid bilayer.
The short timescales and low deformation rates used in previous studies makes 
membrane dilatation negligible~\cite{Rawicz2000, Henriksen2004a}.
Vesicles are also known to have negligible shear rigidity as they do not have a cytoskeletal network or an actin cortex.
We therefore use the Helfrich model~\cite{Helfrich1973} for the membrane:
\begin{equation}
	\mathbb{H} = \oint \frac{\kappa}{2} {(2H)}^2 dA + \oint \sigma dA.
	\label{eq:helfrich}
\end{equation}

In \cref{eq:helfrich}, $\mathbb{H}$ represents the elastic energy of the vesicle membrane, $\kappa$ is the membrane bending modulus, $H$ is the mean curvature, and $\sigma$ is the surface tension.
The surface tension is a spatially varying Lagrange multiplier that ensures local area conservation.
The surface tension enforces  $\nabla_s \cdot \bm{u} = 0$ on the interface, where $\nabla_s = (\bm{I} - \bm{nn}) \cdot \nabla$.
We note that the original Helfrich model includes spontaneous curvature, a parameter to describe a membrane's curvature preference when the sides of the bilayer are chemically different.
Although biological vesicles may have multiple lipid components or chemical 
differences between the inner and outer fluids~\cite{Dobereiner1999, 
Deuling1976, Bagatolli2009}, our experiments focus on simple vesicles with only 
a viscosity difference between the inner and outer fluids, prompting a 
negligible spontaneous curvature.
We further neglect contributions from thermal fluctuations, membrane viscosity, 
and bilayer friction~\cite{Seifert1997, Noguchi2005}.

The force balance at the membrane surface is:

	\begin{align}
		[[\bm{f}]] = [[\bm{T} \cdot \bm{n}]] = \bm{f_t} + \bm{f_b}\\
		\bm{f_t} = (2H\sigma \bm{n} - \nabla_{s} \sigma)\\
		\bm{f_b} = \kappa (4KH - 4H^3 - 2\nabla_{s}^{2} H)\bm{n}
		\label{eq:force_bal}
	\end{align}

where $[[\bm{f}]]$ is the jump in viscous traction across the interface which can be decomposed to the bending ($\bm{f_b}$) and tension ($\bm{f_t}$) contributions, $\boldsymbol{n}$ is the outward-pointing unit normal vector, and $K$ is the Gaussian curvature of the interface.
The mean curvature $H$ is defined to be one for the unit sphere.

The vesicle is placed in a time-dependent extensional flow field described by $\bm{u}^{\infty} = \nabla \bm{u}^{\infty} \cdot \bm{x}$ and defined as:

	\begin{equation*}
		\nabla\pmb{u}^{\infty} = \dot{\epsilon}_0
		\begin{bmatrix}
			-\sin( 2 \pi \omega t) & 0 & 0 \\
			0 & \sin(2 \pi \omega t) & 0 \\
			0 & 0 & 0
		\end{bmatrix}
		\label{eq:grad_U_LAOE}
	\end{equation*}

where ${\omega}$ is the frequency of the oscillatory flow and $\dot{\epsilon}_0$ is the maximum strain rate.

The membrane area ($A$) is maintained constant by the incompressibility constraint while the low permeability of the membrane allows us to assume that the volume ($V$) of the vesicle is constant during the timescale of experiments (minutes).
Therefore, we non-dimensionalize distances by the equivalent radius $a = \sqrt[]{A/(4\pi)}$, time scales by $\kappa / a^3 \mu_{out}$, velocities by $\kappa / a^2 \mu_{out}$, stresses by $\kappa / a^3$, and surface tensions by $\kappa / a^2$.
We obtain four relevant dimensionless groups from the non-dimensionalization:
\begin{equation*}
	\begin{aligned}
	&	\Ca \equiv \frac{\mu_{out}\dot{\epsilon}_0 a^3}{\kappa}, \quad
	\De \equiv \frac{\omega a^3 \mu_{out}}{\kappa} \nonumber \\
	&	\lambda \equiv \frac{\mu_{in}}{\mu_{out}}, \quad
	\nu \equiv \frac{3V}{4\pi a^3}
	\label{eq:non_dim_param}
	\end{aligned}
\end{equation*}
These parameters were previously described in \cref{sec:Stokes_trap} and are elaborated upon here.
The base capillary number $(\Ca)$ compares the viscous stress to the bending stress and corresponds to the non-dimensionalized, maximum extension rate experienced by the vesicle during the flow cycle.
$\De$ is the flow frequency non-dimensionalized by the bending timescale.
When $\De \gg 1$, the fluid flow will have a short cycle time compared to the membrane's bending time.
The viscosity ratio ($\lambda$) is the ratio of inner and outer fluid viscosities.
Cellular systems such as red blood cells (RBCs) commonly have a more viscous inner fluid, and this parameter can be tuned to more closely model the system of choice.
The reduced volume ($\nu$) is a measure of the asphericity of the vesicle, corresponding to its osmotic deflation.
For example, a reduced volume of $\nu = 1$ corresponds to a perfectly spherical vesicle shape, while a value of $\nu = 0.2$ would be highly deflated.
One can experimentally alter the reduced volume of a vesicle by introducing an osmotic pressure difference between the inner and outer membranes, for example by adding sucrose to the outer fluid.

Applying this non-dimensionalization, the external velocity gradient becomes:
\begin{align}
	\nabla\pmb{u}^{\infty} = \Ca
	\begin{bmatrix}
		-\sin( 2 \pi \De \: t) & 0 & 0 \\
		0 & \sin(2 \pi \De \: t) & 0 \\
		0 & 0 & 0
	\end{bmatrix}
	\label{eq:grad_U_LAOE_non_dim}
\end{align}
where all parameters are assumed to be non-dimensional from this point forward.

\subsubsection{Boundary integral formulation}

The Stokes flow assumption enables the use of the boundary integral (Green's function) formulation to simulate vesicle shape dynamics.
The Stokes equations are recast into a boundary integral form:
\begin{align}
  & \frac{1+\lambda}{2} u_j(\bm{x_0}) = u_j^{\infty}(\bm{x_0}) - \frac{1}{8\pi}\int_S G_{ij}(\bm{x},\bm{x_0})[[f_i]](\bm{x})dA(\bm{x}) \nonumber \\
  & + \frac{1 - \lambda}{8\pi}\int_S T_{ijk}(\bm{x}, \bm{x_0}) u_i(\bm{x})n_k(\bm{x})dA(\bm{x}) \label{eq:b_i}
\end{align}
where $u_i^{\infty}$ is the external velocity field, $\bm{x}_0$ is the singularity point, and $[[f_i]]$ is the jump in viscous traction across the interface, given in \cref{eq:force_bal}.
The kernels $G_{ij}(\bm{x},\bm{x_0})$ and $T_{ijk}(\bm{x}, \bm{x_0})$ are the Stokeslet (point force) and stresslet (point dipole) solutions to Stokes flow:
\begin{equation}
	G_{ij}(\bm{x} , \bm{x}_0) = \frac{\delta_{ij}}{r} + \frac{\tilde{x}_i \tilde{x}_j}{r^3}
\end{equation}
\begin{equation}
	T_{ijk}(\bm{x} , \bm{x}_0) = -6 \frac{\tilde{x}_i \tilde{x}_j \tilde{x}_k}{r^5}
\end{equation}
where $\tilde{\bm{x}} = \bm{x} - \bm{x}_0$ and $r = |\tilde{\bm{x}}|$.
Repeated indices are assumed to be summed in the above equations.
These equations are also subject to the membrane incompressibility constraint:
\begin{equation}
	\nabla_s \cdot \bm{u} = 0
\end{equation}

Implementation details for the simulations are similar to prior 
work~\cite{lin2019shape}.
Here, we highlight a few key differences.
In this paper, we start with an icosahedron and subdivide the mesh into 1024 elements for a quasi-spherical vesicle; 5024 elements for the tubular vesicles.
We then use a scaling transformation to deform the mesh into a prolate spheroid with the desired reduced volume $\nu$, followed by relaxing the mesh to its equilibrium (no flow) configuration.
In this way, the vesicle has a prolate spheroid-like shape at the start of any cycle.
It is possible to start with an oblate spheroid or any arbitrary ellipsoid-like 
shape, but it has been shown that the global minimum energy state for a vesicle 
with reduced volume greater than 0.652 is of the prolate shape 
family\cite{Seifert1997}.
Choosing an alternative starting shape does not affect the steady limit cycle behavior.
Vesicle dynamics are simulated in oscillatory flow with a timestep of $10^{-3}$ strain units.

The majority of the analysis in this study is focused on vesicle behavior that has reached a steady limit cycle in time-dependent flow, such that the dynamics are the same regardless of the number of additional strain rate cycles.
The startup dynamics have been simulated but are not elaborated on in this paper.
We simulate vesicles of reduced volumes between $0.60 < \nu < 0.90$ and viscosity ratios $\lambda = 0.1, 1.0$, and $10$ for flows with capillary numbers $0.1 < \Ca < 80$ and Deborah numbers $0.1 < \De < 10$.
Significantly higher capillary numbers ($\Ca \gtrapprox 200$) become numerically intractable as the timestep needed for convergence in our implementation becomes prohibitively small.
Higher and lower $\De$ can be simulated, but the current range of values is sufficient for comparison to the majority of experimental conditions for GUVs in microfluidic devices.

We define the parameter:
\begin{align}
	\Ca_x(t) \equiv -\Ca \Sin{2 \pi \De \cdot t}
	\label{eq:cax}
\end{align}
which represents the time-dependent capillary number in the x-direction.
This will be the measure used for the instantaneous strain rate.
We also define a deformation parameter:
\begin{align}
	D \equiv \frac{l_x - l_y}{l_x + l_y}
	\label{eq:deform_param}
\end{align}
where $l_x$ and $l_y$ are the x- and y-axis lengths of the vesicle respectively.
The deformation parameter ($D$) provides a measure of vesicle shape distortion.
For $D$ values near zero, the vesicle shape projected in the x-y plane will be circular.
Positive values of $D\approx$ 0.50 correspond to prolate spheroid like shapes along the x-axis, while negative values correspond to the same shapes along the y-axis.

\section{Results and Discussion}

\subsection{Dynamical regimes\label{sec:dynamical_regimes}}

\begin{figure}
	\centering
	\textbf{Symmetrical}

	\vspace{-3mm}

	\captionsetup{width=0.3\linewidth}
	\subfloat[Ca = 28.8, De = 6.40, $\nu$ = 0.85\label{fig:deform_comp_sym}]{
		\gTrim{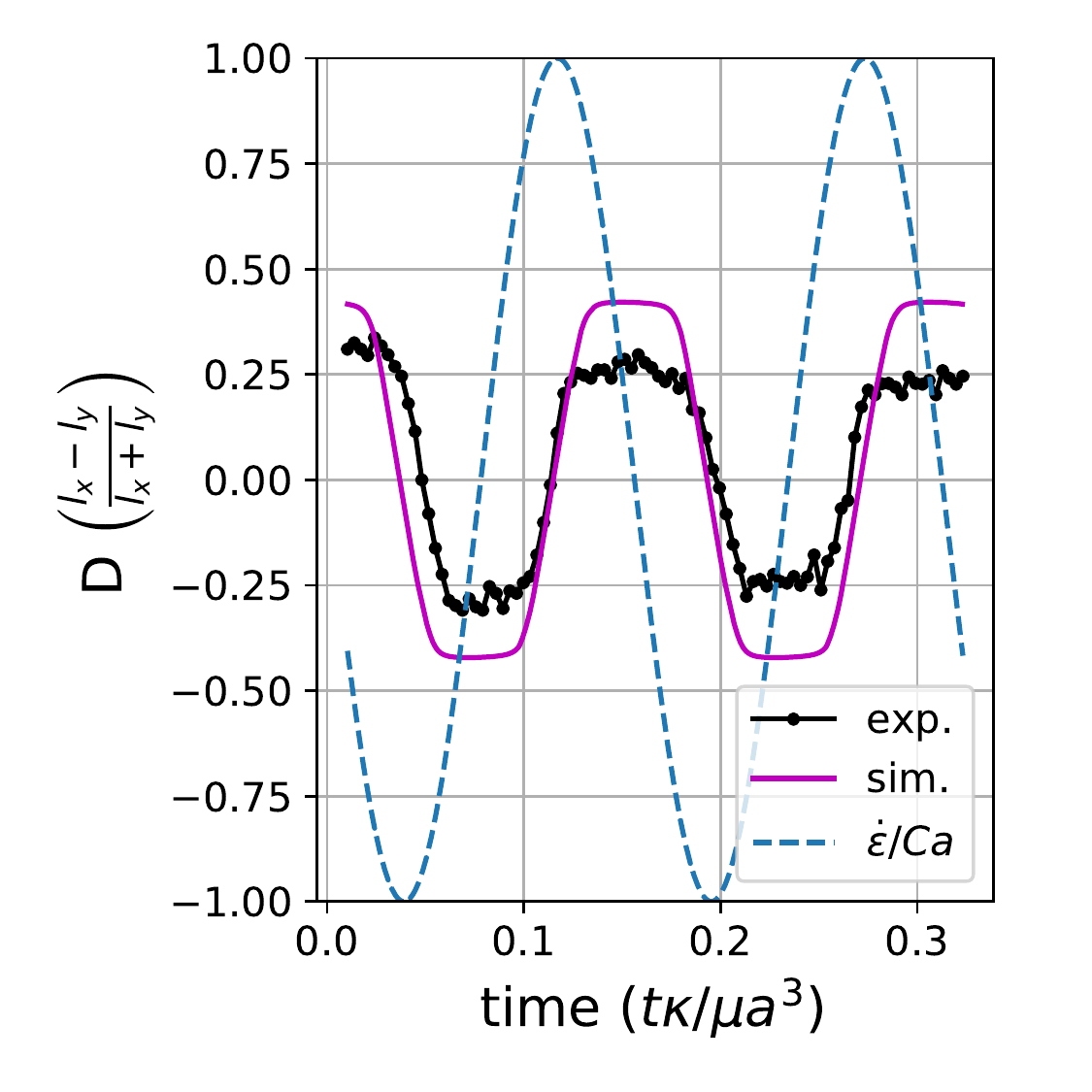}
	}
	\subfloat[Ca = 18.2, De = 3, $\nu$ = 0.88]{
		\gTrim{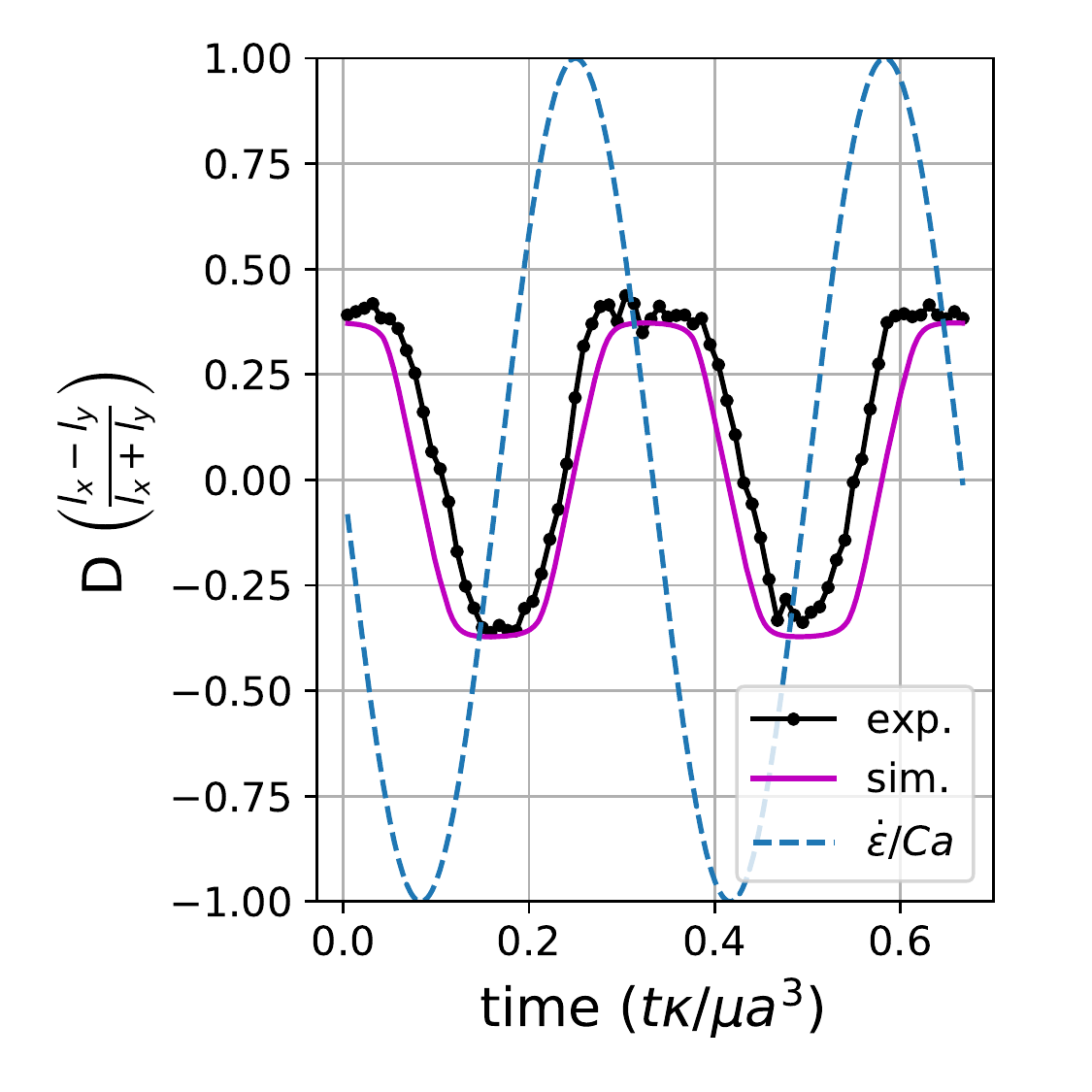}
	}
	\subfloat[Ca = 17.9, De = 6, $\nu$ = 0.91]{
		\gTrim{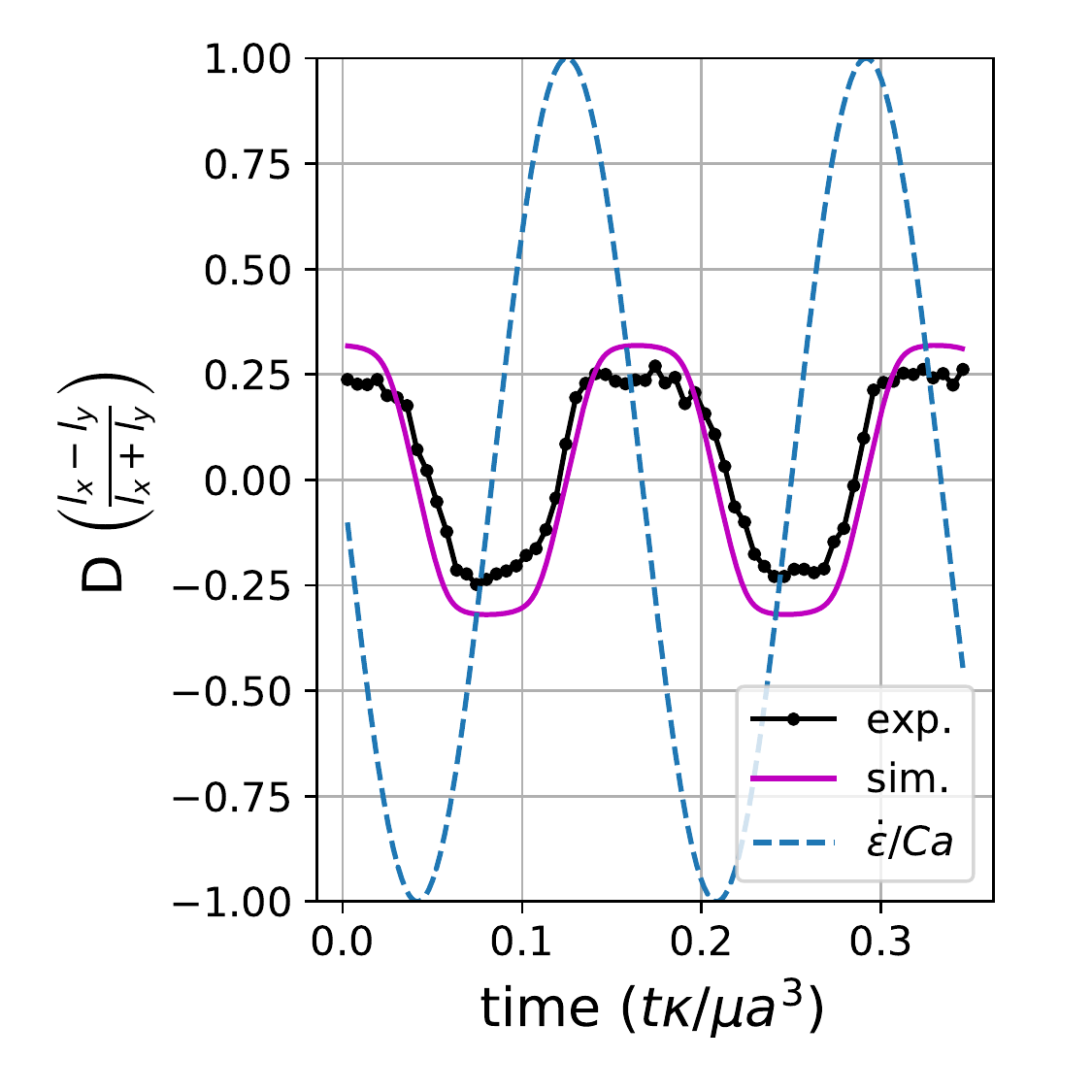}
	}\\

	\hrule
	\vspace{2mm}

	\textbf{Reorienting}

	\vspace{-3mm}

	\subfloat[Ca = 28.8, De = 12, $\nu$ = 0.85\label{fig:deform_comp_reori}]{
		\gTrim{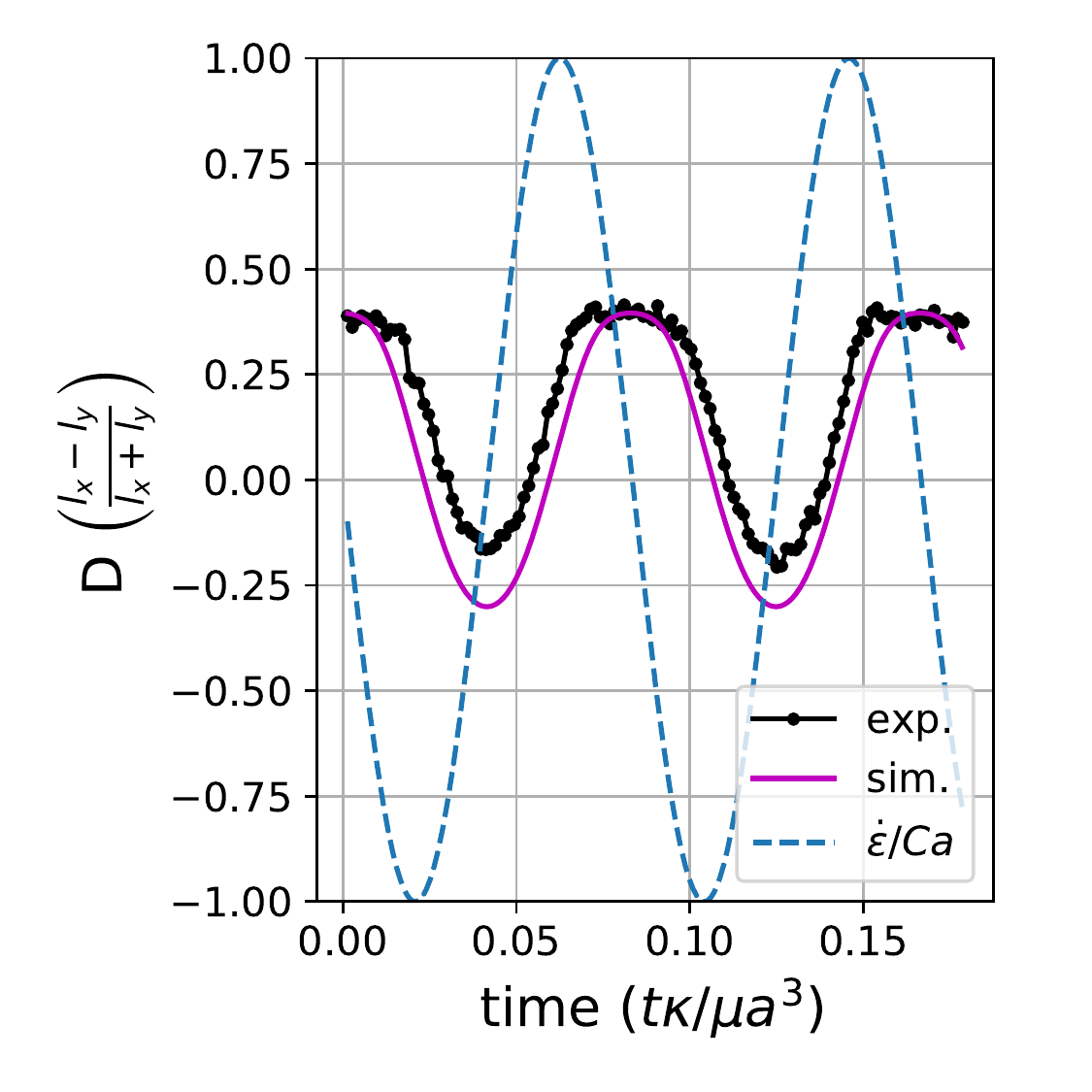}
	}
	\subfloat[Ca = 18.2, De = 4.5, $\nu$ = 0.88]{
		\gTrim{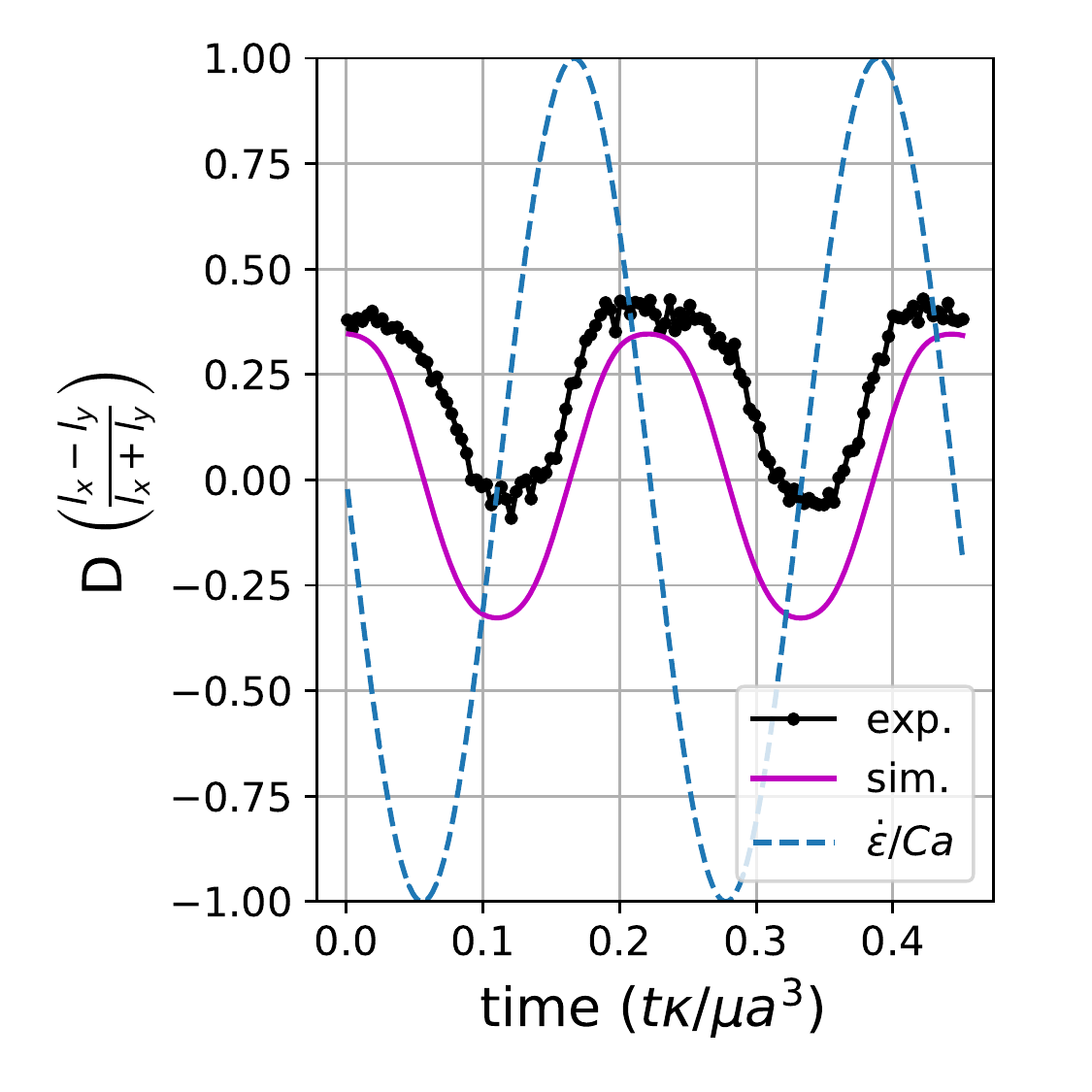}
	}
	\subfloat[Ca = 17.9, De = 14.9, $\nu$ = 0.91]{
		\gTrim{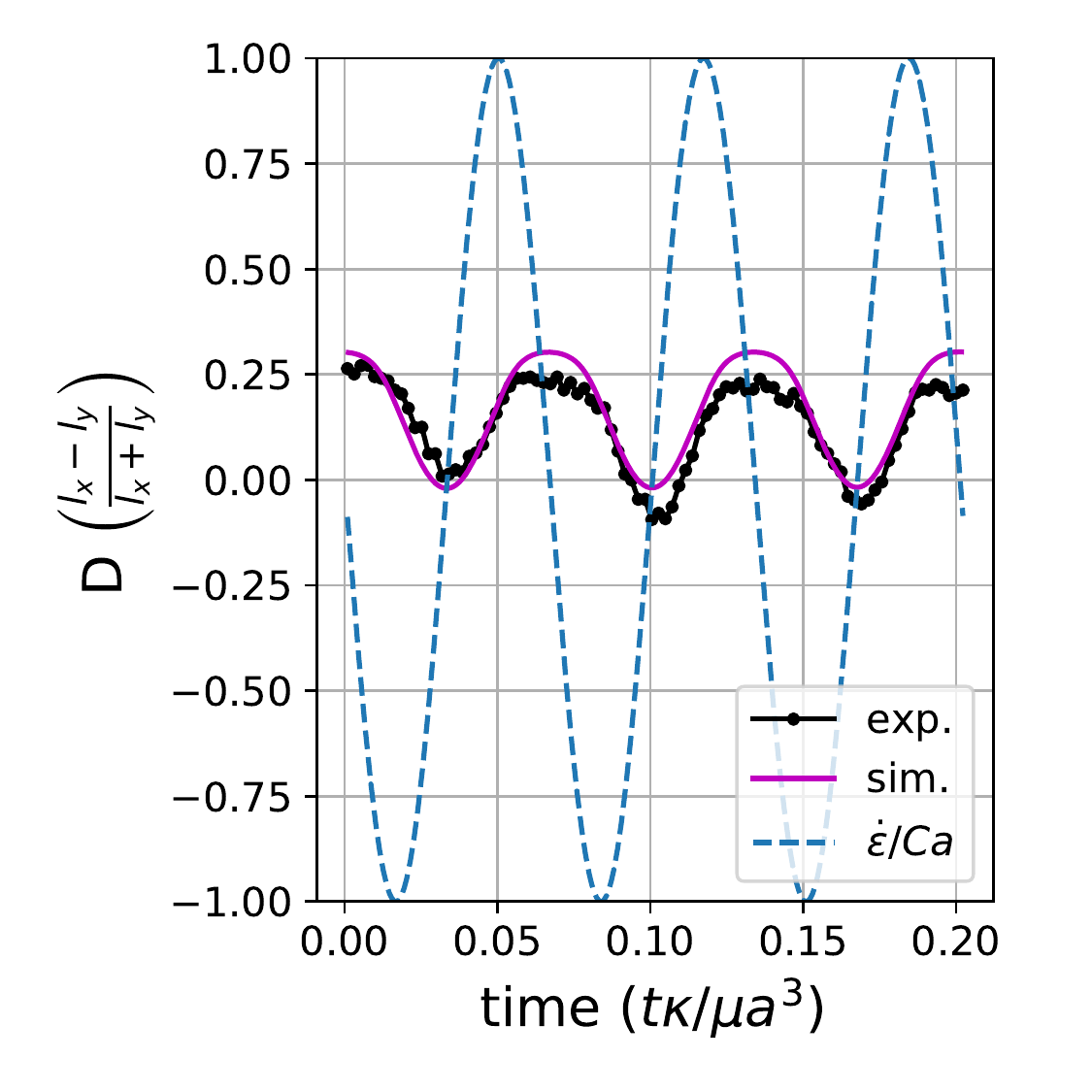}
	}\\

	\hrule
	\vspace{2mm}

	\textbf{Pulsating}

	\vspace{-3mm}

	\subfloat[Ca = 28.8, De = 48, $\nu$ = 0.85\label{fig:deform_comp_puls}]{
		\gTrim{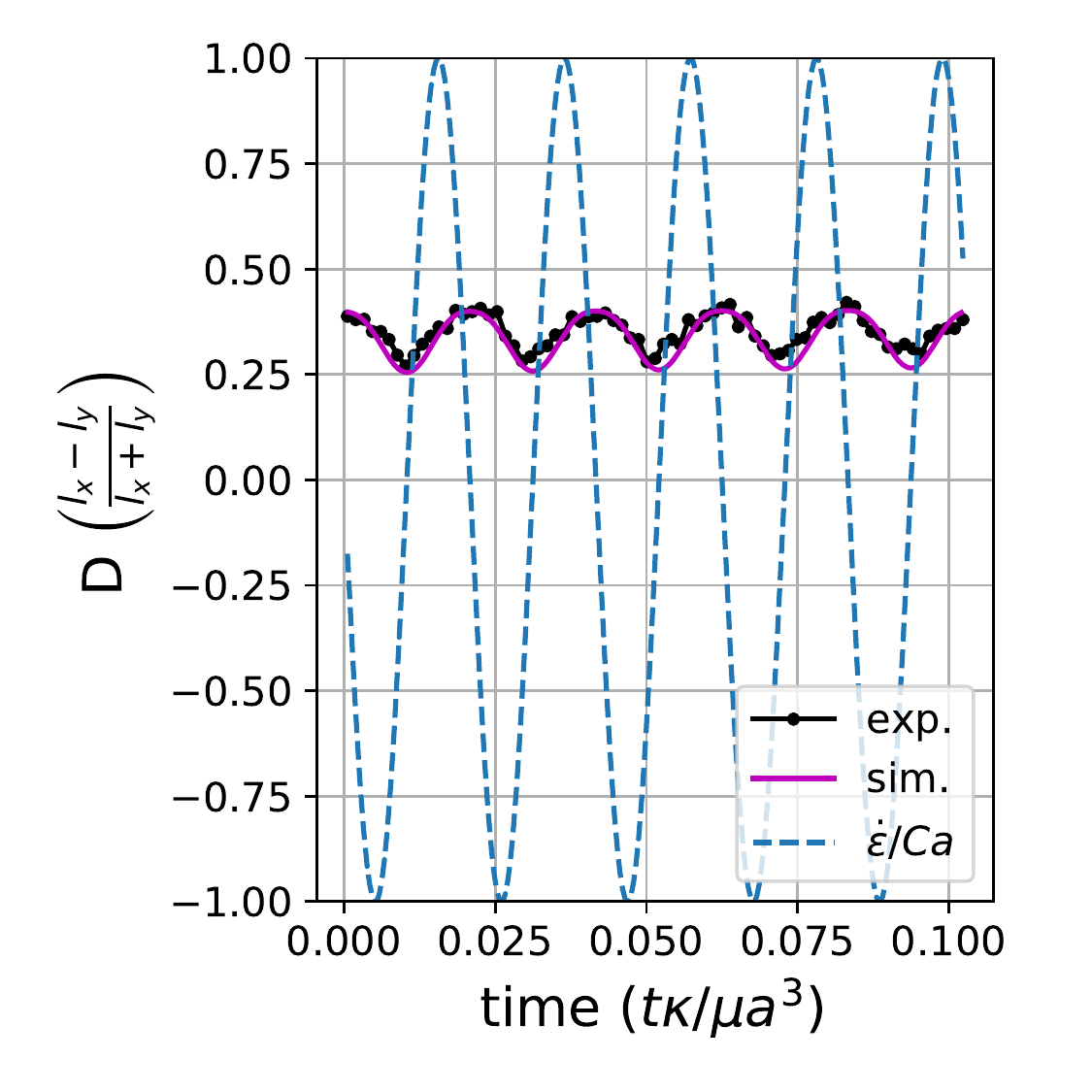}
	}
	\subfloat[Ca = 18.2, De = 18.2, $\nu$ = 0.88]{
		\gTrim{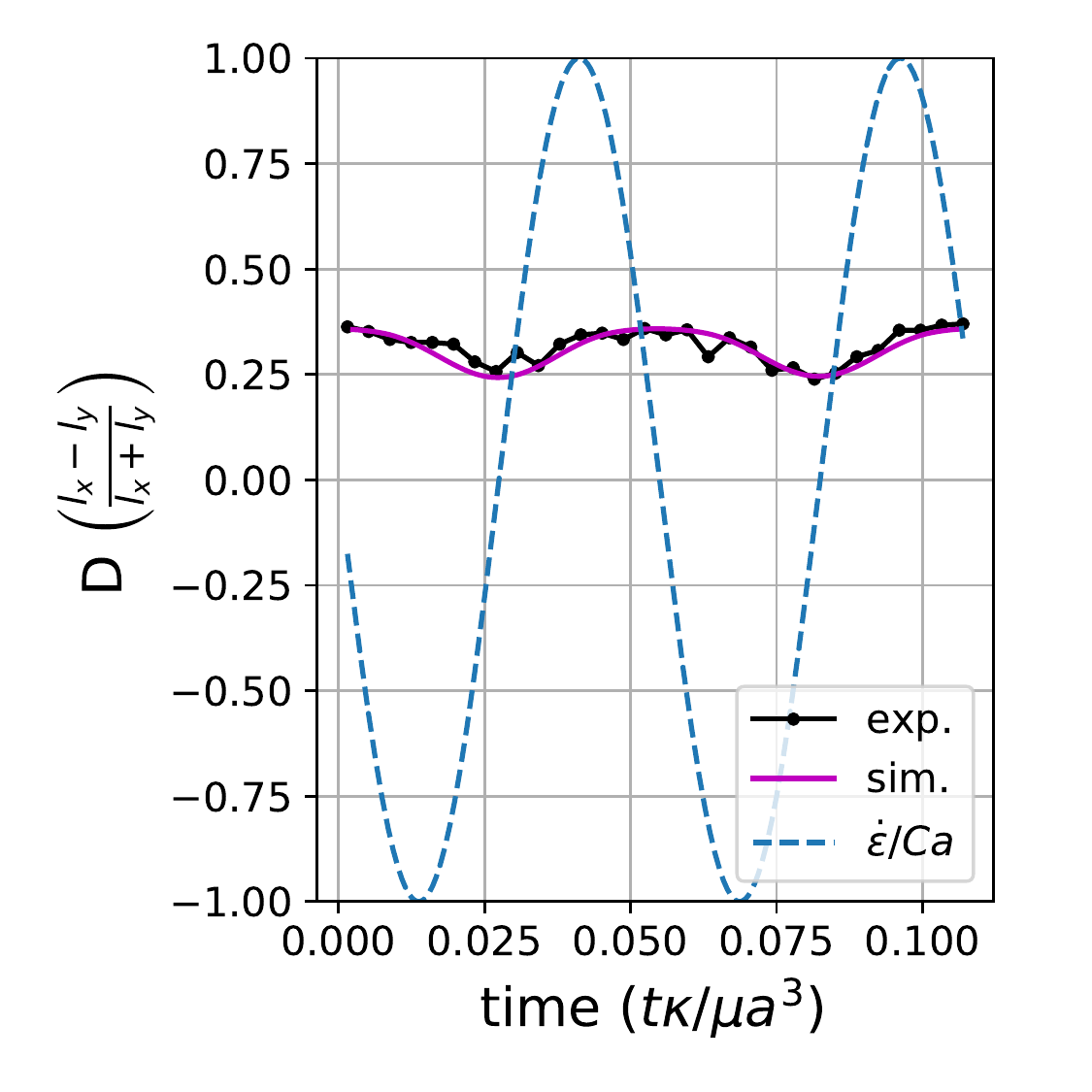}
	}
	\subfloat[Ca = 17.9, De = 29.9, $\nu$ = 0.91]{
		\gTrim{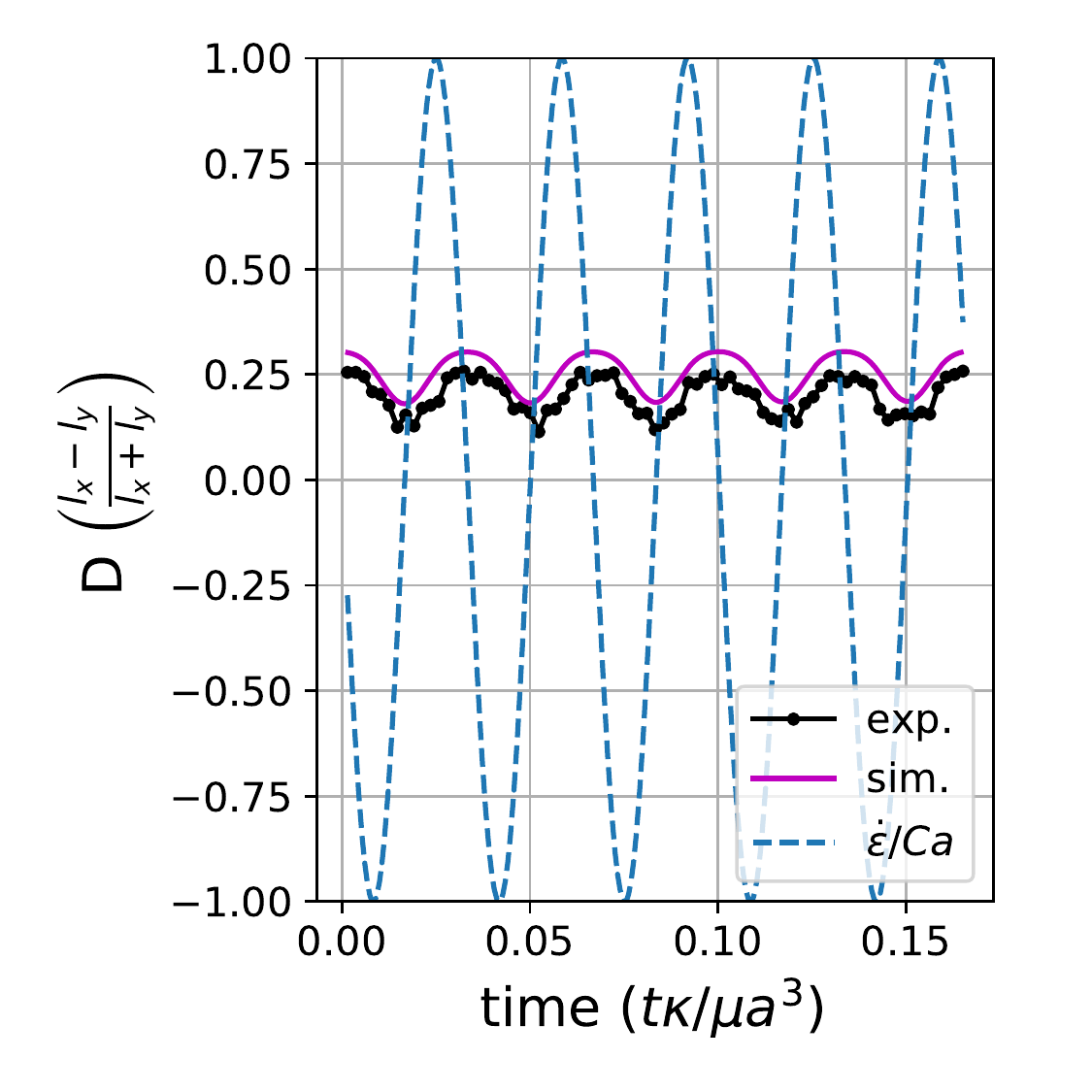}
	}
	\captionsetup{width=1.0\linewidth}
	\caption{%
		Transient deformation parameter $D$ for vesicle dynamics in time-dependent LAOE from experiments and simulations.
		All vesicles have a viscosity ratio of $\lambda = 1.0$.
		The $\dot{\epsilon}/\Ca$ line is the instantaneous strain rate of the external flow along the x-axis.
		A negative $\dot{\epsilon}/\Ca$ value is compression along the x-axis.
	}%
	\label{fig:deform_comp}

\end{figure}

\begin{figure}
	\centering
	\textbf{Symmetrical}

	\vspace{-3mm}

	\captionsetup{width=0.3\linewidth}
	\subfloat[Ca = 28.8, De = 6.4, $\nu$ = 0.85\label{fig:deform_liss_comp_sym}]{%
		\gTrim{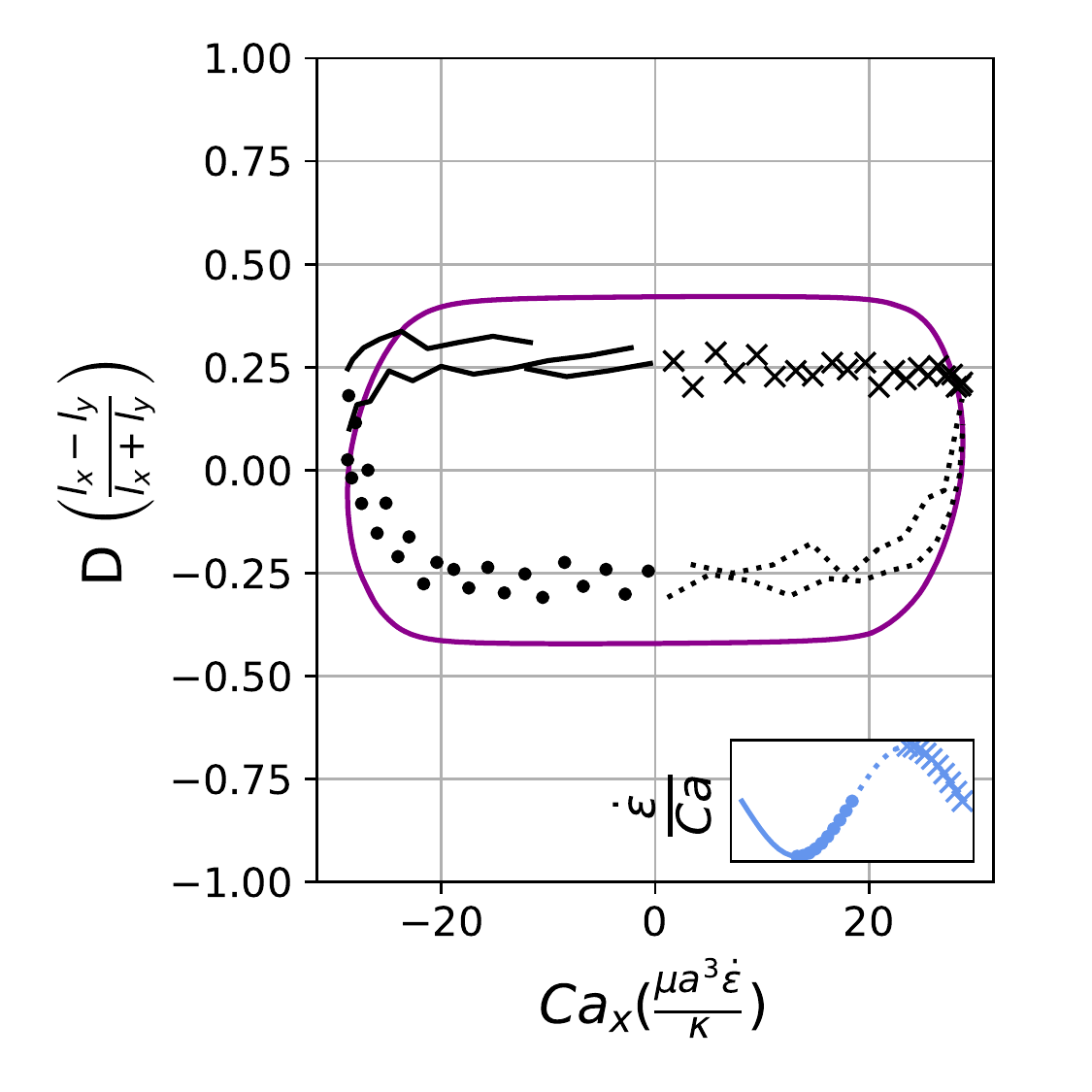}
	}
	\subfloat[Ca = 18.2, De = 3, $\nu$ = 0.88]{
		\gTrim{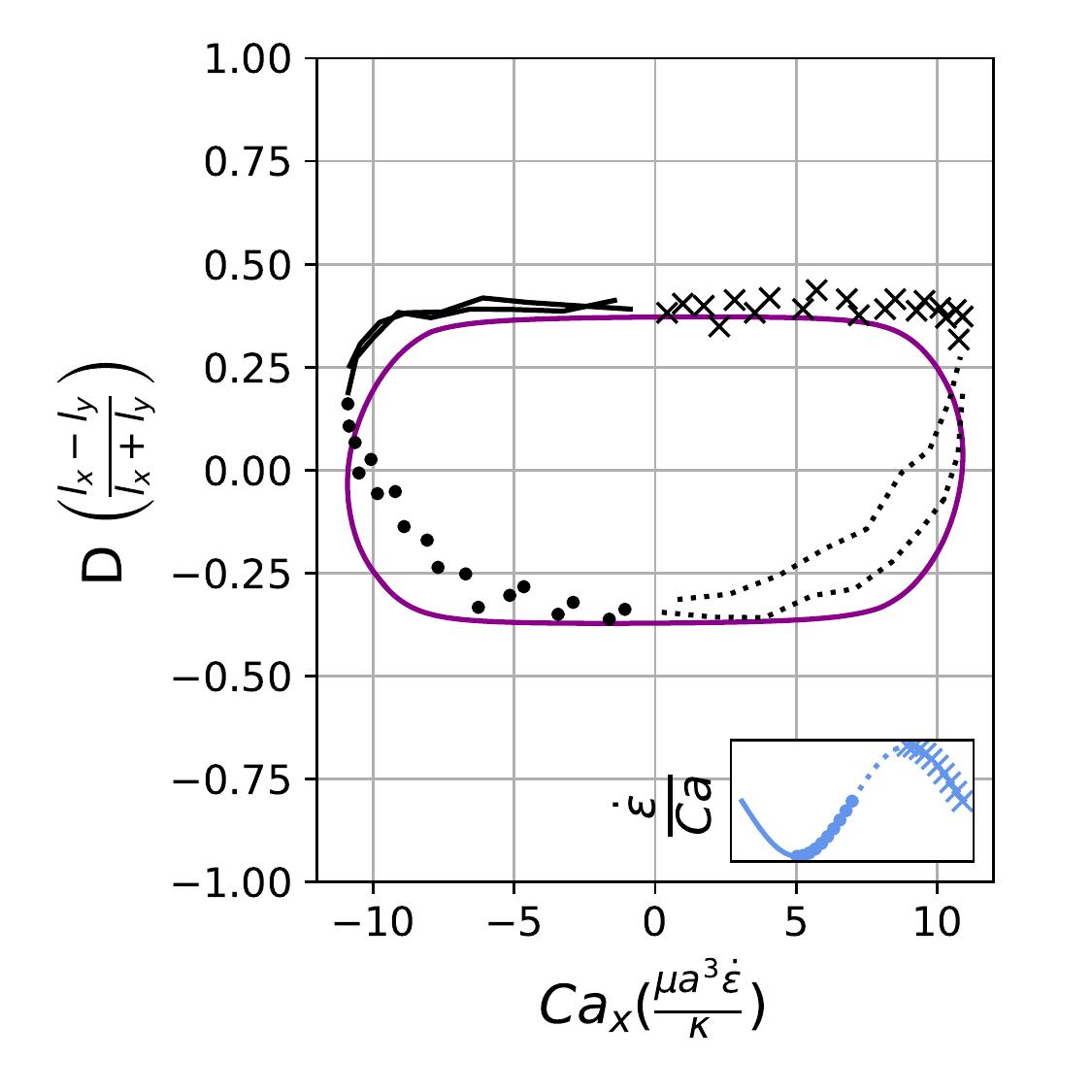}
	}
	\subfloat[Ca = 17.9, De = 6, $\nu$ = 0.91]{
		\gTrim{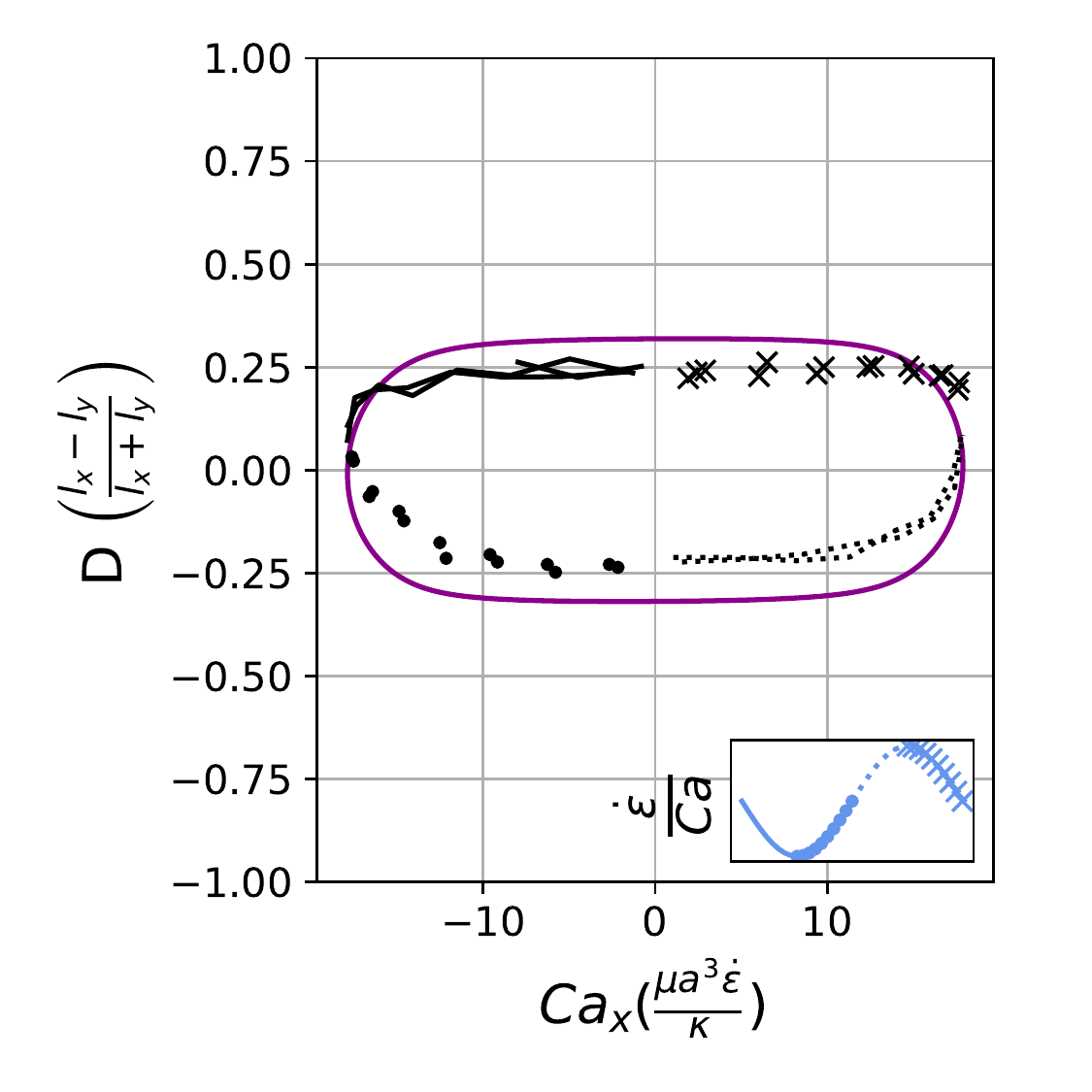}
	}\\

	\hrule
	\vspace{2mm}

	\textbf{Reorienting}

	\vspace{-3mm}

	\subfloat[Ca = 28.8, De = 12, $\nu$ = 0.85\label{fig:deform_liss_comp_reori}]{
		\gTrim{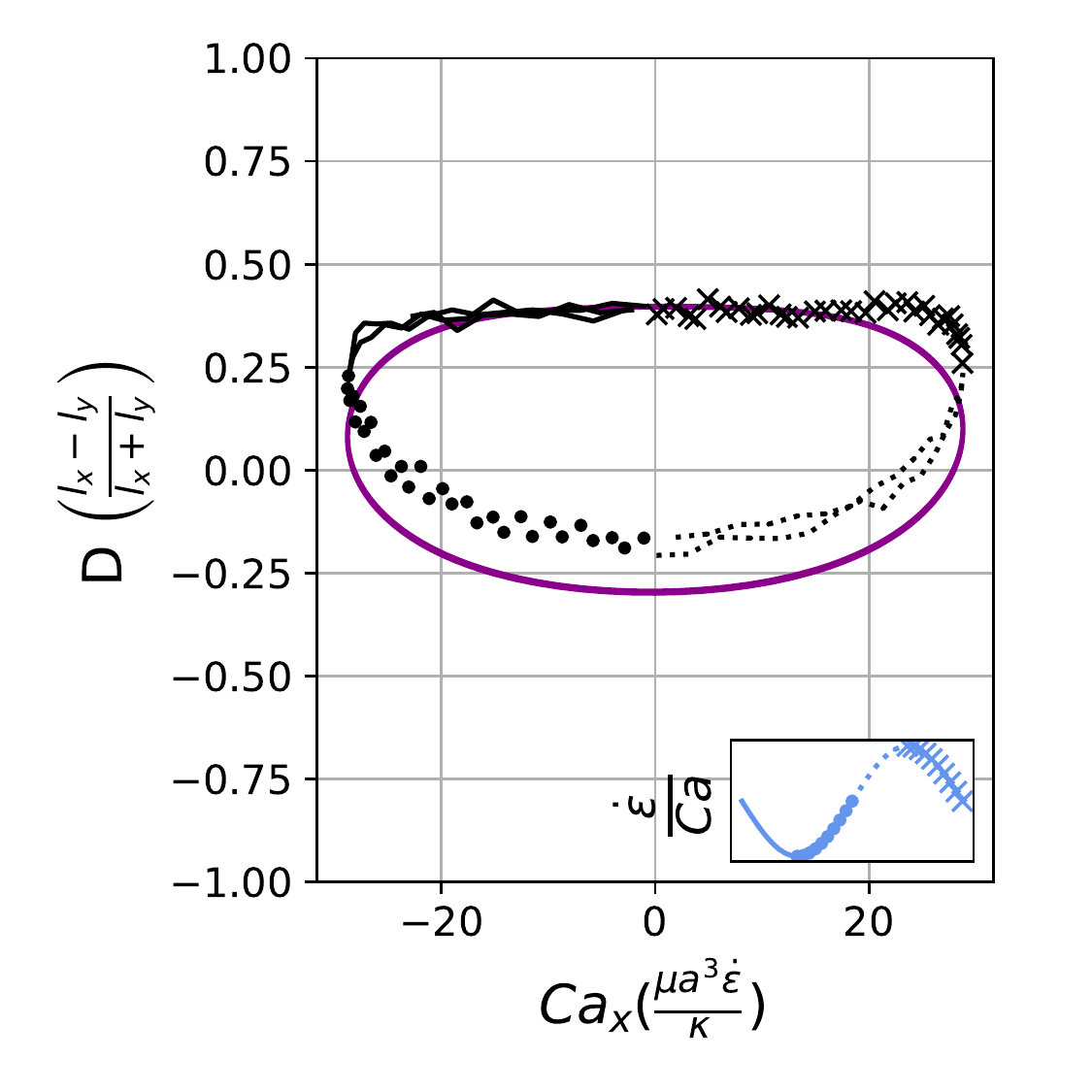}
	}
	\subfloat[Ca = 18.2, De = 4.5, $\nu$ = 0.88]{
		\gTrim{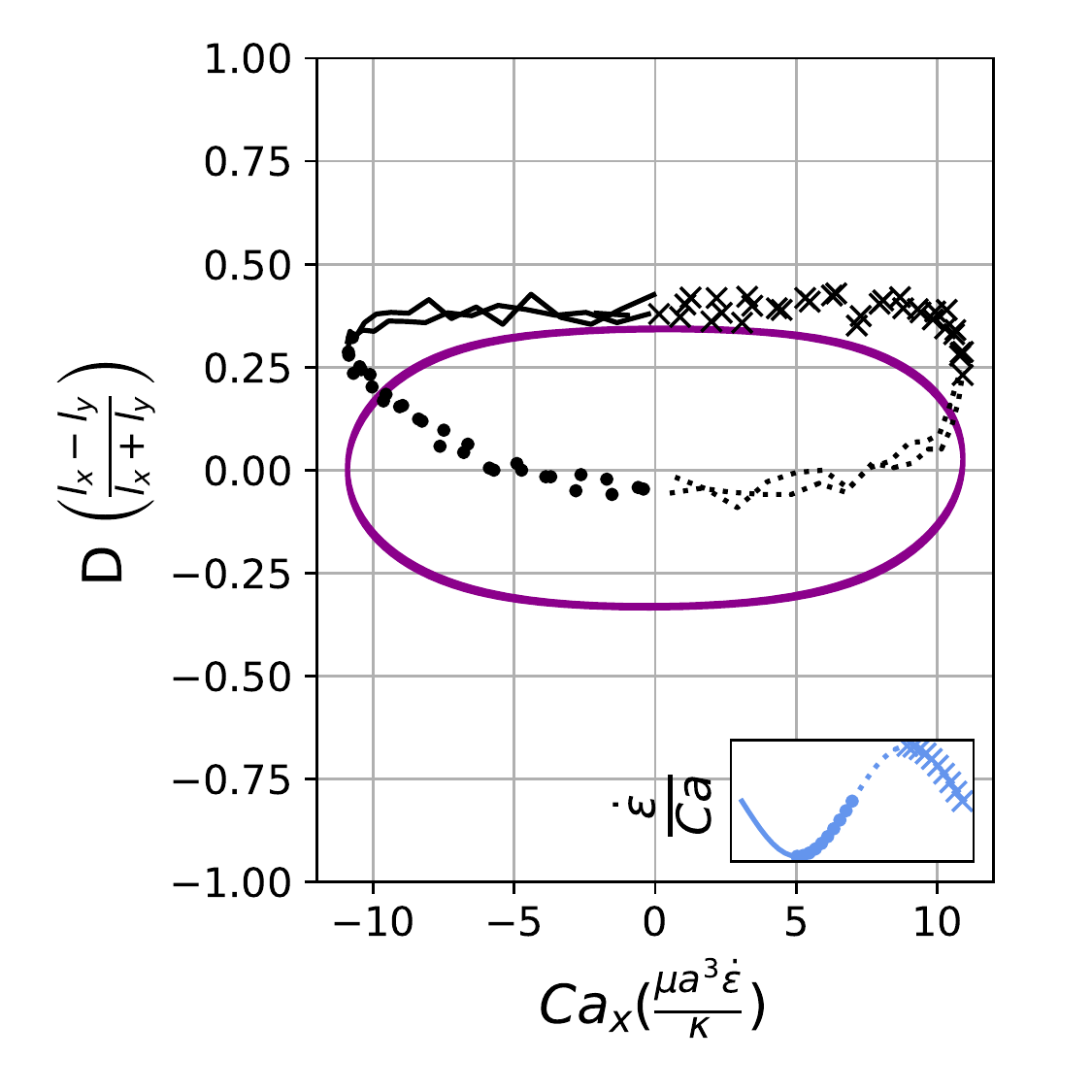}
	}
	\subfloat[Ca = 17.9, De = 14.9, $\nu$ = 0.91]{
		\gTrim{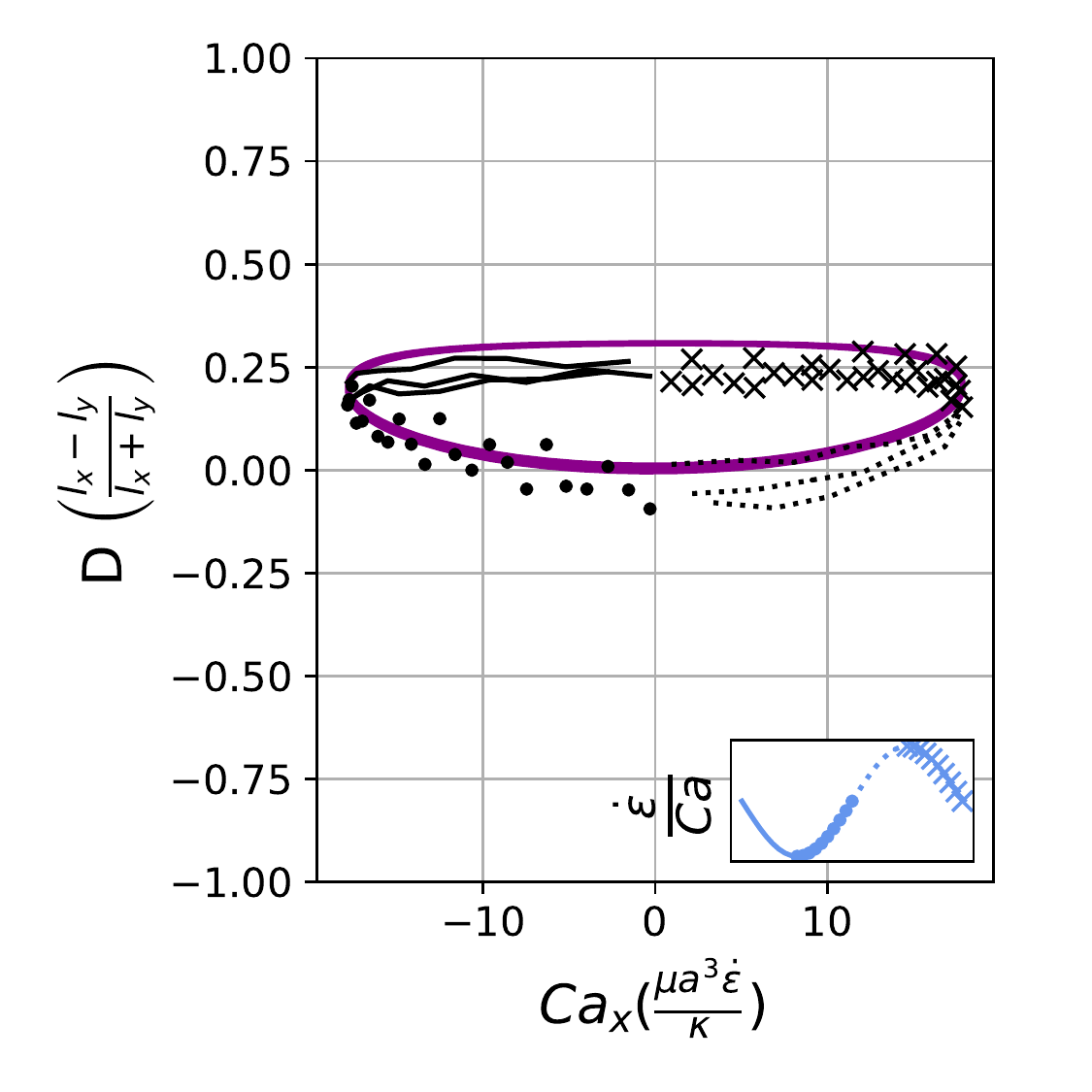}
	}\\

	\hrule
	\vspace{2mm}

	\textbf{Pulsating}

	\vspace{-3mm}

	\subfloat[Ca = 28.8, De = 48, $\nu$ = 0.85]{
		\gTrim{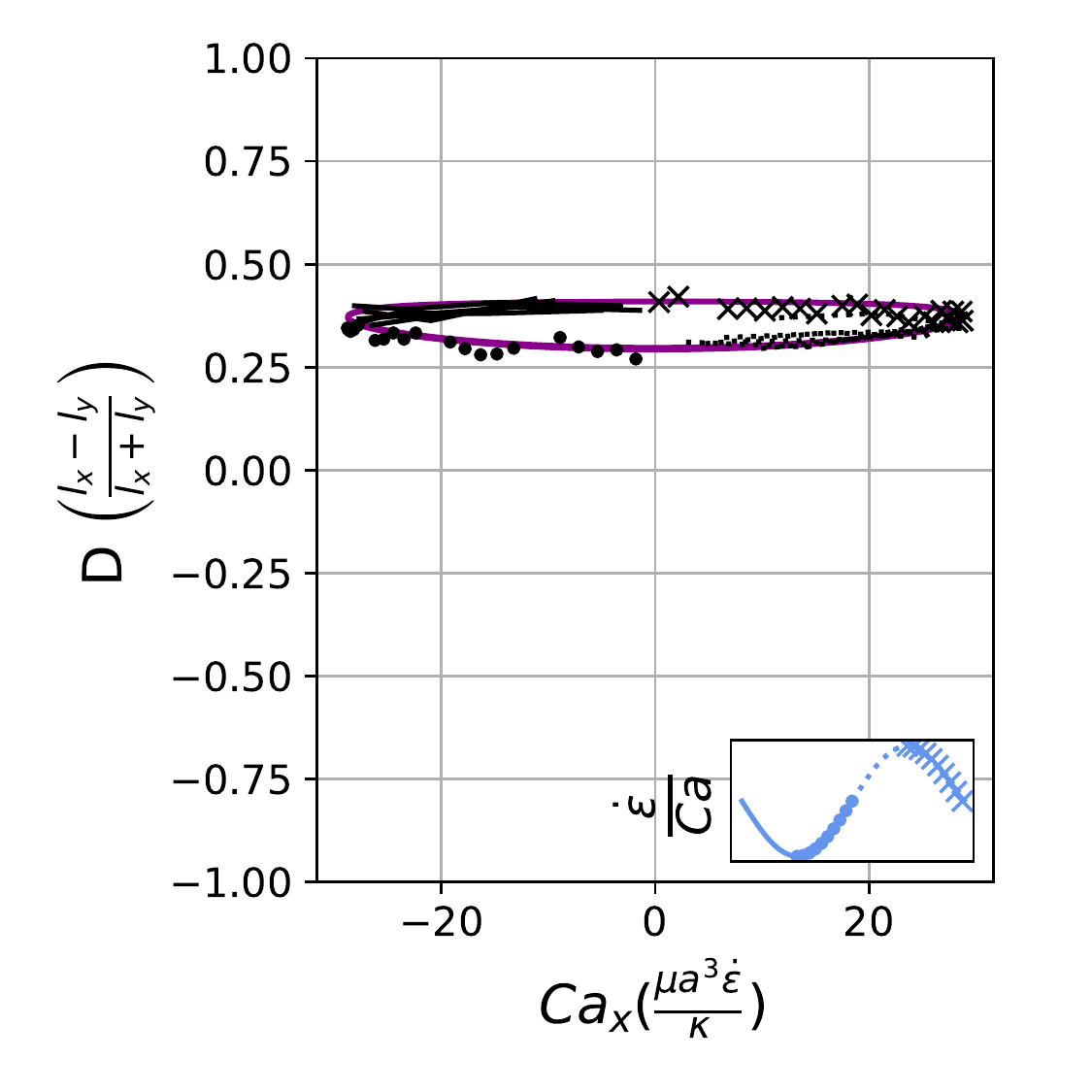}
	}
	\subfloat[Ca = 18.2, De = 18.2, $\nu$ = 0.88]{
		\gTrim{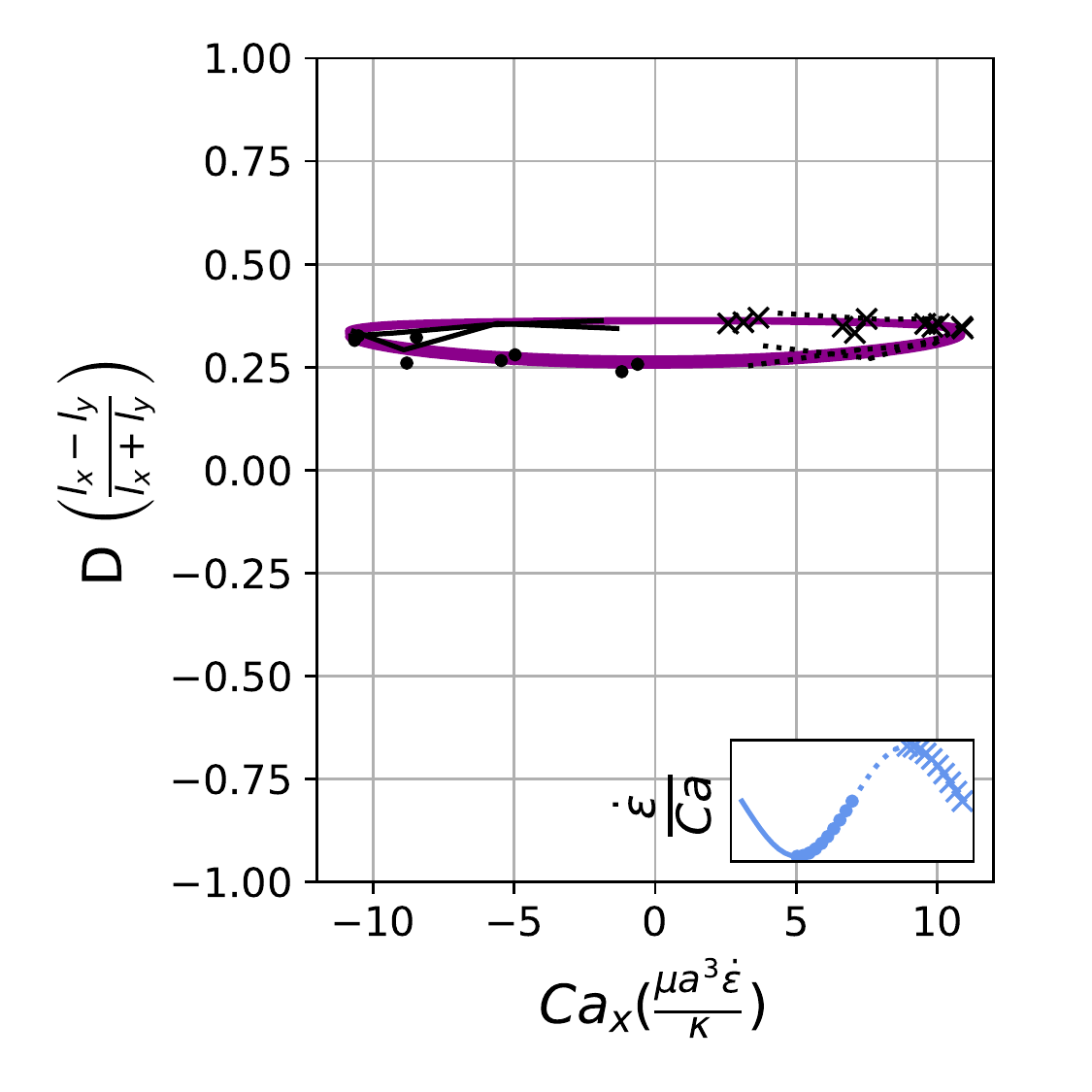}
	}
	\subfloat[Ca = 17.9, De = 29.9, $\nu$ = 0.91\label{fig:deform_liss_comp_puls}]{
		\gTrim{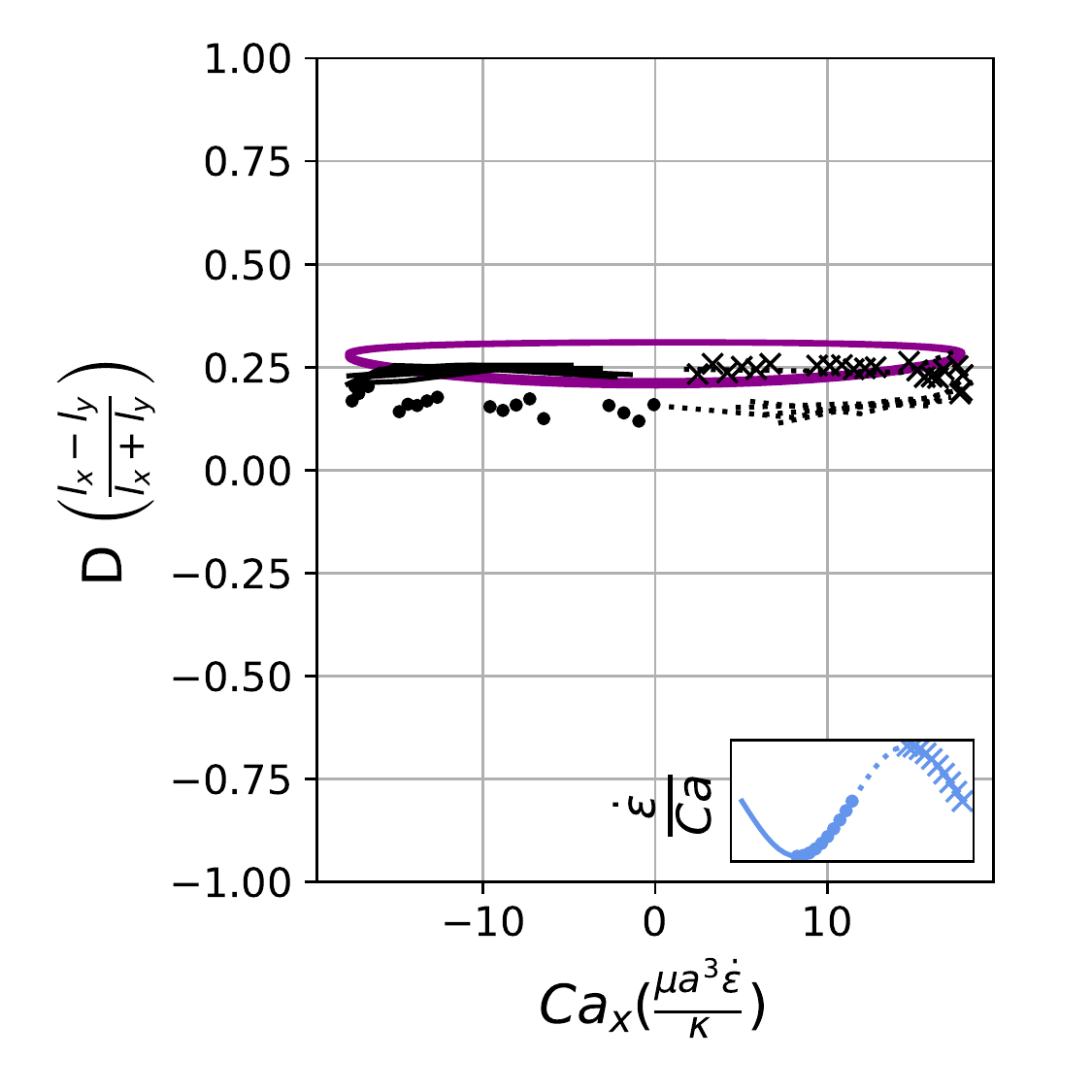}
	}
	\captionsetup{width=1.0\linewidth}
	\caption{%
		Lissajous-type curves of the deformation parameter $D$ versus the dimensionless instantaneous strain rate.
		All vesicles have a viscosity ratio of $\lambda = 1.0$.
		Black data points are experimental data; purple data points show numerical data.
		The oscillatory strain rate cycle is separated into four parts that have been noted with different markers, as shown in the legend in the bottom right hand corner.
	}%
	\label{fig:deform_liss_comp}
\end{figure}

\begin{figure}
	\centering
	\includegraphics[width=1.0\linewidth]{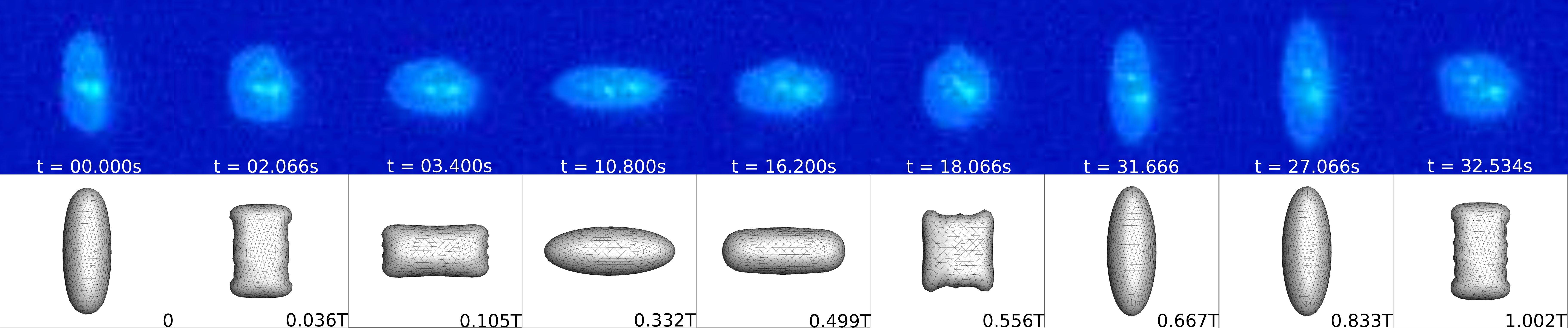}
	\caption{%
		Comparison of the experimental and simulation vesicle shapes in the symmetrical regime over one flow cycle at the same conditions of Ca = 39.5, De = 1.7, $\nu$ = 0.83, $\lambda$ = 1.00.
		The times in the figure are in seconds for the experimental video.
		Shapes from the simulations at the same non-dimensional times are shown below.
		$T$ is the non-dimensional period, defined as $T = 1/ \De$.
	}%
	\label{fig:exp_vs_sim_snapshots}
\end{figure}

Experiments were performed in the range of approximately $10 < \Ca < 1000$ and $0.5 < \De < 100$, whereas the majority of the simulations are in the range of $1 < \Ca < 40$ and $1 < \De < 10$.
Simulations were performed for several vesicles matching the conditions in the experiments, as discussed in the following section (\cref{fig:deform_comp}, \cref{fig:deform_liss_comp} and \cref{fig:exp_vs_sim_snapshots}).
It is possible to perform additional simulations at $\Ca \approx 100$, but current results suggest that the vesicle dynamics do not significantly change at higher Ca for quasi-spherical vesicles.

We observe three distinct dynamical regimes of vesicle dynamics based on the ratio between Capillary number and Deborah number.
We refer to these regimes as symmetrical, reorienting, and pulsating based on the deformation characteristics of vesicles in each case.
Vesicles in all three regimes can experience significant non-linear stress responses.
Snapshots of vesicle shapes from simulations and experiments for each of these regimes over a full strain rate cycle are shown in \cref{fig:regimes_snapshots} and \cref{fig:exp_regimes_snapshots}.

We quantitatively compare the simulations and experiments by plotting the deformation parameter $D$ (defined in \cref{eq:deform_param}) and instantaneous strain rate $\Ca_x$ (defined in \cref{eq:cax}) as a function of time, as shown in \cref{fig:deform_comp,fig:deform_liss_comp}.
Experimental trajectories are generally limited to 2--4 strain rate cycles due to the photobleaching of the vesicle membrane during fluorescence imaging experiments.
Observing vesicle deformation over more strain rate cycles is experimentally feasible, however, we generally opted to observe dynamics under different experimental parameters (Ca, De) for the {\em same vesicle} in a series of subsequent experiments.
For the numerical data, we simulated vesicle dynamics over at least 10 strain rate cycles.

\subparagraph{\textbf{Symmetrical regime:}}
Starting with the symmetrical regimes results, we find the symmetrical regime occurs under flow conditions where the vesicle deformation timescale is shorter or exactly equal to half of a strain rate cycle.
Based on our simulations, this occurs approximately when $\Ca \geq 3.33 \:\De$ for a vesicle with a reduced volume $\nu = 0.80$.
Our experiments and simulations show that vesicle dynamics in the symmetrical regime are described by two common characteristics (\cref{fig:exp_vs_sim_snapshots}).
First, the vesicle reaches approximately the same maximum length twice during one strain rate cycle, regardless of $\Ca$.
The observation of a maximum length is reasonable for quasi-spherical vesicles, 
as it has been shown that vesicles with $\nu > 0.75$ have a stable steady-state 
shape at infinite $\Ca$, regardless of viscosity ratio~\cite{Narsimhan2014}.
Second, vesicle membranes exhibit transient wrinkling when vesicles are exposed to the compressional cycle of the oscillatory extensional flow.
The transient wrinkling behavior is examined later in this section.
These features are illustrated in \cref{fig:exp_vs_sim_snapshots}, where a characteristic time series of images of vesicle shape in LAOE is qualitatively compared to the equivalent numerical simulation.
In general, vesicle shapes determined from experiments are in good agreement with those determined from numerical simulations.
Turning to the deformation parameter plots (\cref{fig:deform_comp} and \cref{fig:deform_liss_comp}), we see the simulations and experiments agree well at the majority of the tested parameters.
Some of the experimental datasets show fluctuations in the deformation over the strain rate cycles and disagreement between the simulations on the maximum deformation.
These discrepancies likely occur due to challenges in imaging a three-dimensional object in a two-dimensional plane and because the experiments are limited to a few strain rate cycles.
Nevertheless, we generally observe good agreement between simulations and experiments in terms of the deformation parameter in transient flows.

Transient wrinkling dynamics were first reported by \cite{Kantsler2006} for a 
single cycle of suddenly reversed extensional flow and subsequently elaborated 
upon by \cite{turitsyn2008wrinkling} and \cite{li2005numerical}.
Wrinkling behavior is caused by a negative surface tension created during vesicle compression.
Moreover, a critical compression rate exists below which thermal fluctuations dominate the observed wrinkling.
In our work, we study vesicle dynamics in an extensional flow with smoothly varying sinusoidal strain rate dependence, rather than an abrupt step-function reversal of compressional/extensional axes.
We still observe qualitatively the same membrane wrinkling features as those reported in prior work.
Additional experimental snapshots of vesicles showing wrinkling dynamics are included in the supplementary materials (Fig. S2, Fig. S3, Fig. S4 and Fig. S5).

\subparagraph{\textbf{Reorienting regime:}}
At lower $\Ca / \De$ ratios, vesicles no longer reach the same length twice during a strain rate cycle, and the dynamics are no longer described by the symmetrical regime.
The reorienting deformation regime occurs when the vesicle deformation rate and flow frequency are comparable ($\Ca \approx 2 \De$ for $\nu = 0.80$).
Surprisingly, the vesicle deforms to unequal maximum lengths between the horizontal and vertical orientations, producing a vertically asymmetrical deformation parameter plot, as seen in \cref{fig:deform_comp_reori}.
In contrast, prior work on droplets in oscillatory extensional flow report 
symmetric deformation (equal deformation in x- and y-orientations), regardless 
of flow strength or oscillatory frequency~\cite{li2005drop, li2005numerical}.
Single polymers in LAOE also deform symmetrically between the two half cycles 
for the range of Weissenberg and Deborah numbers studied in prior 
work~\cite{zhou2016single}.
The phenomenon of asymmetric stretching of vesicles along the two axes arises due to the enclosed membrane for fluid-filled vesicles.
In particular, we posit that the asymmetrical deformation occurs because the vesicle has an anisotropic, energetically preferred shape at equilibrium.
Under quiescent conditions, the global energy minimum for quasi-spherical 
vesicles is a prolate spheroid shape.~\cite{Seifert1997}.

\subparagraph{\textbf{Pulsating regime:}}
At even lower $\Ca / \De$, the vesicle no longer reorients and simply pulsates along one axis during LAOE.
We refer to this dynamical regime as the pulsating regime, which approximately occurs when $\Ca \le 2 \De$ for $\nu = 0.80$.
Note that the strain in the pulsating regime is not necessarily infinitesimal.
As shown in \cref{fig:deform_comp_puls}, the deformation parameter curve illustrates that vesicles are generally oriented along the x-axis and can deform significantly in this regime.
It is possible to probe the small amplitude oscillatory extension regime by keeping the $\De$ constant and reducing the $\Ca$.
In the small amplitude regime, vesicles do not deform appreciably, and the Lissajous curve approaches a constant value, thereby informing on the linear viscoelastic rheology of vesicle suspensions.
Similar behavior occurs when increasing the $\De$ and keeping $\Ca$ constant at small values.
In this case, the membrane does not have appreciable time to reorient during the time at which the strain rate changes.

\subsection{Quasi-spherical initial shape considerations\label{sec:quasi_initial_shape}}

\begin{figure}
	\begin{minipage}[c]{0.45\linewidth}
		\subfloat[Symmetrical regime\label{fig:regimes_snapshots_sym}]{\includegraphics[width=1.0\linewidth]{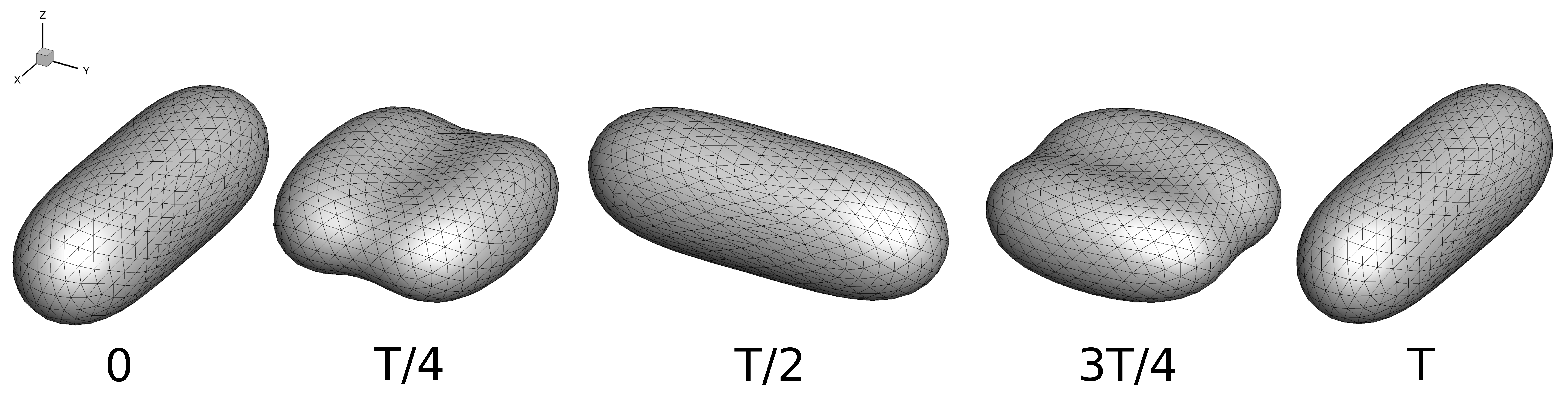}}\\
		\subfloat[Reorienting regime\label{fig:regimes_snapshots_reori}]{\includegraphics[width=1.0\linewidth]{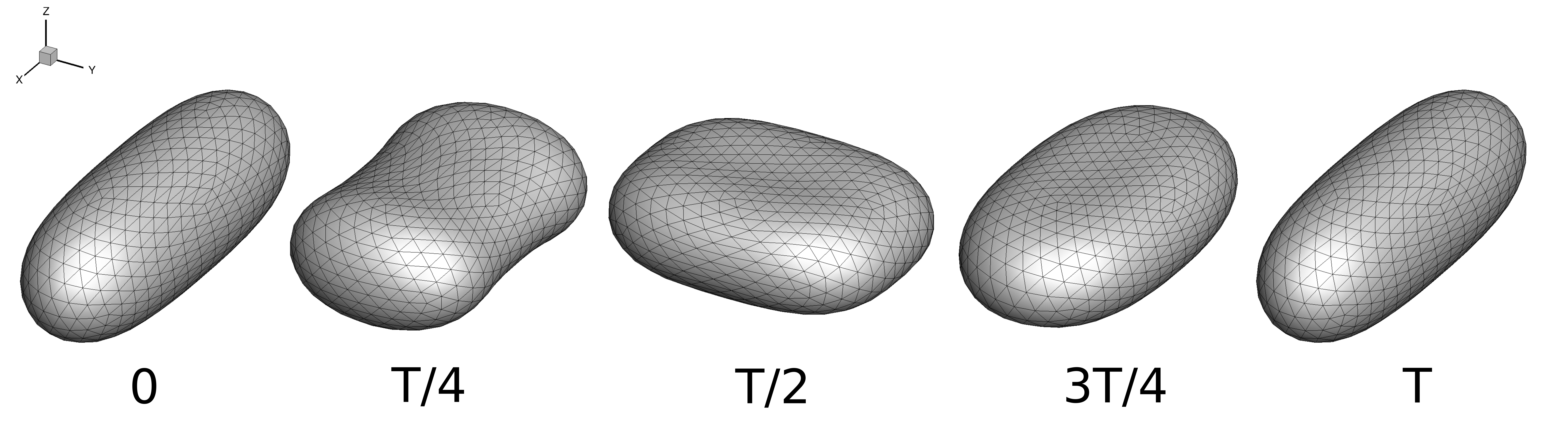}}\\
		\subfloat[Pulsating regime\label{fig:regimes_snapshots_pulse}]{\includegraphics[width=1.0\linewidth]{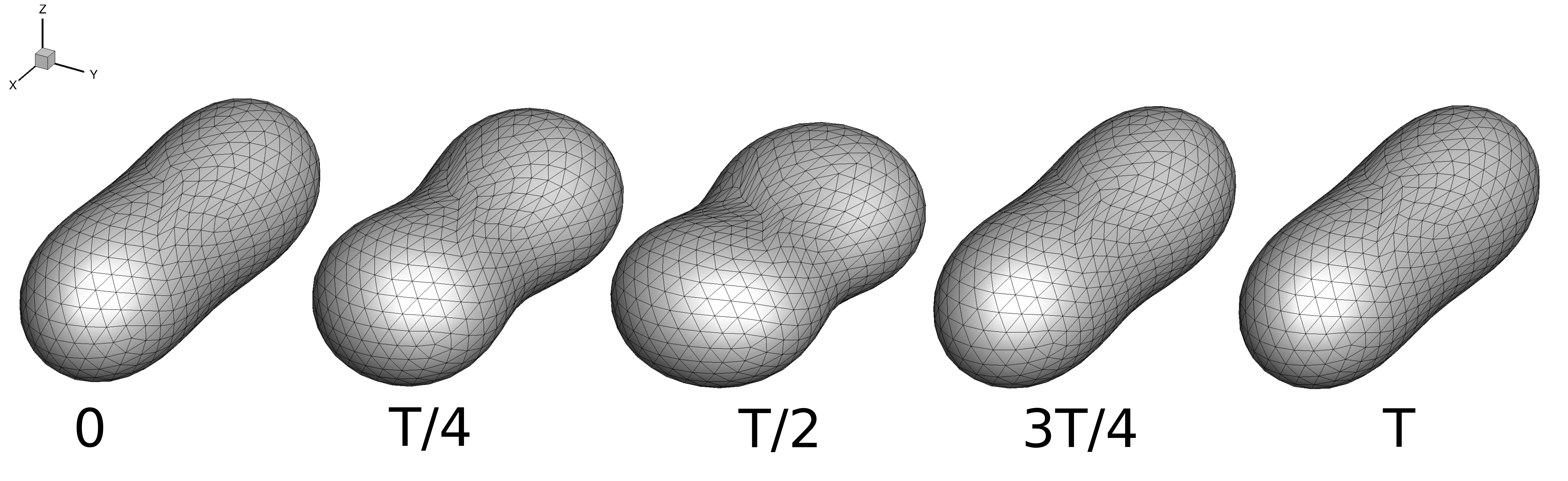}}
		\captionsetup{width=1.0\linewidth}
		\caption{%
			Snapshots of vesicle shapes from simulations over a flow cycle for the three dynamical regimes.
			The values under the figures are fractions of a strain rate period defined as $T = 1/\De$.
		}%
		\label{fig:regimes_snapshots}
	\end{minipage}
	\hfill
	\begin{minipage}[c]{0.45\linewidth}
		\includegraphics[width=1.0\textwidth]{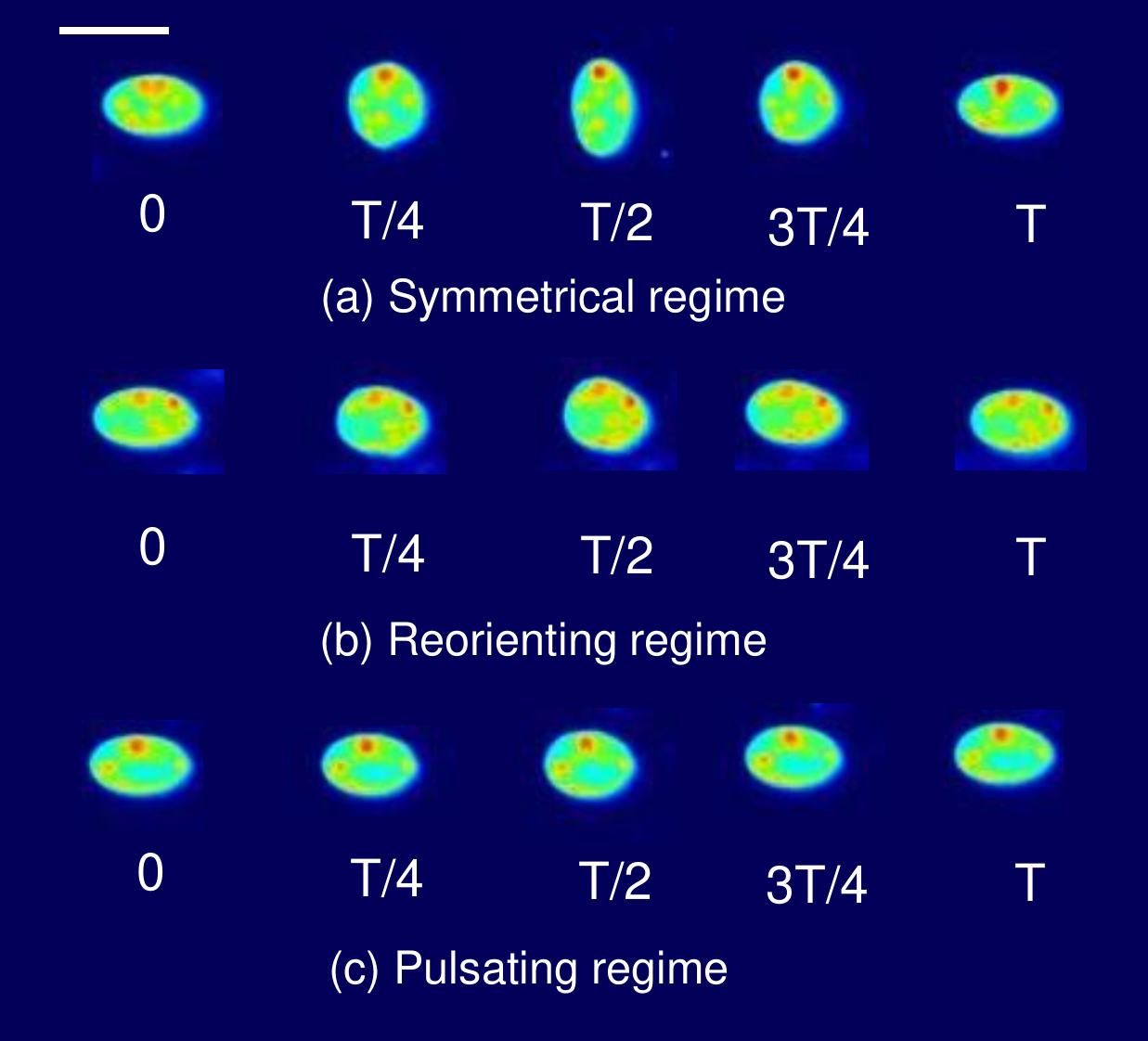}
		\captionsetup{width=1.0\linewidth}
		\caption{%
			Snapshots of vesicle shapes from experiments over a flow cycle for the three dynamical regimes.
			The values under the figures are fractions of cycle time $T$ in seconds. Scale bar is 20 $\mu$m. False color is applied to the grayscale images for enhancing the resolution.
		}%
		\label{fig:exp_regimes_snapshots} 
	\end{minipage}%
\end{figure}

\begin{figure}
	\centering
	\subfloat[oblate initial \label{fig:oblate_lissajous_ob}]{\includegraphics[width=0.49\linewidth]{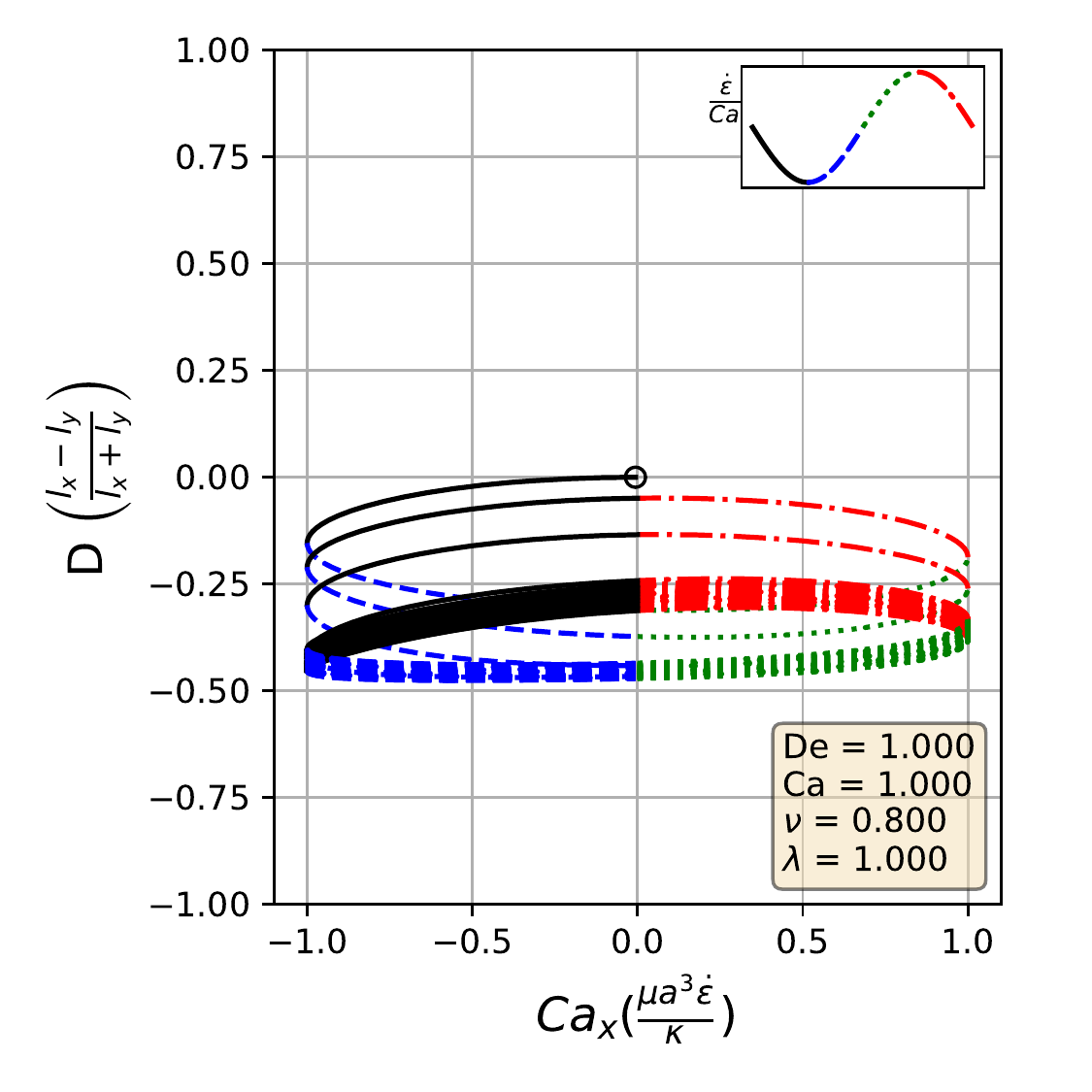}}
	\subfloat[prolate initial \label{fig:oblate_lissajous_pro}]{\includegraphics[width=0.49\linewidth]{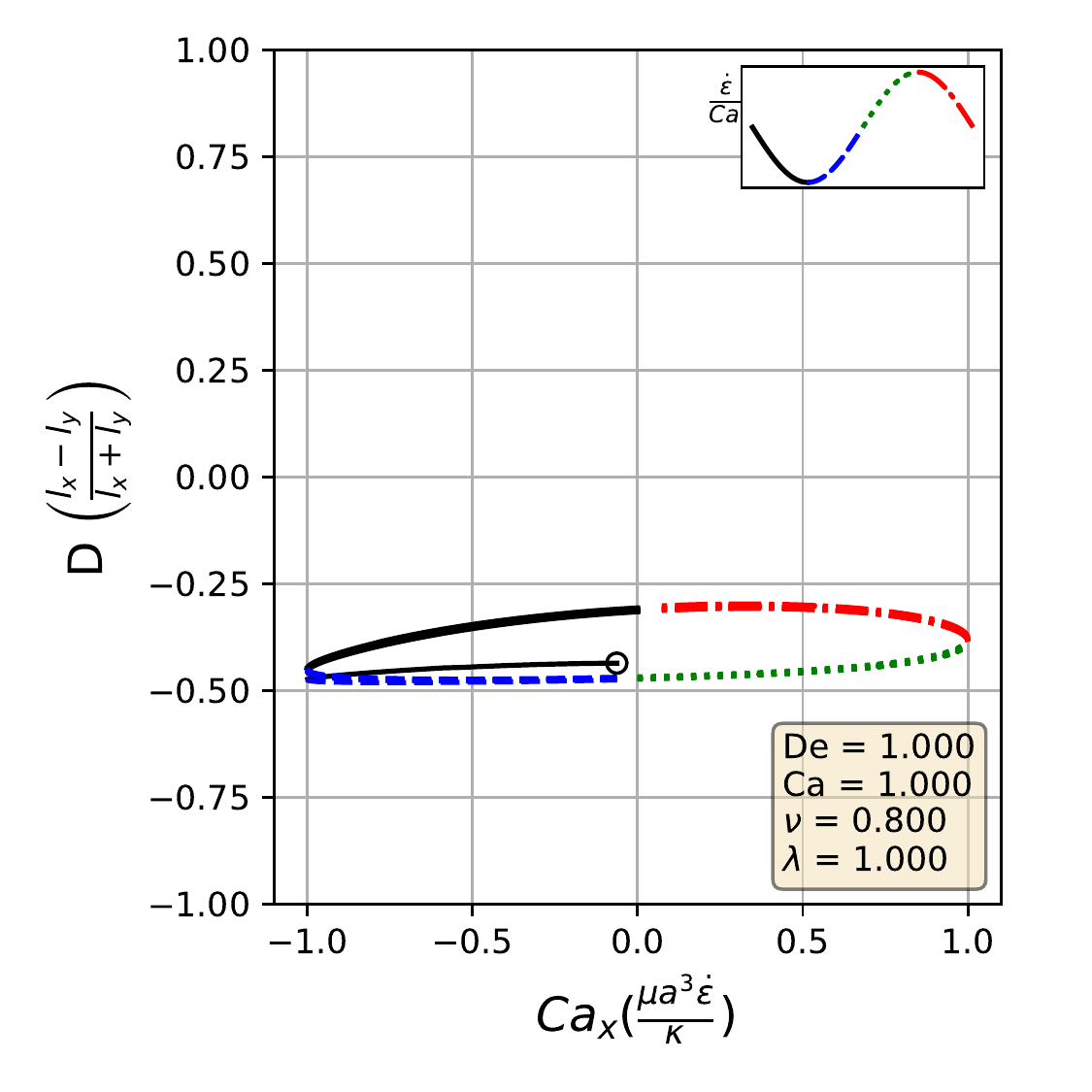}}
	\caption{%
		Lissajous type deformation parameter curves from an oblate shape initial condition and a prolate shape initial condition.
		Top right legend indicates color coding for the strain rate cycle.
		The black circle marks the deformation parameter of the initial shape.
	}%
	\label{fig:oblate_lissajous}
\end{figure}

Overall, the simulations discussed up to this point (including results in 
\cref{fig:regimes_snapshots} and \cref{fig:deform_comp}) were performed using a 
prolate-like initial shape, because it is the global equilibrium shape for 
reduced volumes $\nu \geq 0.652$~\cite{Seifert1997}.
These results suggest that the unequal stretching observed in the pulsating and reorienting regimes occurs during the steady limit cycle, for this particular initial shape.
However, there are other local minimum energy shapes for vesicles, such as the oblate shape family.
To determine whether the pulsating and reorienting regimes are possible with a different initial condition, we performed simulations using an oblate shape such that the initial deformation parameter was set to zero.
We examined this initial condition because vesicle shape is isotropic in the x-y plane, where an image obtained through optical microscopy would show a circle.
The oblate initial condition simulations test if the anisotropic deformations will still occur if the vesicle starts with a shape isotropic in the x-y plane rather than an initially anisotropic shape.
Simulation results for the oblate initial condition are plotted in \cref{fig:oblate_lissajous}, which shows that vesicle dynamics during the steady limit cycle for the oblate initial condition (\cref{fig:oblate_lissajous_ob}) are the same as that observed from the prolate-like initial condition (\cref{fig:oblate_lissajous_pro}).
We repeated these simulations at several other capillary numbers and Deborah numbers, observing no dependence of the dynamics on the initial conditions.

\subsection{Quasi-spherical phase diagrams\label{sec:phase_diagrams}}

\begin{figure}
	\centering
	\subfloat[\label{fig:phase_diagram_vol080}]{\includegraphics[width=0.49\linewidth]{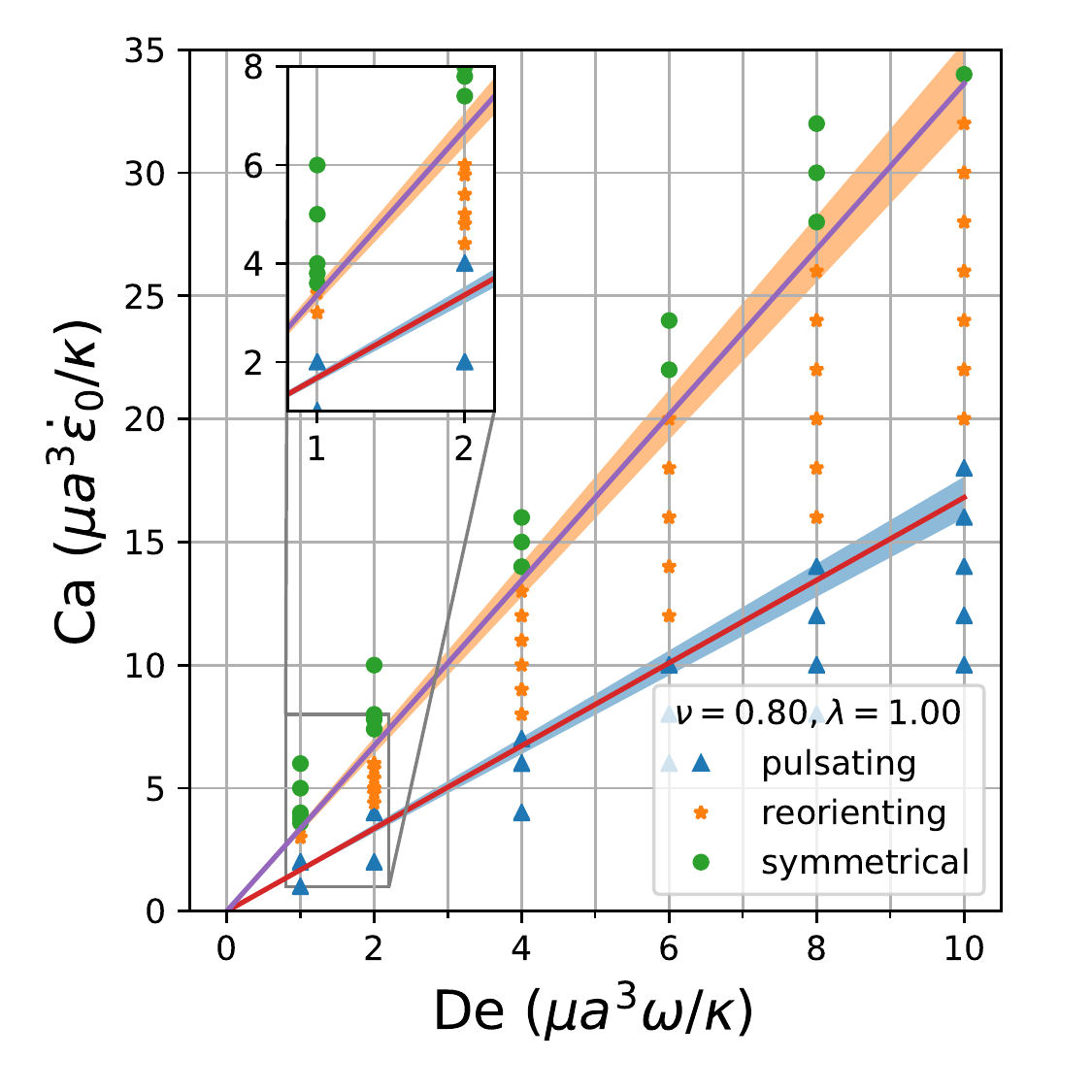}}
	\subfloat[\label{fig:phase_diagram_vol090}]{\includegraphics[width=0.49\linewidth]{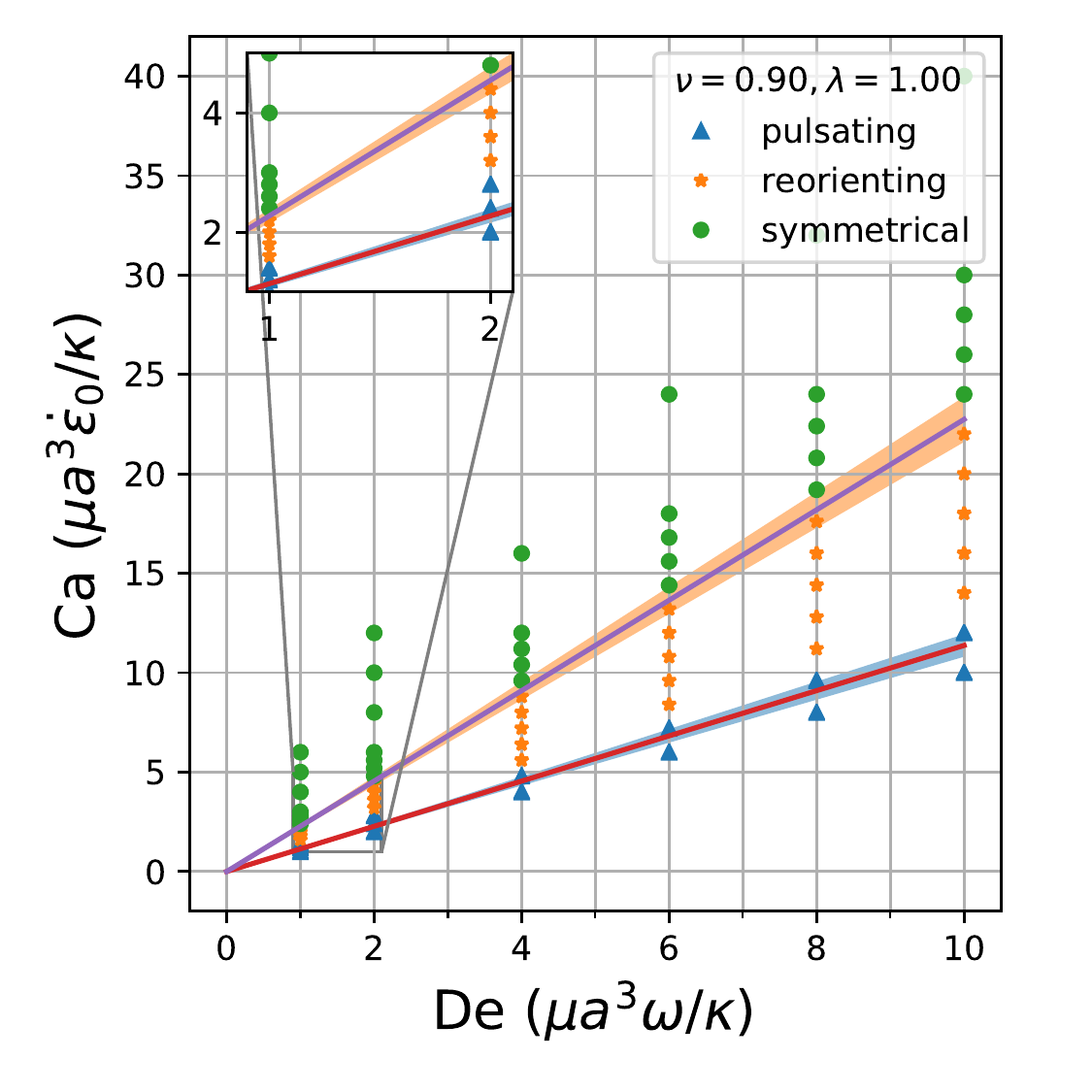}}
	\caption{%
		Phase diagrams for the low to medium to high deformation regimes for vesicles of reduced volume $\nu = 0.80$ and $\nu = 0.90$.
		Lines in the diagrams are from the semi-analytical theory presented near the end of \cref{sec:dynamical_regimes}.
		Due to uncertainty in determining the $D_{o}$ value, a 5\% error has been included on the lines.
	}%
	\label{fig:phase_diagrams}
\end{figure}

By comparing the deformation parameter results for each simulation, we can plot a phase diagram of different dynamical regimes observed during oscillatory flows.
Which regime a vesicle experiences can be quantitatively determined by assessing the minimum and maximum deformation parameter over a cycle.
If both the minimum and maximum deformation parameter are positive, the vesicle dynamics are classified as the pulsating regime, reflecting that the vesicle does not change orientation.
If the vesicle has a positive maximum $D$ and a negative minimum $D$, we check if the differences in magnitudes are within a threshold value of 0.01.
Should they be within 0.01 of each other, the vesicle is in the symmetrical regime, since the vesicle reaches the same maximum length twice a cycle.
This threshold value was chosen heuristically to reflect the discretization accuracy.
If the magnitudes are not within this threshold value, vesicle dynamics are classified as the reorienting regime.
Results from this analysis is plotted in \cref{fig:phase_diagrams}

The phase boundaries appear to be mostly linear, suggesting that the dynamics result from a simple interaction between the flow frequency and the strain rate, $\Ca/\De=\dot{\epsilon}_0/\omega$.
Here, we derive the phase boundaries in the limit of a quasispherical vesicle 
\cite{Vlahovska2006}. For small excess area ($\Delta=4\pi(\nu^{-2/3}-1)\ll 1$), 
the vesicle shape is characterized by a perturbation series in terms of 
spherical harmonics \cite{Vlahovska2006}.
In a planar extensional flow, there are only two modes excited for the leading order correction to the vesicle shape.  When one solves for the deformation parameter as defined in \cref{eq:deform_param}, one obtains
\begin{equation}
    \begin{aligned}
        D(t) &= (L_{\infty}-1)\\
        & \left(1-2\left(1+\left(\frac{1+A_o}{1-A_o}\right)\exp\left(\frac{60}{\pi(32+23\lambda)}\frac{\Ca}{\De}\frac{1}{L_{\infty}-1}\left[\cos(2\pi \De~ t)-1\right]\right)\right)^{-1}\right),
    \end{aligned}
\end{equation}
where parameters $L_{\infty}=1+\sqrt{15/8}\left(\nu^{-2/3}-1\right)^{1/2}$ and $A_o = (\nu^{-1/3}l_x^{max}/2-1)/(L_{\infty}-1)$; $l_x^{max}$ is the maximum x-axis length of the vesicle. For the detailed derivation of these results, one can refer to the supporting information.

Following the definitions of the phase boundaries discussed previously, we can derive the two phase boundaries in the limit of $A_0 \ll 1$, i.e. $\ln\left(\frac{1+A_o}{1-A_o}\right)\approx\frac{1}{(L_{\infty}-1)}\ln\left(\frac{1+D_o}{1-D_o}\right)$,
\begin{equation}
	\label{eq:CaDe1}
	\frac{\Ca}{\De}=\frac{\pi(32+23\lambda)}{120}\log\left(\frac{1+D_o}{1-D_o}\right)\quad \text{for pulsating/reorienting phases},\\
	\end{equation}
	
	\begin{equation}
	\label{eq:CaDe2}
	\frac{\Ca}{\De}=\frac{\pi(32+23\lambda)}{60}\log\left(\frac{1+D_o}{1-D_o}\right) \quad \text{for reorienting/symmetrical phases}    
	\end{equation}
In the above equations, $D_0$ is the maximum deformation parameter during the 
LAOE cycle.  Note that the value of $D_o$ is determined by our numerical runs 
at the highest $\Ca$ and $\De$ numbers. Based on the quasispherical vesicle 
theory, the deformation phase boundaries depend on the viscosity ratio, where 
the factor ${(23\lambda+32)}^{-1}$ is related to the relaxation time of the 
quasi-spherical vesicle~\cite{Vlahovska2006}.
\cref{fig:phase_diagram_vol080} shows the phase boundaries are accurately calculated by using \cref{eq:CaDe1} and \cref{eq:CaDe2} when the reduced volume is $\nu=0.8$.
Increasing $\nu$ from 0.80 to 0.90 shifts the phase boundaries downwards, but maintains a similar linear relation (\cref{fig:phase_diagram_vol090}).
We also simulated viscosity ratio $\lambda = 10$ and found that higher viscosity ratios shift the boundaries to higher capillary numbers. We include the dynamics evolution of $l_x$ and $l_y$ (simulations vs. analytical solutions) and $\lambda = 10$ results in the supplementary materials for brevity.

\begin{figure}
	\centering
	\captionsetup{width=0.4\linewidth}
	\subfloat[Symmetrical regime\label{fig:stress_response_high}]{\fTrim{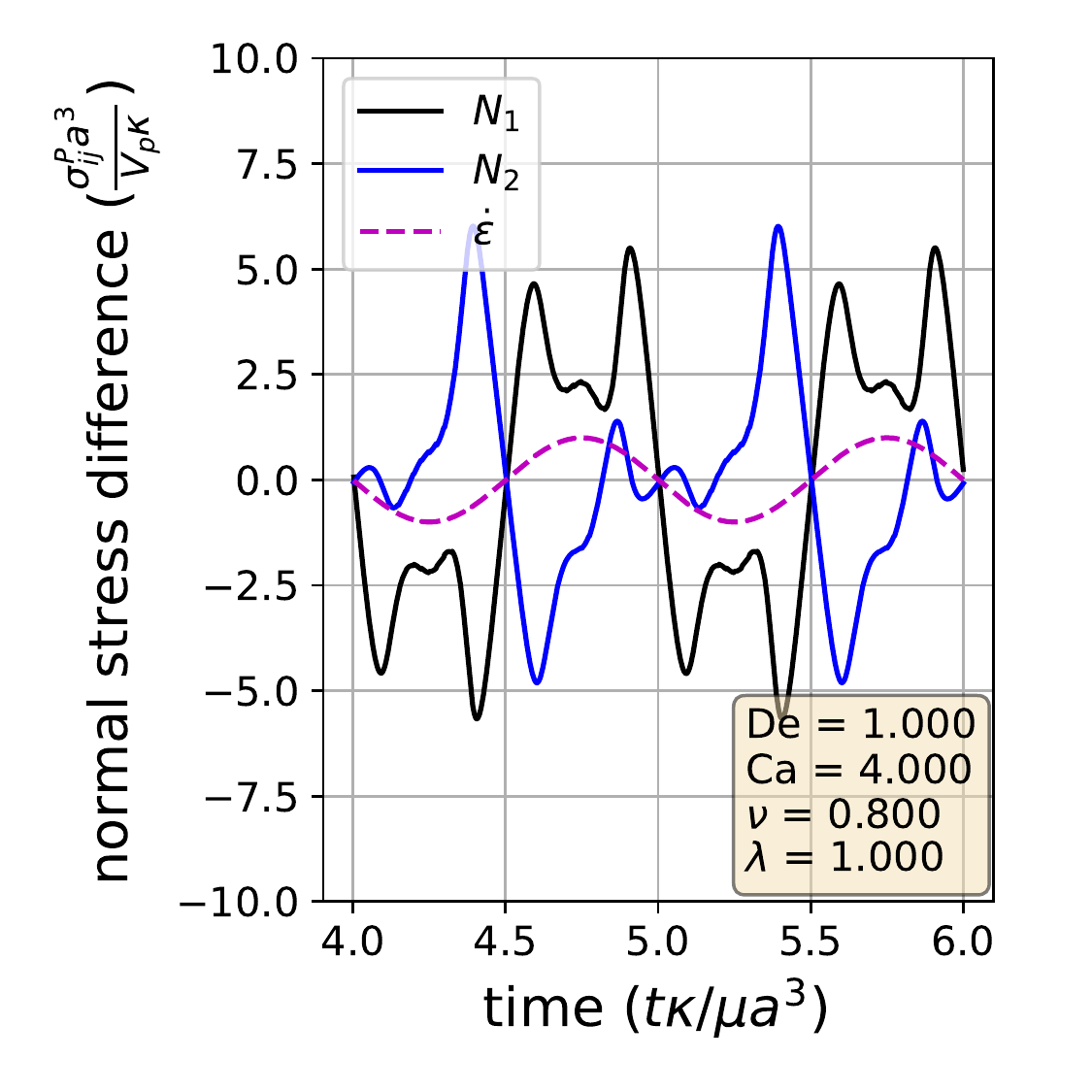}}\\
	\subfloat[Reorienting regime\label{fig:stress_response_med}]{\fTrim{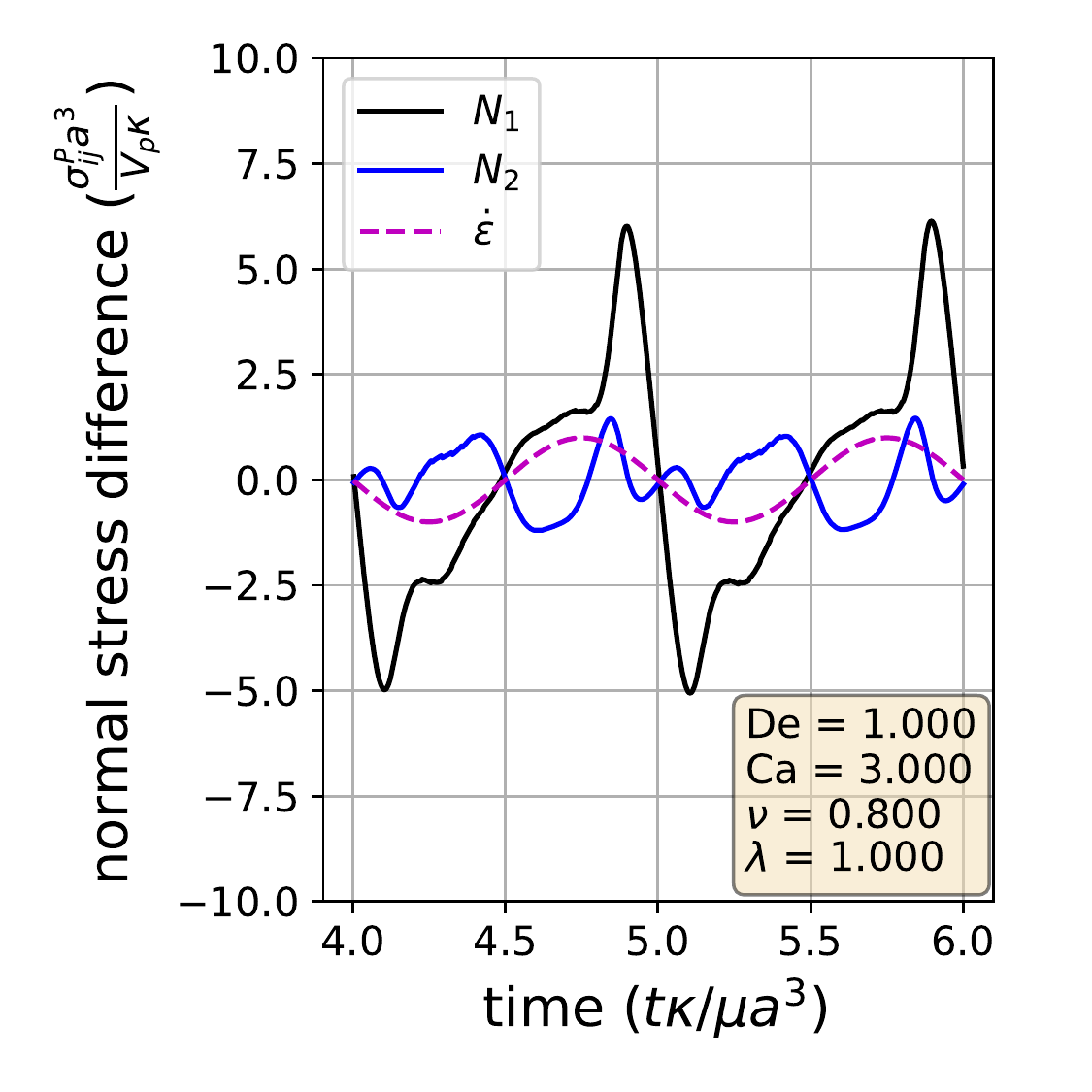}}
	\hfill
	\subfloat[Pulsating regime\label{fig:stress_response_low}]{\fTrim{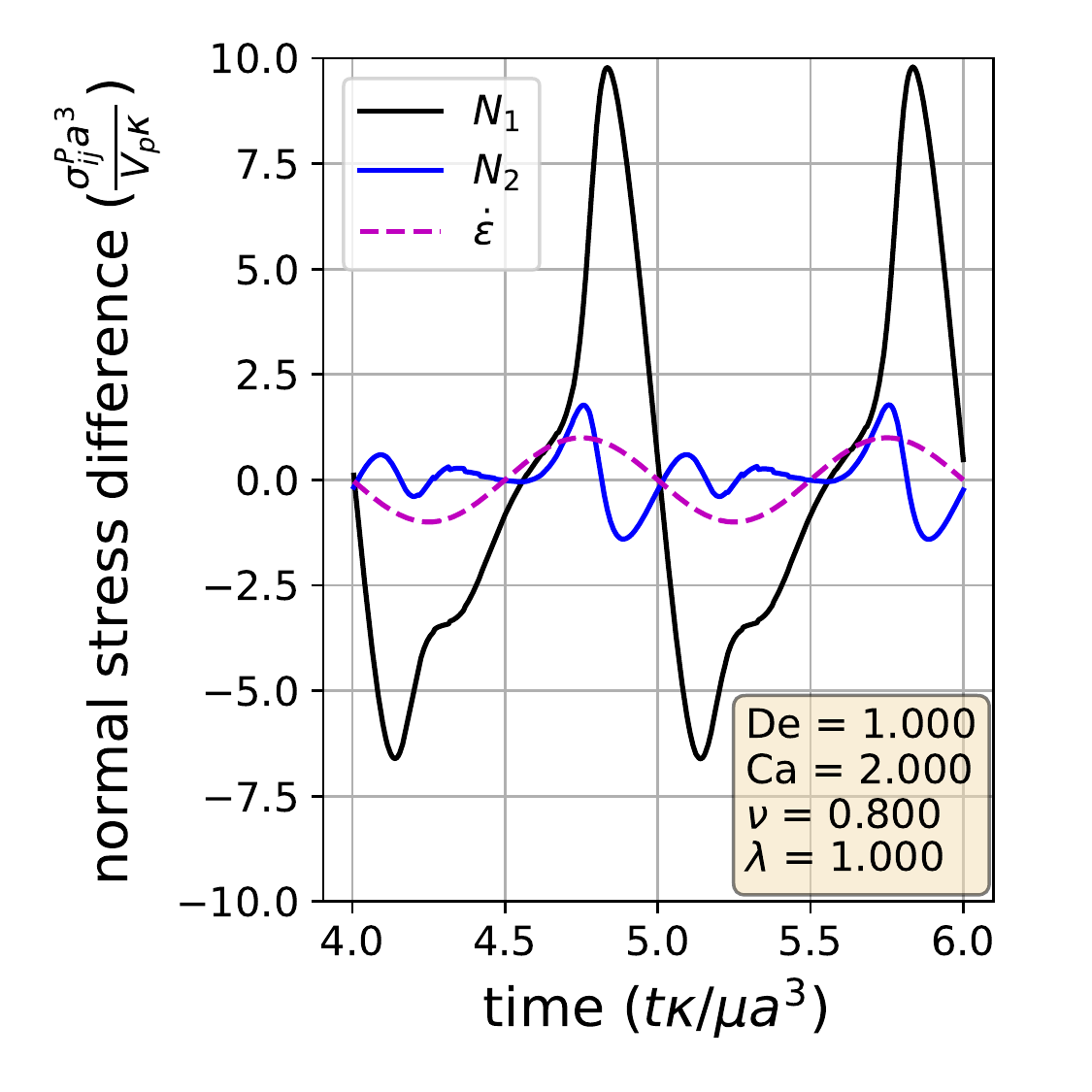}}
	\captionsetup{width=0.9\linewidth}
	\caption{%
		Normal stress differences versus time for simulations in the pulsating, reorienting, and symmetrical regimes.
		Data over two strain rate cycles is plotted.
		The $\dot{\epsilon}/\Ca$ dotted line is the strain rate of the external flow; it is used to show the directionality of the flow.
		Parameters used are included in the figure legends.
	}%
	\label{fig:stress_reponses}
\end{figure}

\subsection{Stress response and dilute suspension rheology\label{sec:stress_response}}

For dilute vesicle suspensions where the macroscopic length scale is large in comparison to the size of the vesicles, the extra stress (the bulk stress contribution from the particles) is the product of the number density of particles and the particle stresslet: $\sigma_{ij}^{P} = n \: \tilde{S}_{ij}^{P}$.
Using the boundary integral formulation, we calculate the particle 
stresslet~\cite{Pozrikidis1992}:
\begin{equation}
	\tilde{S}_{ij}^P = \Int{\frac{1}{2} ([[f_i]] x_j + [[f_j]] x_i)}{S, D} -
	\Int{(1 - \lambda) \mu_{\mathrm{out}} (v_i n_j + v_j n_j)}{S, D}
\end{equation}
where $[[\bm{f}]]$ is the surface traction, $\lambda$ is the viscosity ratio, $\mu_{\mathrm{out}}$ is the outer viscosity, $\bm{v}$ is the velocity, and $\bm{n}$ is the normal vector.
We define the dimensionless particle coefficient of stresslet as:
\begin{align}
	S_{ij} = \frac{\tilde{S}_{ij}^P}{\dot{\epsilon} \mu_{\mathrm{out}} V_p}
\end{align}
where $V_p$ is the vesicle volume and $\dot{\epsilon}$ is the strain rate.
Similarly the normal stress differences are defined as:
\begin{align}
	N_1 = S_{xx} - S_{yy} \\
	N_2 = S_{yy} - S_{zz}
\end{align}
Comparing the normal stress differences to the strain rate, we can derive the 
rheological characteristics of a dilute vesicle suspension, such as the 
effective viscosity and bulk normal stresses~\cite{Danker2008}.
For extensional flow rheology, a key quantity of interest is the extensional viscosity of a solution.
Extensional viscosity is often characterized using a quantity known as a Trouton ratio (ratio of extensional to shear viscosity), which for a planar extensional flow is a multiple of $N_1$.
For a planar flow, the extensional viscosity is
\begin{align}
    \eta_E = \frac{\sigma_{11} - \sigma_{22}}{\dot{\epsilon}}.
\end{align}
The planar Trouton ratio is
\begin{align}
    \frac{\eta_E}{\eta} = 4 + \phi * N_1,
\end{align}
where $\phi$ is the volume fraction of vesicles in the suspensions, and $N_1$ is the first normal stress difference.
Our simulations have focused on rather large deformations of the vesicle shape, therefore the stress response analysis will reflect the non-linear viscoelasticity.

Using the definitions of the particle stresslet and normal stress differences, we determine the vesicle stress as a function of time in extensional flow.
In \cref{fig:stress_reponses}, we show the stress response over two cycles for three sets of parameters; one from each of the three dynamical regimes discussed before.
A linearly viscoelastic material will show purely sinusoidal normal stress differences for this type of plot, as there is a simple linear relation between the strain rate and the stress.
On the other hand, for non-linear viscoelasticity, the normal stress differences will display more complex behaviors.

\begin{figure}
	\centering
	\captionsetup{width=0.40\linewidth}
	\subfloat[Symmetrical regime\label{fig:Lissajous_stress_reponse_sym}]{\fTrim{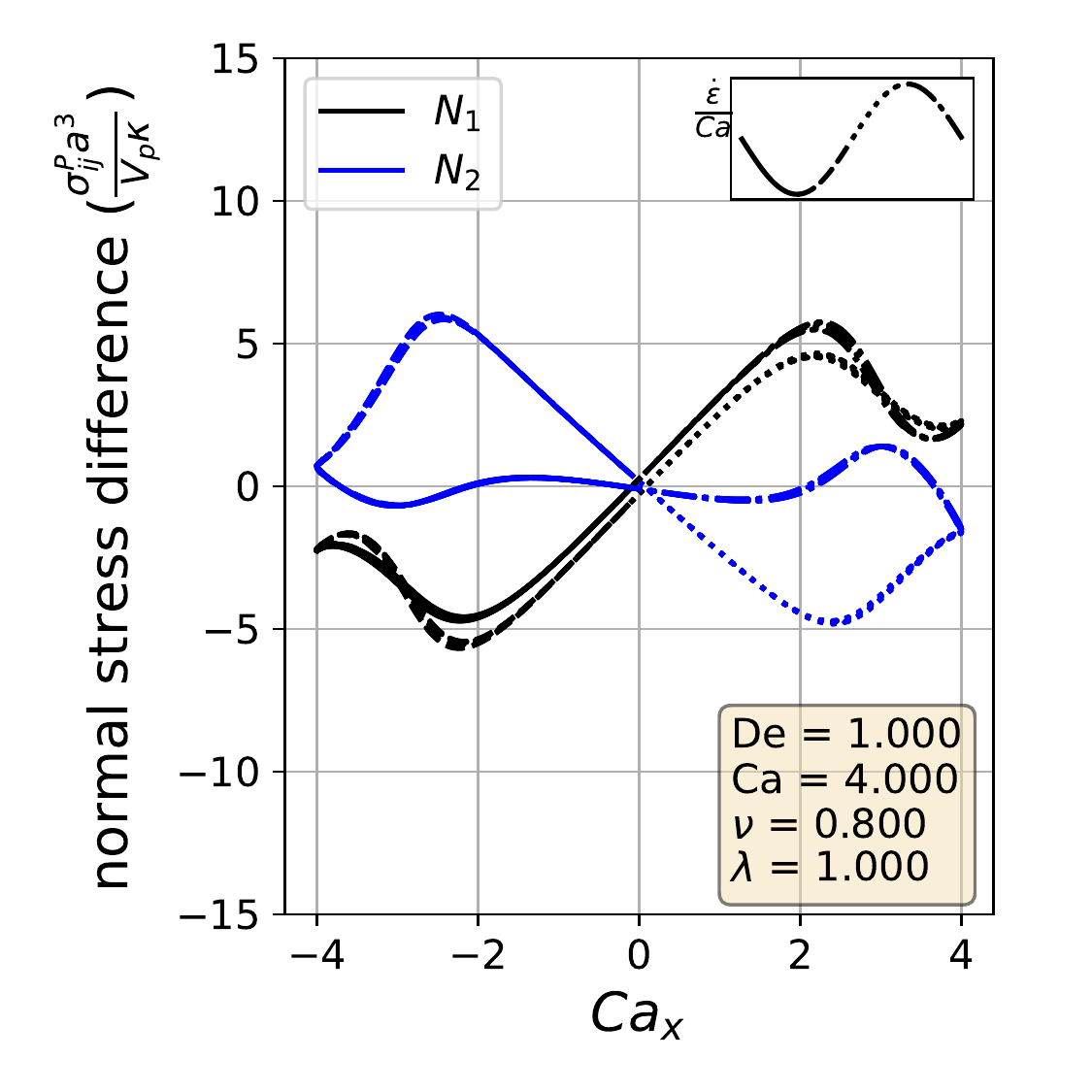}}\\
	\subfloat[Reorienting regime\label{fig:Lissajous_stress_reponse_reori}]{\fTrim{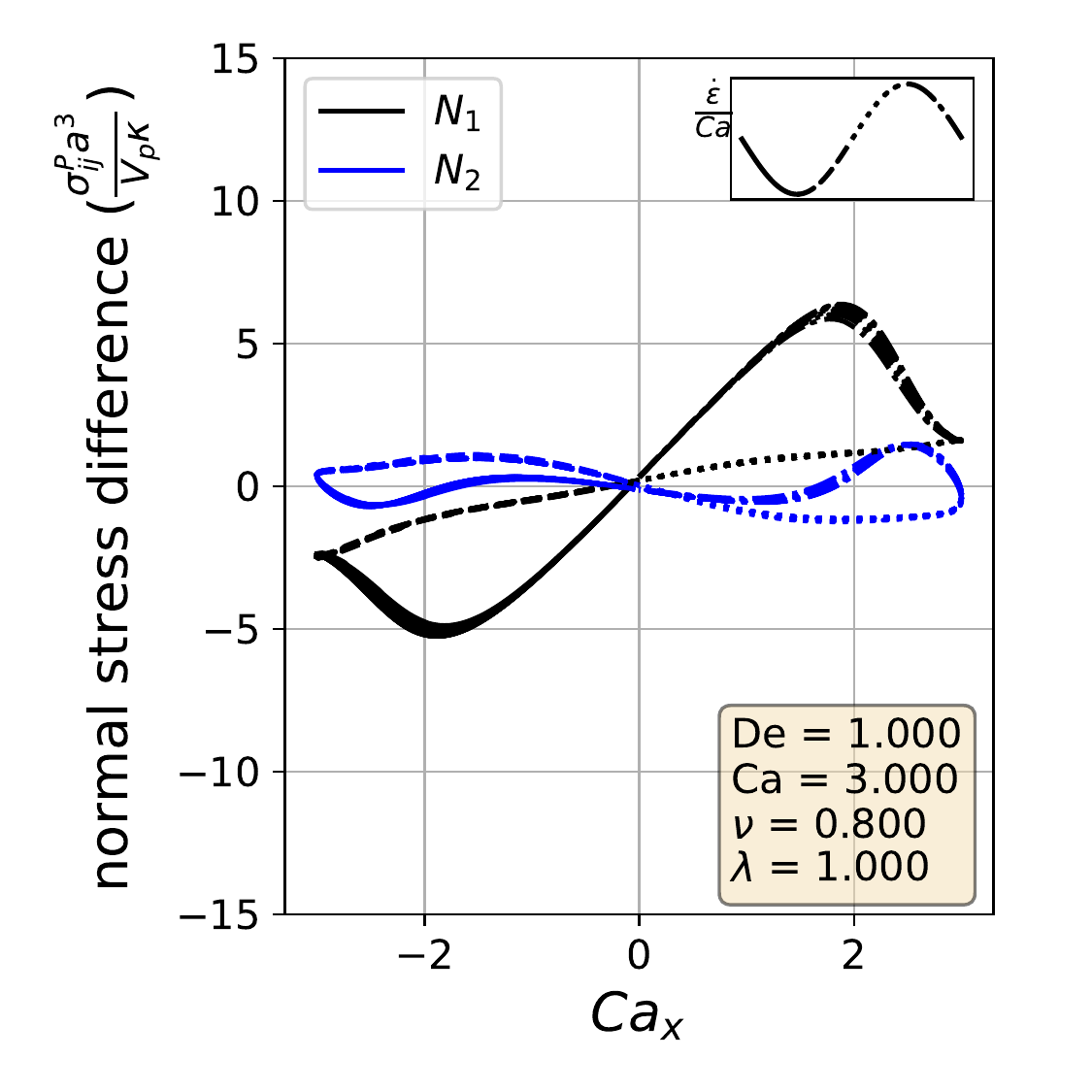}}
	\hfill
	\subfloat[Pulsating regime\label{fig:Lissajous_stress_reponse_puls}]{\fTrim{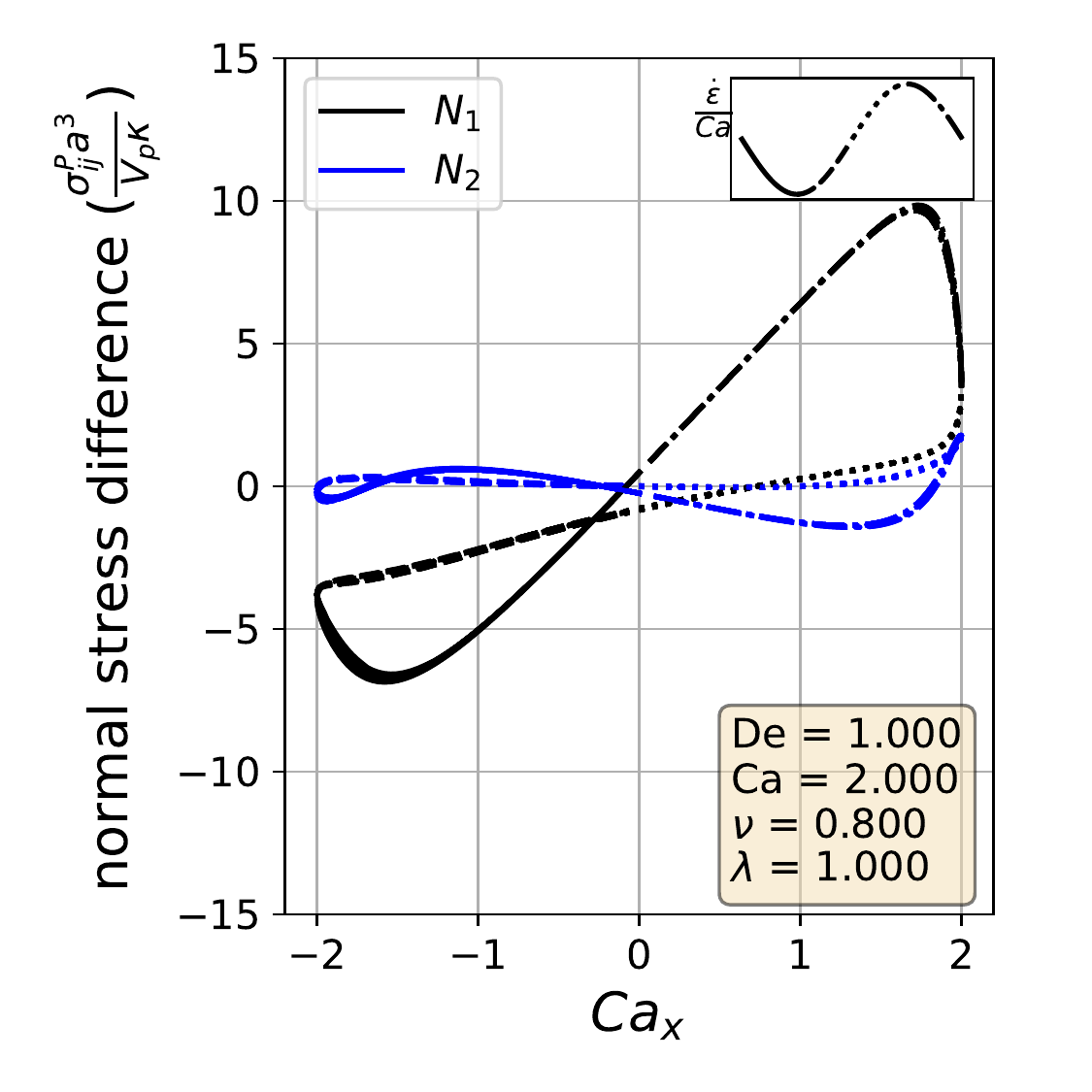}}
	\captionsetup{width=0.9\linewidth}
	\caption{%
		Lissajous-type normal stress difference versus strain rate ($\Ca_x$) curves for simulations in the pulsating, reorienting, and symmetrical regimes.
		The strain rate cycle is separated into four periods demarcated by the line formatting.
		Parameters used are included in the figure legends.
	}%
	\label{fig:Lissajous_stress_reponses}
\end{figure}

\cref{fig:stress_reponses} shows that vesicle dynamics in the three regimes (symmetrical, reorienting, and pulsating) have non-linear characteristics.
To analyze these stress responses, we re-plot the data from \cref{fig:stress_reponses} into a Lissajous-type form with the instantaneous strain rate ($\Ca_x$) on the x-axis and the stress response on the y-axis (\cref{fig:Lissajous_stress_reponses}).
For this type of plot, a purely viscous material would display a straight line, whereas a purely elastic material would produce an elliptical curve.
For example, the first and second normal stress difference for Newtonian flow around a rigid sphere corresponds to the lines: $N_1 = 10 \cdot \Ca_x/\Ca$ and $N_2 = -5 \cdot \Ca_x / \Ca$.
Here, we focus on $N_1$ because it is related to the extensional viscosity of the solution (Trouton ratio).
We also discuss the $N_2$ stress differences for completeness.

In the symmetrical regime (\cref{fig:Lissajous_stress_reponse_sym}), we observe that $N_1$ is symmetric across the origin and that the lines for increasing and decreasing strain rate are nearly the same for $-2 < \Ca_x <2$.
On the other hand, $N_2$ differs significantly depending on the directionality of the flow.
The $N_1$ curve is mostly linear in the region $-2 < \Ca_x < 2$ and is approximately equal to zero when the strain rate is zero, suggesting that the vesicle contributes a purely viscous response in that region.
We further examine this region in more detail by comparing the vesicle deformation to the stress response.
From the simulation video and the Lissajous-type deformation parameter curve (\cref{fig:deform_liss_comp}), we know that the vesicle retains a prolate spheroid like shape and only changes marginally for the $-2 < \Ca_x < 2$ region.
The relatively small amount of deformation that occurs in the $-2 < \Ca_x < 2$ region suggests that the vesicle acts like a rigid particle there, explaining the close to linear stress response for $N_1$ in the region.
In the other strain rate regions, the stress differences shift rapidly in accordance to the vesicle's large deformations and reorientation.

In the reorienting and pulsating regimes (\cref{fig:Lissajous_stress_reponse_puls,fig:Lissajous_stress_reponse_reori}), the $N_1$ curves are no longer symmetric across the origin, and the stress responses for increasing and decreasing strain rate are distinct.
The maximum $N_1$ response is larger in magnitude than the minimum for both regimes; this is likely due to the unequal amounts of deformation between the two strain rate period halves (\cref{fig:deform_liss_comp}).
For this analysis, qualitative differences between the shape of the reorienting and pulsating regime curves correspond to the extent of asymmetry in the $N_1$ response.
Moreover, we observe vesicles in the pulsating regime can have a non-zero normal stress difference when the time-dependent strain rate is zero, as seen in \cref{fig:Lissajous_stress_reponse_puls}.

For a more quantitative analysis, we decomposed the stress responses into a Fourier series.
This decomposition is commonly applied to large amplitude oscillatory shear (LAOS) experiments and is known as Fourier transform (FT) rheology.
FT rheology is commonly performed using oscillatory shear flows on polymeric 
liquids to probe the shear stress response in the non-linear 
regime~\cite{hyun2011review, wilhelm2002fourier}.
The computation is straightforward and relies on taking the Fourier transform of the $N_1$ or $N_2$ stress difference:
\begin{equation}
	f(k) = \Int{N_{1, 2}(t) e^{-2 \pi i t k}}{t, -\Infinity, \Infinity},
\end{equation}
In this way, the periodic stress signal is transformed into frequency space.
Because the external flow field is sinusoidal, the strain rate ($\dot{\epsilon}$) and strain ($\epsilon$) are proportional to sine and cosine functions.
Therefore, the Fourier transformed data provide a description of how the stress depends on different orders of the strain and strain rate.
If the stress response was purely linear order, the Fourier transformation would show a single peak at the first mode.
A non-linear stress response would have additional peaks at higher modes.

\begin{figure}
	\centering
	\captionsetup{width=0.40\linewidth}
	\subfloat[Symmetrical regime]{\fTrim{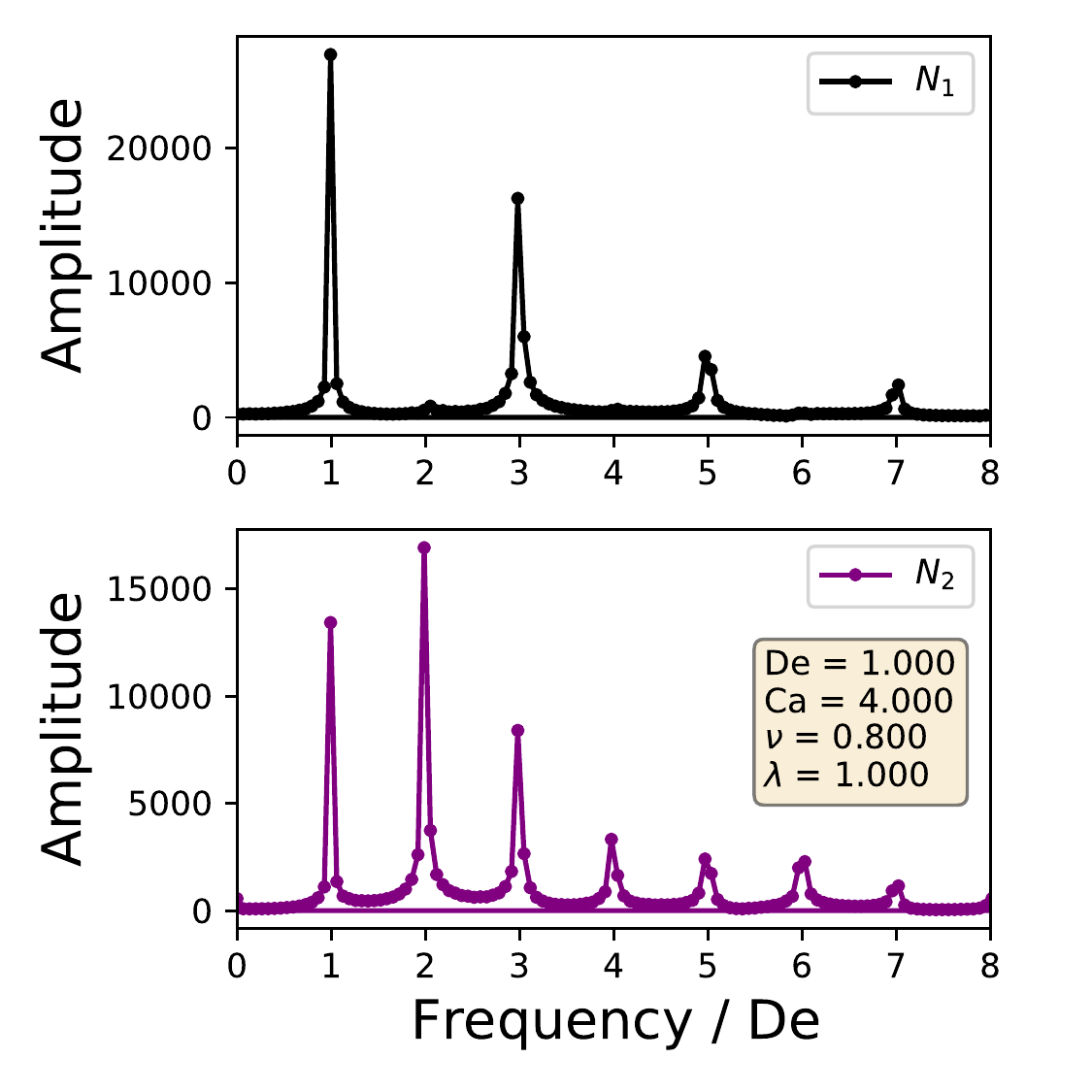}}\\
	\subfloat[Reorienting regime]{\fTrim{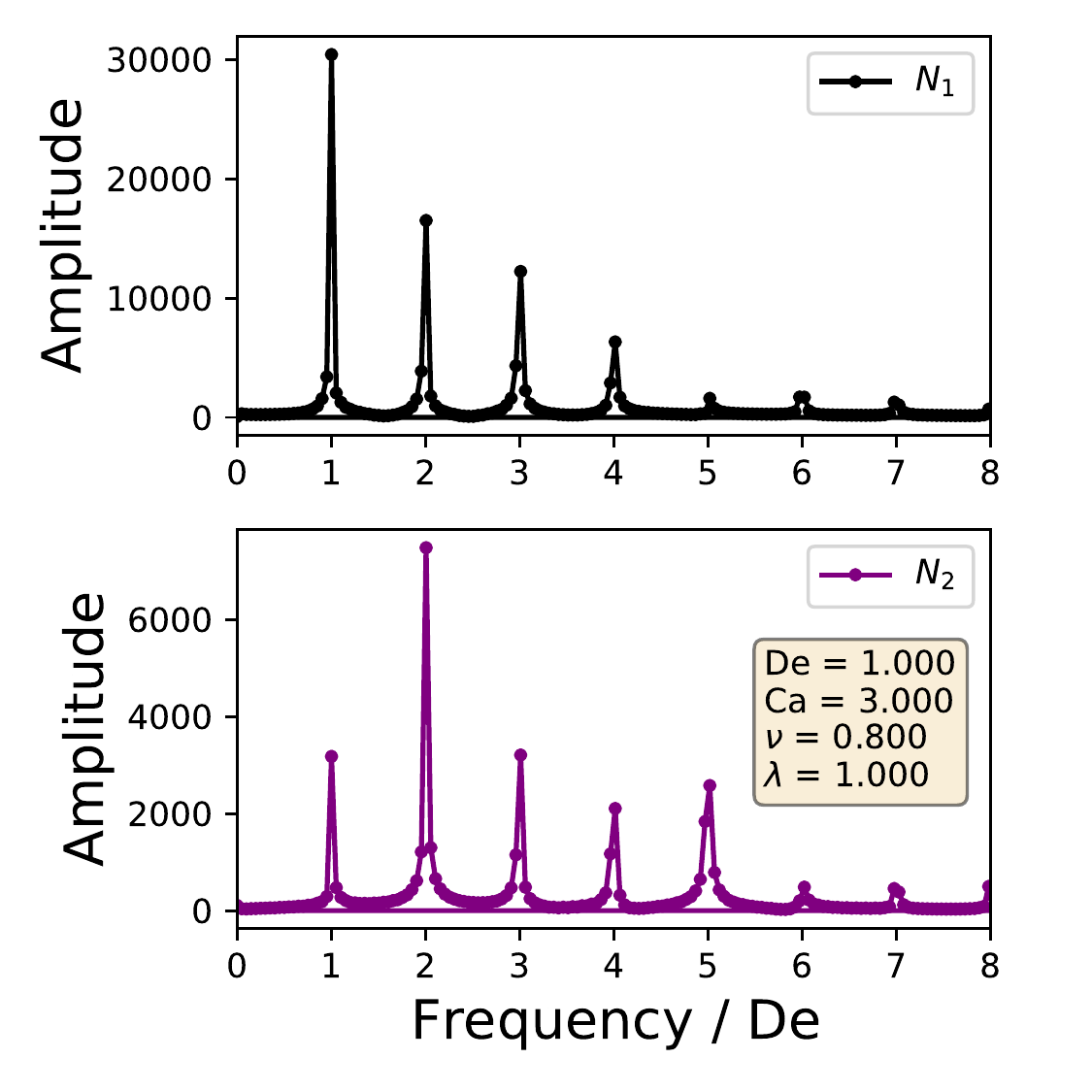}}
	\hfill
	\subfloat[Pulsating regime]{\fTrim{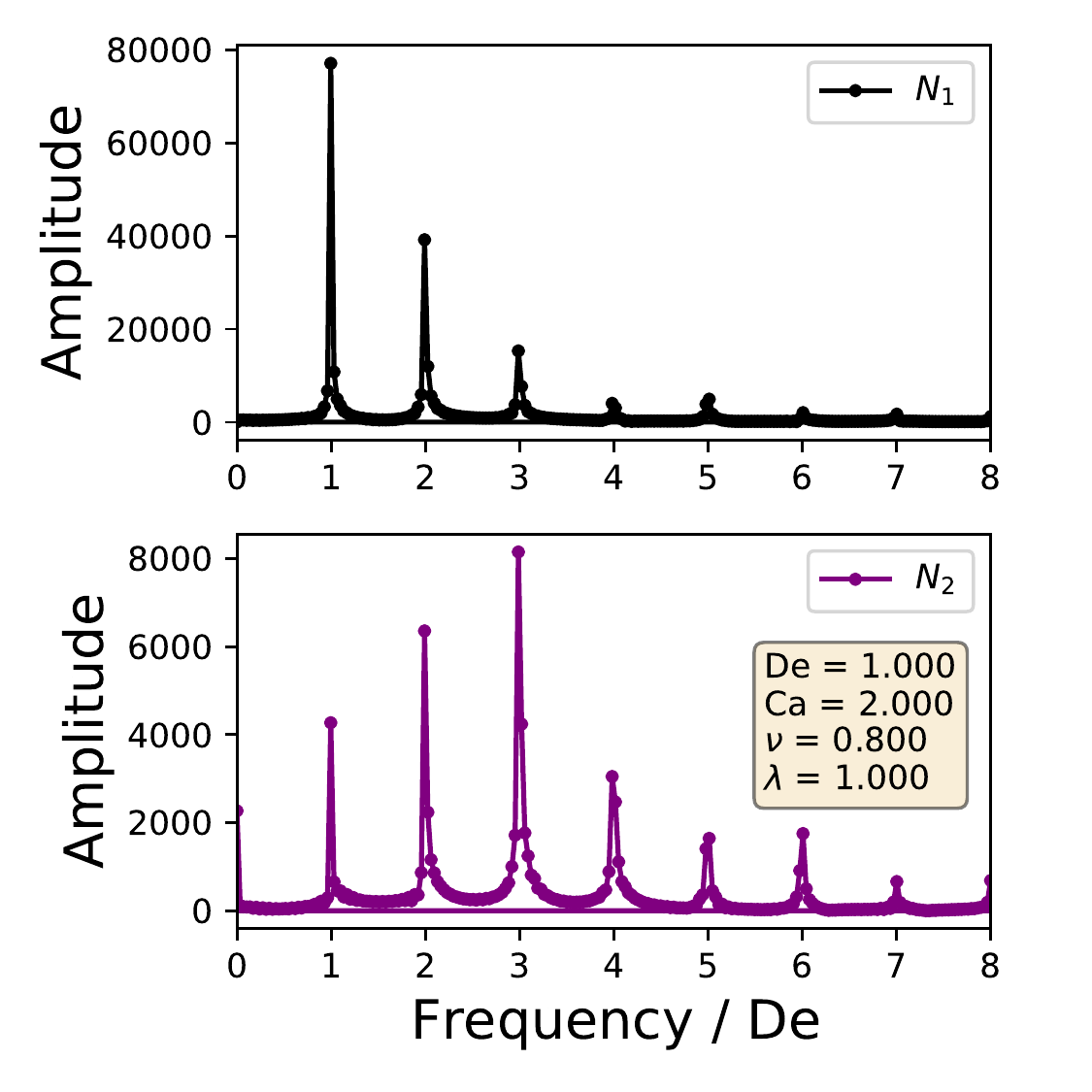}}
	\captionsetup{width=0.90\linewidth}
	\caption{%
		Fourier decompositions of the stress responses for indicative parameter sets in each of the dynamical regimes.
	}%
	\label{fig:fourier_stress_reponses}
\end{figure}

The Fourier decompositions for both $N_1$ and $N_2$ are shown in \cref{fig:fourier_stress_reponses}, where it is clear that all three regimes show higher order behavior.
For all regimes, we observe the expected behavior of the linear order mode being the highest amplitude with the higher order modes decreasing monotonically for $N_1$.
On the other hand, the highest amplitude mode for $N_2$ is not the linear order mode, with the highest generally being the second or third mode.
Comparing the $N_1$ decompositions between the dynamical regimes, we observe that the symmetrical regime does not have even order modes, whereas the reorienting and pulsating regimes have even higher order modes.
This change in FT rheology is consistent with the phase boundary defined in \cref{sec:dynamical_regimes}, and this transition can be used instead of the deformation parameter analysis to demarcate the phase boundary.

In large amplitude oscillatory shear (LAOS), the typical macroscopic stress 
response shows that the stress is an odd function of the direction of 
shearing\cite{hyun2011review}.
Such a restriction is not necessarily expected in an extensional flow, but would be related to whether the microstructure of the fluid stretches symmetrically during these flows.
In the symmetrical regime, both the vesicle stress response and deformation are time symmetric, leading to only odd order Fourier modes.
The time symmetry does not hold for the reorienting or pulsating regimes, allowing for even order modes.
Based on the currently available results, we do not expect droplets to have 
even order Fourier modes in LAOE, regardless of flow rate or flow 
frequency~\cite{li2005numerical}.
Broadly speaking, our results show that membrane-bound vesicles are an interesting example of how anisotropic microstructural deformations can lead to complex rheology.

\subsection{Transient dynamics of tubular vesicles in large amplitude oscillatory extension\label{sec:tubular_vesicles}}

\begin{figure}
	\begin{center}
		\includegraphics[width=0.6\textwidth]{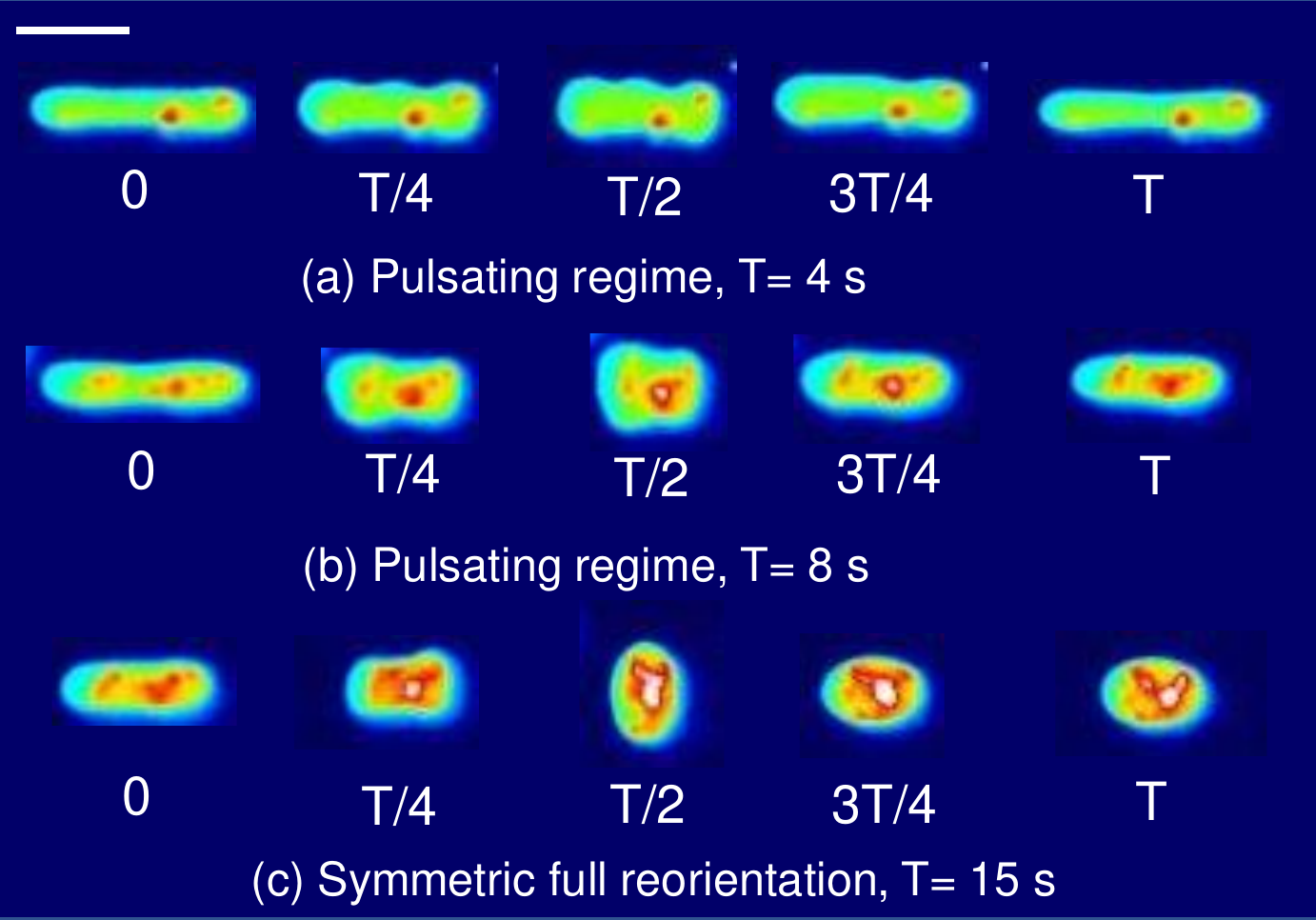}
		\caption{\label{fig:tubular_nu0.63} Dynamics of a tubular vesicle with reduced volume $\nu = 0.64 \pm 0.02$ in LAOE.
			(a) Snapshots showing pulsating dynamics of a vesicle over one sinusoidal strain rate input cycle with time period $T= 4$ s at $Ca=21.3$ and $De=17.7$.
			(b) Snapshots showing pulsating dynamics with wrinkles of a vesicle over one sinusoidal strain rate input cycle with time period $T= 8$ s at $Ca=21.3$ and $De=8.9$.
			(c) Snapshots showing change in 2D shape of a vesicle over one flow cycle with time period $T= 15$ s at $Ca=21.3$ and $De=4.7$. Scale bar is 20 $\mu$m. False coloring is applied to the grayscale images for resolution enhancement.
		}
	\end{center}
\end{figure}

We also investigated the transient dynamics of tubular vesicles in large amplitude oscillatory extension (\cref{fig:tubular_nu0.63}).
In general, we find that tubular vesicles undergo wrinkling/buckling instabilities during the compression phase of the flow cycle similar to quasi-spherical vesicles.
However, we occasionally observe buckling instabilities that induce unexpected shape changes.
In these situations, the vesicle’s initial, tubular shape is not recovered at the end of the flow cycle.

\begin{figure}
	\centering
	\captionsetup{width=0.3\linewidth}
	\subfloat[Ca = 21.3, De = 17.7, $\nu$ = 0.64]{
		\gTrim{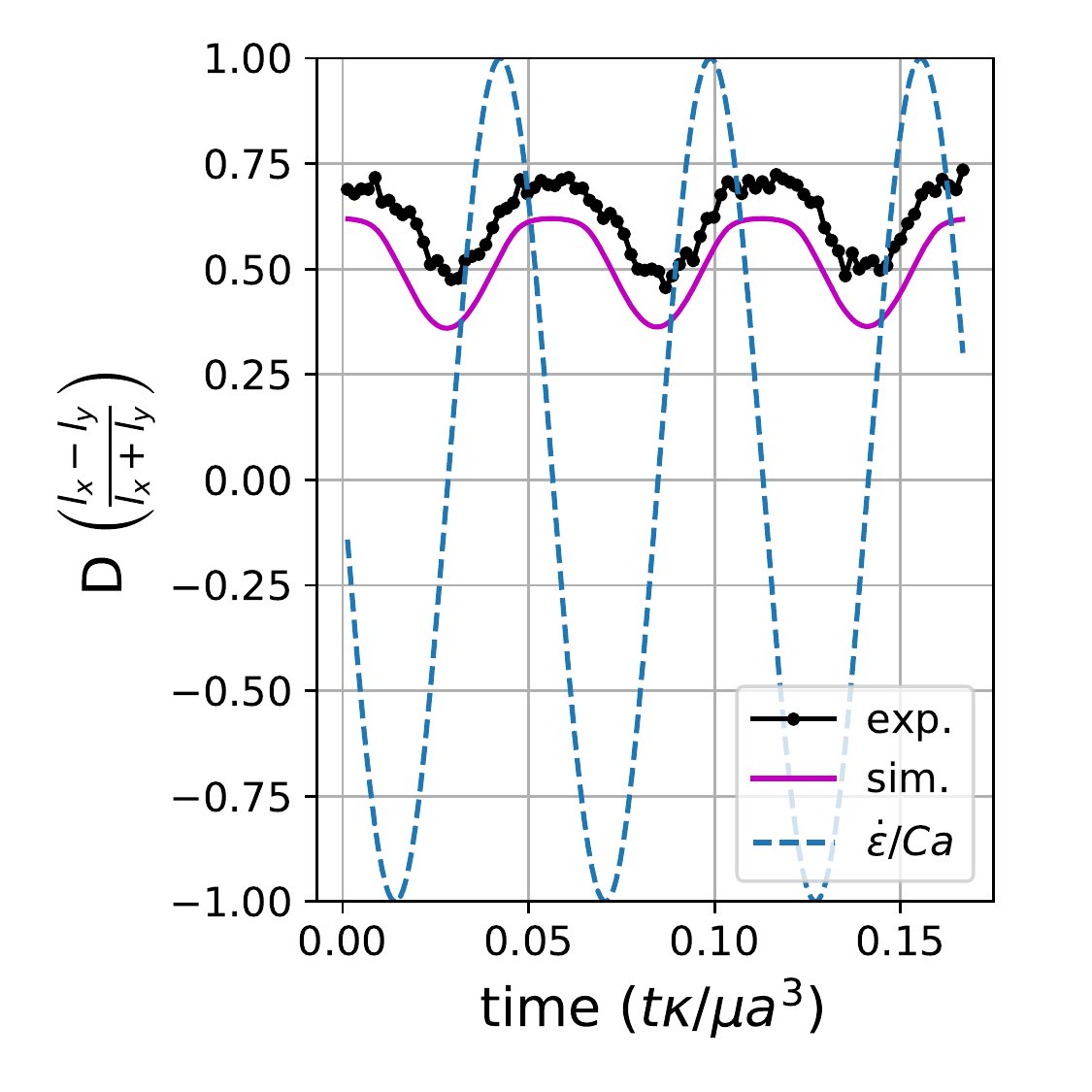}
	}
	\subfloat[Ca = 21.3, De = 8.9, $\nu$ = 0.64]{
		\gTrim{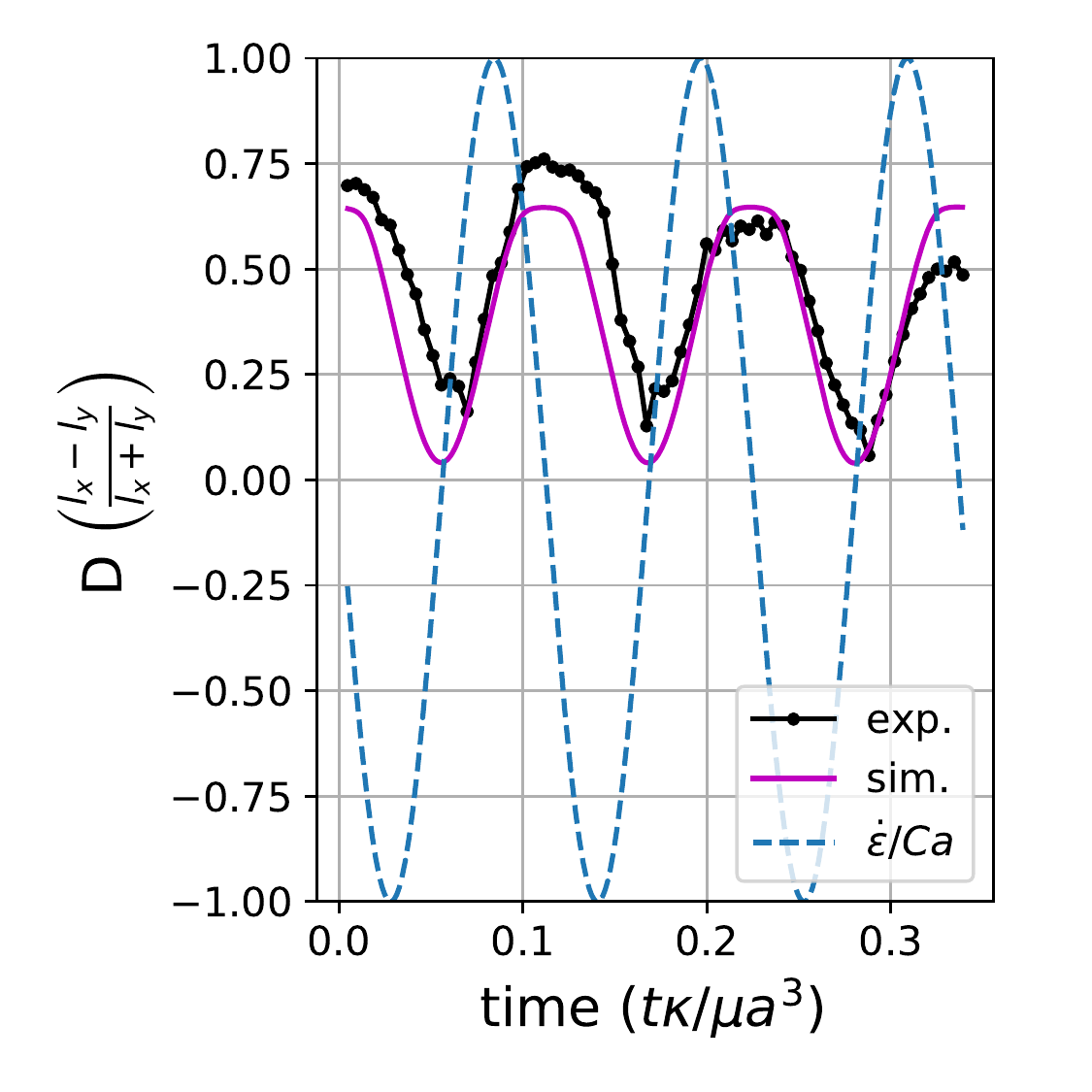}
	}
	\subfloat[Ca = 21.3, De = 4.7, $\nu$ = 0.64]{%
		\gTrim{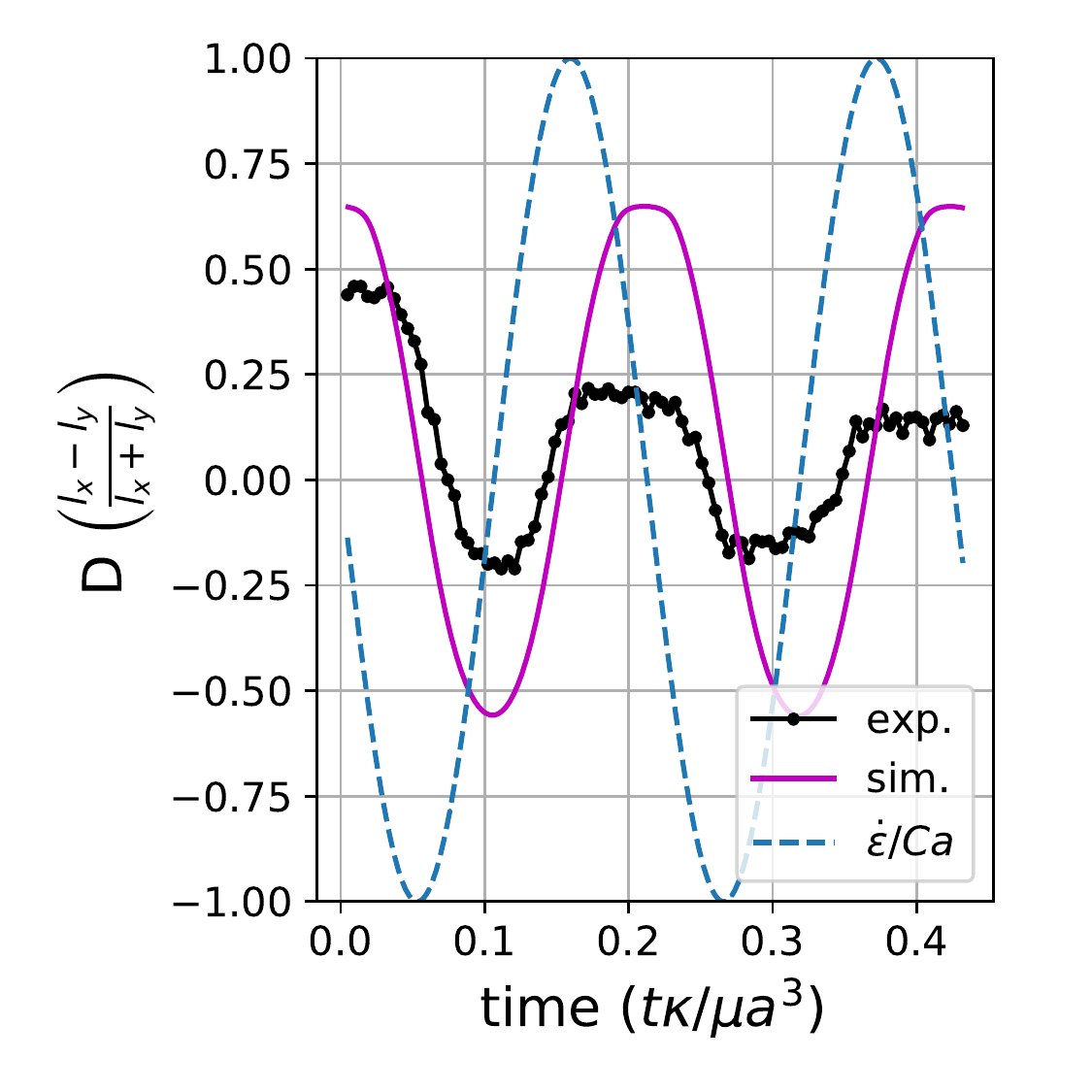}
	}\\

	\subfloat[Ca = 21.3, De = 17.7, $\nu$ = 0.64]{
		\gTrim{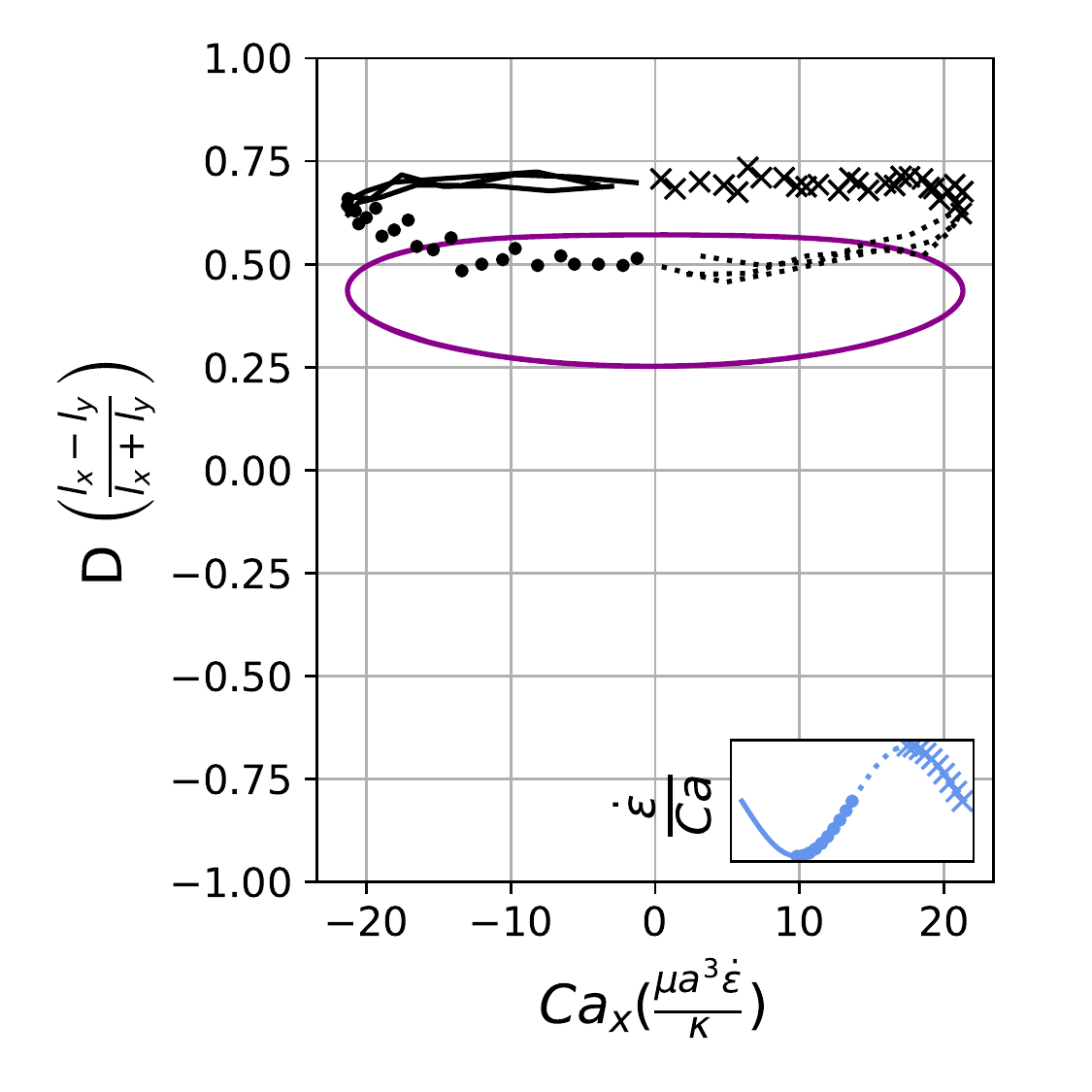}
	}
	\subfloat[Ca = 21.3, De = 8.9, $\nu$ = 0.64]{
		\gTrim{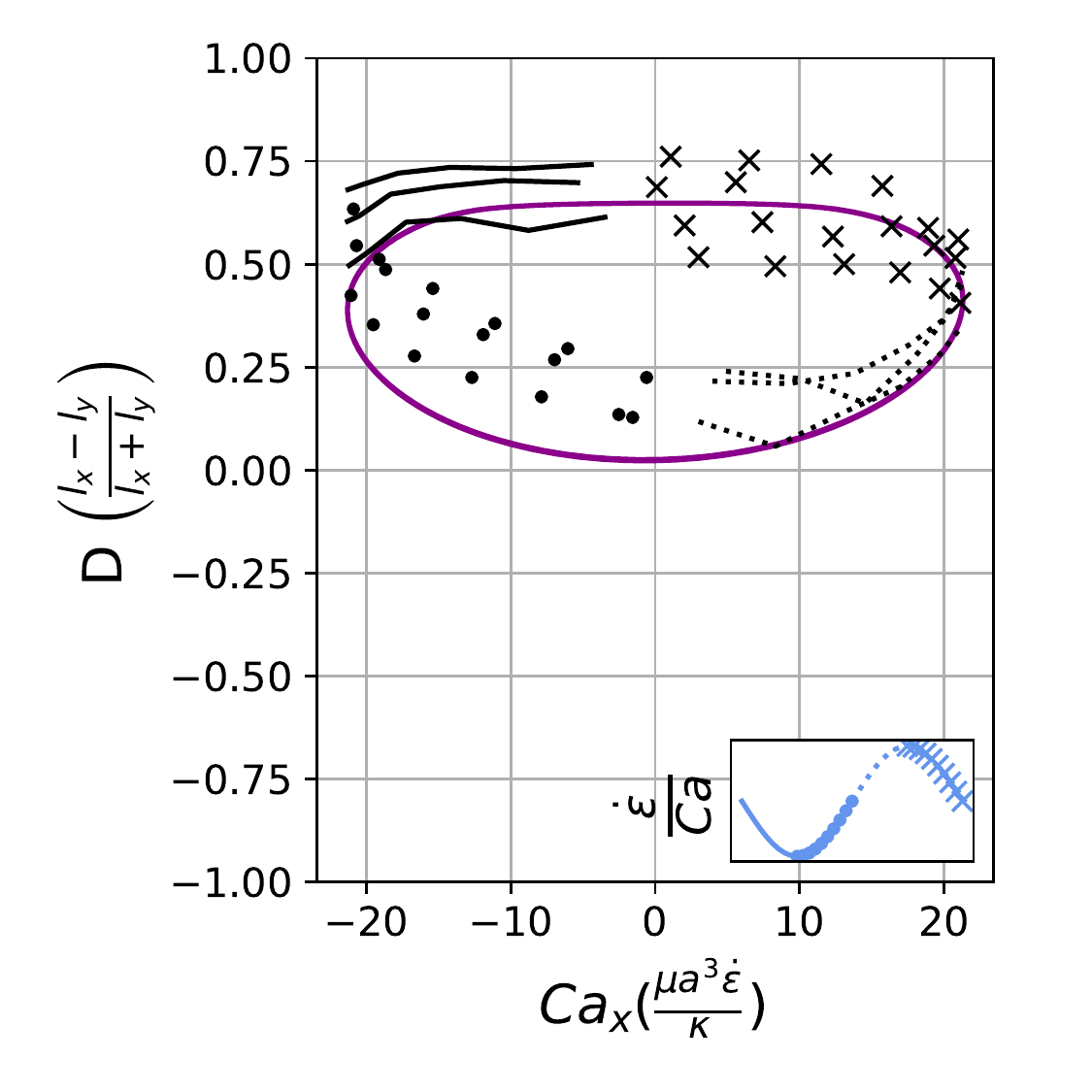}
	}
	\subfloat[Ca = 21.3, De = 4.7, $\nu$ = 0.64]{%
		\gTrim{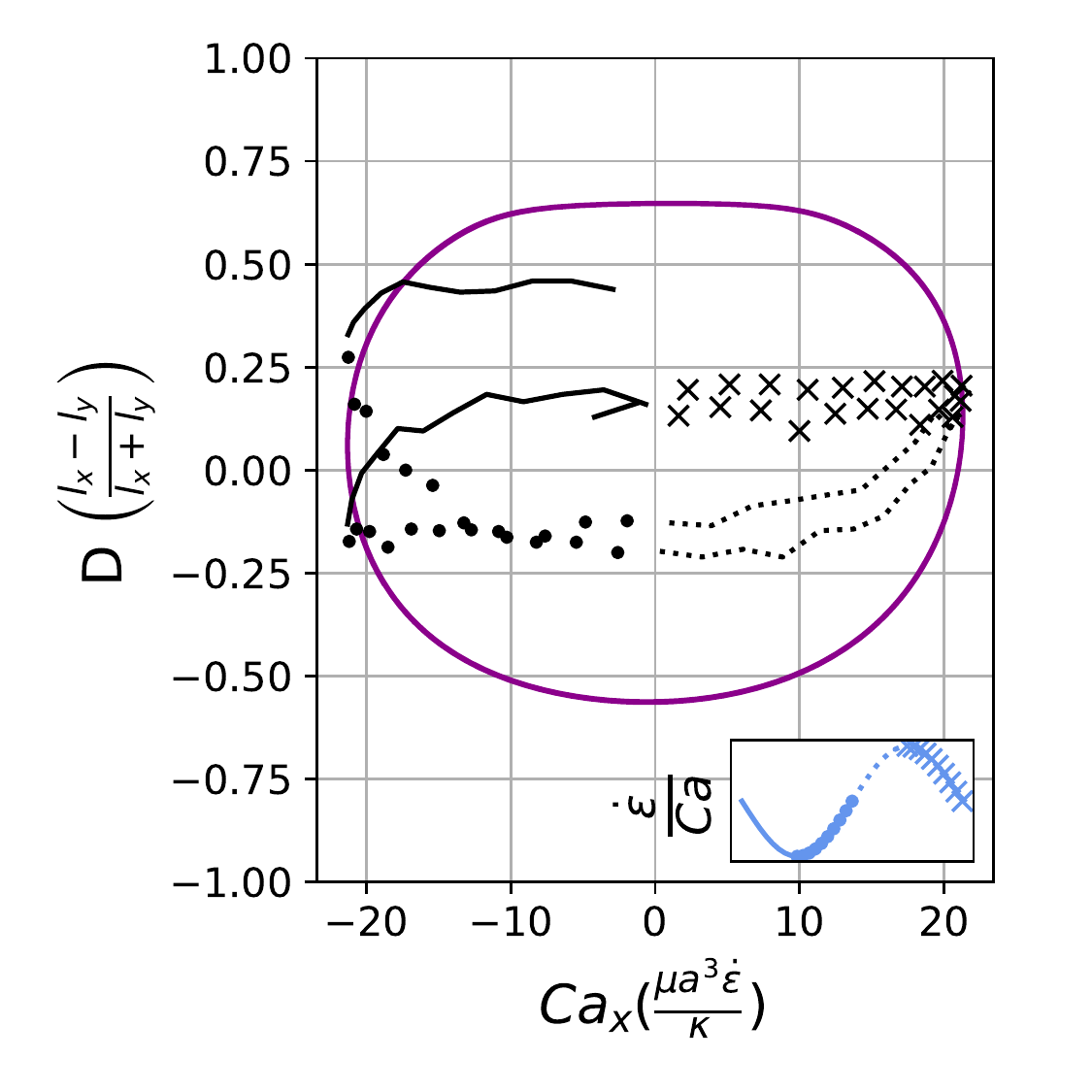}
	}
	\captionsetup{width=0.8\linewidth}
	\caption{%
		Experimental and simulation single vesicle Lissajous curves and deformation plots for $\nu=0.64$.
	}
	\label{fig:tubular_c2e}
\end{figure}

\cref{fig:tubular_nu0.63}a shows experimental snapshots of a tubular vesicle with reduced volume $\nu=0.64\pm0.02$ exposed to a sinusoidal strain rate at $\Ca = 21.3$ and $\De = 17.7$.
In this situation, the vesicle exhibits pulsating motion along the x-axis with buckles during the compressional part of the flow cycle.
The vesicle's starting, tubular shape is recovered at the end of the LAOE cycle.
To further demonstrate this behavior, we construct single vesicle Lissajous curves (\cref{fig:tubular_c2e}(d)) defined as plot of deformation parameter as a function of $\Ca$, and deformation parameter as a function of time (\cref{fig:tubular_c2e}(a)).
These plots show the vesicle reaches the same value of deformation parameter $D \approx 0.7$ at the end of each of the three repeated flow cycles, implying that the vesicle conformation is fully recovered after deformation.
There is decent agreement in the qualitative dynamics between the simulations and experiments in this region, but the simulated deformation parameters appear to be lower than the ones measures experimentally.

When the same vesicle is exposed to a flow cycle at a lower frequency ($\De=8.9$), the membrane has more time to deform in response to the flow.
Here, the vesicle undergoes pulsating motion with wrinkles (\cref{fig:tubular_nu0.63}b) and we observe appreciable deformation along $y$ axis in both the simulations and experiments, as shown in \cref{fig:tubular_c2e}b,e.
Surprisingly, the experimental results show the vesicle deformation parameter reducing with each subsequent LAOE cycle.
The deformation at the end of first cycle is $D\approx0.7$ and it decreases to $D\approx0.6$ at the end of second cycle, and further to $D\approx0.5$ at the end of third cycle.
Experimentally, it seems that the vesicle conformation changes over each LAOE cycle while our simulations predict no change over the strain rate cycles.
By the end of third repeated cycle, we experimentally observe that the 2D shape of vesicle appears to be more spheroidal than tubular.
Interestingly, the vesicle did not recover its original tubular shape even when relaxed for $\approx$ 2 min.
It is noteworthy that we did not observe any reduction in deformation parameter at the higher flow frequency discussed previously (De = 17.7).
These observations suggest that for a given $\Ca$, there appears to be a critical $\De$ below which the change occurs.

Finally, the same vesicle is exposed to LAOE flow cycle with an even lower frequency (De = 4.7).
We observe that the vesicle undergoes full reorientation from the $x$ axis to $y$ axis, undergoes a wrinkling instability during compression and the initial spheroidal shape changes to a more spherical shape at the end of the  first periodic cycle (\cref{fig:tubular_nu0.63}c).
The deformation behavior seen experimentally during the second repeated cycle is symmetric and follows similar dynamics as those observed for quasi-spherical vesicles.
This behavior is more apparent in \cref{fig:tubular_c2e}c,f which shows a slight reduction in deformation at the end of first cycle.
We observe a large difference in deformation between the simulations and experiments at these parameters.
Where the simulations predict the vesicle stretching to $D \approx 0.63$, the experiments only reach $D \approx 0.25$.
Additionally, the simulations show that the vesicle does not deform symmetrically at these parameters, reaching $D \approx -0.5$ and $D \approx 0.6$.
The experiments were performed sequentially from the higher to lower $\De$ on the same vesicle in the experiments, and it seems that the gradual change in vesicle deformation carried over from the previous experiments.

In summary, the experimental data in \cref{fig:tubular_nu0.63} and \cref{fig:tubular_c2e} shows that the maximum deformation of tubular vesicles may decrease in repeated LAOE cycles and the initial tubular shape may not be recovered.
In contrast, the quasi-spherical vesicles always recover a prolate shape following repeated LAOE deformation cycles.
We conjecture that the observation of shape transition from prolate tubular to 
oblate spheroid during LAOE deformation in \cref{fig:tubular_nu0.63}b,c can be 
explained in the context of the area-difference elasticity 
model~\cite{Seifert1997}.
Briefly, the negative membrane tension on the vesicle membrane during the 
compressional phase of LAOE flow leads to a decrease in area per lipid which 
reduces the preferred monolayer area 
difference~\cite{avital2015small,sakashita2012three}.
The decrease in monolayer area difference triggers the shape transition from a 
prolate tubular shape to an oblate spheroid in accordance with the ADE 
model~\cite{ziherl2005nonaxisymmetric, Seifert1997}.
This hypothesis is consistent with prior observations where the prolate to 
oblate transition was triggered by chemical modification of the ambient 
environment of vesicles~\cite{kodama2018migration}.
Resolving what exactly is occurring during compressional flow requires additional experiments, likely with 3D confocal microscopy to obtain the full three dimensional vesicle shape.

Additional experimental data on dynamics of highly deflated vesicles ($\nu=0.35$) is included in the Supplementary Information (Fig. S6 and Fig. S7).

\begin{figure}
	\begin{center}
		\includegraphics[width=0.7\textwidth]{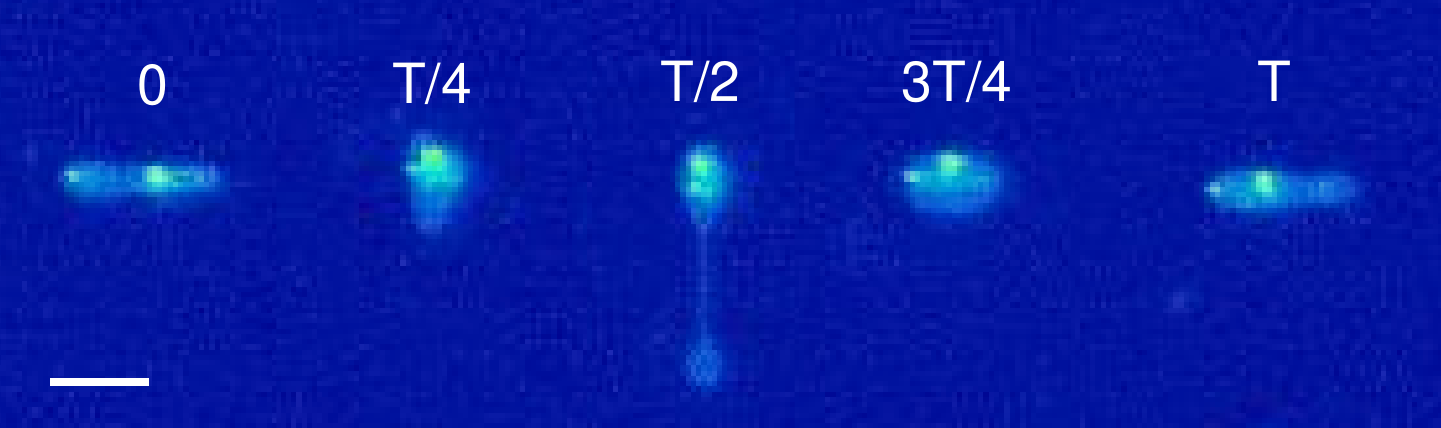}
		\caption{\label{fig:asym_dumbbell} Asymmetric dumbbell formation in
			a vesicle with reduced volume $\nu=0.69$ exposed to LAOE flow at
			$\Ca=52.5$
			and $\De=1.2$.
		Scale bar is 10 $\mu$m.}
	\end{center}
\end{figure}

In steady extensional flow with $\De=0$, the critical capillary number required 
to trigger dumbbell shape transition is a function of reduced volume and the 
comprehensive phase diagram in $\Ca-\nu$ space has been reported in an earlier 
work~\cite{kumar2020conformational}.
\cref{fig:asym_dumbbell} qualitatively demonstrates how oscillatory extensional flow alters these shape instabilities.
At $\De=1.2$, we observe that the critical capillary number $\Ca$ required to induce asymmetric dumbbell is much higher compared to steady extensional flow at $\De=0$.
For instance, the critical $\Ca$ required to generate asymmetric dumbbell in 
steady extension for $\nu=0.69$ is $\approx$ 5.3 \cite{kumar2020conformational}.
However, in LAOE flow at $\De=1.2$, the transition to dumbbell shape occurs at 
$\Ca=52.5$ which is approximately ten times higher than the critical $\Ca$ for 
steady flow. This observation can be rationalized by considering the 
competition between flow cycle time $T$ and inverse of the predicted growth 
rate of asymmetric instability from linear stability 
analysis~\cite{Narsimhan2015}. Briefly, the presence of flow oscillations ($\De 
> 0$) prevents any instability formation which requires a time scale larger 
than cycle time $T$. Thus, a large $\Ca$ is needed to reduce the time scale of 
instability sufficiently to observe the dumbbell formation within the flow 
cycle time $T$. While it is possible to explore the phase diagram describing 
conformation change to asymmetric/symmetric dumbbell on $\Ca-\De$ space for the 
entire range of reduced volumes using the Stokes trap, the parameter space is 
vast and it remains a ripe area for future numerical simulations.

\section{Conclusions}
In this work, we examined the dynamics of vesicles in large amplitude oscillatory extensional (LAOE) flow using both experiments and numerical simulations.
The experiments were carried out using the Stokes trap experimental technique while the simulations were done with the boundary element method.
For quasi-spherical vesicles, the simulations are found to capture the transient wrinkling dynamics as well as the overall vesicle shapes from experiments.
We have identified three dynamical regimes based on their deformation characteristics and named them the symmetrical, reorienting, and pulsating regimes.
Based on these results, we generated a phase diagram in capillary number and Deborah number space for the dynamical regimes; our data suggest that the phase boundaries are linear.
The unique deformation observed in the pulsating and reorienting regimes also has interesting effects on the stress response in that the time symmetry of the stress does not hold.
Additional analysis of the stress response and confirmation by experimental studies is required for a better idea of the dynamics. Finally, we presented results on highly deflated tubular vesicles which shows that lower reduced volume vesicles tend to undergo a shape change following repeated LAOE deformation.
From a broad perspective, we have shown through experiments and simulations that the vesicle system shows interesting dynamics in extensional oscillatory flows.
We have also shown how microstructural changes from extensional and compression of a cell-like suspension can affect the overall rheology.
Similar dynamics might be observed in other cell-like systems such as red blood cells or single-celled organisms, prompting additional study into time dependent flows for these systems.

\section{Acknowledgements}
This work was funded by the National Science Foundation (NSF) through grant CBET PMP 1704668 for C.M.S. and by a PPG-MRL Graduate Research Fellowship from the Illinois Materials Research Lab and Almar T. Widiger Fellowship for D.K.

\section*{Declaration of Interest}
The authors declare no conflict of interest.

\bibliography{ref}

\end{document}